\documentclass{article}
\usepackage{iclr2027_conference,times}
\usepackage[hyphens]{url}
\usepackage{hyperref}
\usepackage{fontawesome5}
\usepackage{graphicx}
\usepackage{caption}
\usepackage{algorithm}
\usepackage{algorithmic}
\usepackage{newfloat}
\usepackage{listings}
\usepackage[most]{tcolorbox}
\definecolor{promptblue}{HTML}{2B6F82}
\definecolor{promptbluebg}{HTML}{F1F7F8}
\newtcolorbox{promptbox}[1]{%
  enhanced,
  breakable,
  colback=promptbluebg,
  colframe=promptblue,
  colbacktitle=promptblue,
  coltitle=white,
  fonttitle=\bfseries,
  fontupper=\footnotesize,
  title={#1},
  title after break={#1 (continued)},
  boxrule=0.7pt,
  arc=2mm,
  outer arc=2mm,
  left=3mm,
  right=3mm,
  top=2mm,
  bottom=2mm,
  before skip=7pt,
  after skip=7pt}
\DeclareCaptionStyle{ruled}{labelfont=normalfont,labelsep=colon,strut=off}
\floatstyle{ruled}
\newfloat{listing}{tb}{lst}{}
\floatname{listing}{Listing}
\usepackage{booktabs}
\usepackage{colortbl}
\usepackage{hhline}
\usepackage{longtable}
\usepackage{placeins}
\usepackage{microtype}
\usepackage{siunitx}
\AtBeginDocument{\setlength{\baselineskip}{11pt plus 0.3pt}}

\newcommand{\agent}[1]{\textsc{#1}}

\definecolor{overallcolumnbg}{HTML}{E9F5F4}
\definecolor{tablegroupbg}{HTML}{F4F8F8}

\title{SoftwareGym: Scaling Computer-Use Agent Training with Diverse Synthesized Software Environments}
\iclrfinalcopy
\hypersetup{
  hidelinks,
  pdftitle={SoftwareGym: Scaling Computer-Use Agent Training with Diverse Synthesized Software Environments},
  pdfauthor={Tian Jin, Wenhao Wang, Jing Kang, Haoting Shi, Rui Ye, Yijun Lu, Siheng Chen}
}
\author{%
\makebox[\dimexpr\textwidth-2\tabcolsep\relax][c]{%
\begin{tabular}[t]{@{}c@{}}
\textbf{Tian Jin}$^{1,*}$ \quad
\textbf{Wenhao Wang}$^{2,*}$ \quad
\textbf{Jing Kang}$^{1}$ \quad
\textbf{Haoting Shi}$^{1}$ \quad
\textbf{Rui Ye}$^{1}$ \quad
\textbf{Yijun Lu}$^{1}$ \\
\textbf{Siheng Chen}$^{1,\dagger}$ \\[0.4em]
\normalfont $^{1}$Shanghai Jiao Tong University \\
\normalfont $^{2}$Zhejiang University \\[0.25em]
\normalfont\footnotesize $^{*}$Equal contribution. \quad $^{\dagger}$Corresponding author. \\[0.15em]
\normalfont\footnotesize Correspondence: \href{mailto:sihengc@sjtu.edu.cn}{\texttt{sihengc@sjtu.edu.cn}}
\end{tabular}%
}%
}

\begin{document}
\maketitle
\lhead{arXiv preprint}

\begin{abstract}
Computer-use agents (CUAs) are rapidly improving, yet synthesizing diverse and
reliable software environments at scale remains challenging. Native application
environments require substantial construction and integration effort, which
limits application coverage and increases the risk of application-specific
overfitting. Browser-hosted environments provide limited interaction depth and
workflow complexity. These limitations motivate \mbox{SoftwareGym}, an automated
pipeline for synthesizing GUI software environments with diverse interfaces,
executable functionality, and runtime-verified behavior. It generates
structurally diverse designs, realizes them as runnable applications, and uses
GUI evaluation with iterative repair to ensure functional reliability. Using
SoftwareGym, we synthesize 182 applications across twelve categories and achieve
a mean GUI task pass rate of 91.5\%. We construct SoftwareGym-Bench, a 120-task
evaluation suite spanning the same twelve categories, and use trajectories
collected from these environments for CUA training. The resulting model improves
Pass@3 from 41.39\% to 56.97\% on the strictly out-of-domain OSWorld benchmark
and achieves 40.0\% on SoftwareGym-Bench. These results demonstrate the
potential of synthesized software environments as scalable and transferable
supervision for CUA training.
\end{abstract}

\begingroup
\urlstyle{same}
\noindent\makebox[\textwidth][c]{%
\begin{tabular}{@{}c@{\hspace{0.6em}}l@{\hspace{1em}}l@{}}
\raisebox{-0.1em}{\faGithub} & \textbf{Code} &
\href{https://github.com/MrKiing-Jin/SoftwareGym}{\textcolor{blue}{\nolinkurl{https://github.com/MrKiing-Jin/SoftwareGym}}} \\[2pt]
\raisebox{-0.15em}{\includegraphics[height=1em]{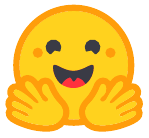}} & \textbf{Model} &
\href{https://huggingface.co/GothamLegend/SoftwareGym}{\textcolor{blue}{\nolinkurl{https://huggingface.co/GothamLegend/SoftwareGym}}}
\end{tabular}}%
\par
\endgroup

\section{Introduction}
\label{sec:intro}
Training capable computer-use agents (CUAs) requires interaction data from
executable software environments \citep{qin2025uitars}. To support generalization,
these environments must provide diverse, realistic, and reliable trajectories.
Constructing such environments at scale has therefore become a central bottleneck
for CUA training.

Current data synthesis either generates tasks and trajectories within existing
desktop and mobile applications
\citep{sun2025osgenesis,xie2026agentsynth,jung2026procua} or constructs
browser-hosted software environments
\citep{cao2026guigenesis,liu2026scalewob}. Native
applications preserve realistic states and complex workflows, but the cost of
integrating them into reproducible training environments limits application
coverage and increases the risk of application-specific overfitting.
Browser-hosted environments support scalable
deployment and state control, but their shared substrate limits interaction
depth and workflow complexity. Existing approaches do not jointly satisfy
three requirements for CUA training: interface and interaction
\textbf{diversity} to support generalization, functional
\textbf{reliability} to produce valid, complex trajectories,
and \textbf{scalability} to expand the environment pool with limited
engineering effort. Achieving these three requirements introduces three
corresponding challenges: interface and interaction diversity is difficult
because unconstrained generation within the same software category often
converges toward similar layouts and interaction patterns; functional
reliability requires every specified function to remain usable through complete
GUI workflows; scalability requires these diversity controls and reliability
checks to be applied across a large software pool, creating a verification
burden that manual development cannot sustain.

To address these challenges, we design \textbf{SoftwareGym}, an automated
multi-agent pipeline for GUI software synthesis. For diversity, SoftwareGym
creates category-level functional specifications and variant-specific interface
and interaction designs. For reliability, it combines source-level verification
with checkpoint-based GUI evaluation and repair; only applications that complete
both source-level and GUI verification enter the software library. For scalability, it automates the complete
process with minimal human input, enabling applications to be generated in
batches with substantially lower time and monetary costs than manual
development. To realize these goals, synthesis proceeds in
three stages. The design stage expands minimal human input into a detailed
functional specification, diverse UI blueprints, and corresponding interaction
schemes. The implementation stage converts each design package into an
executable application and verifies its source-level functionality. The
evaluation stage executes checkpoint-based GUI tests and guides iterative
implementation repair. These stages enable SoftwareGym to produce diverse and
functionally reliable software environments with realistic multi-step workflows
at scale.

\begin{figure*}[t]
\centering
\includegraphics[width=\textwidth]{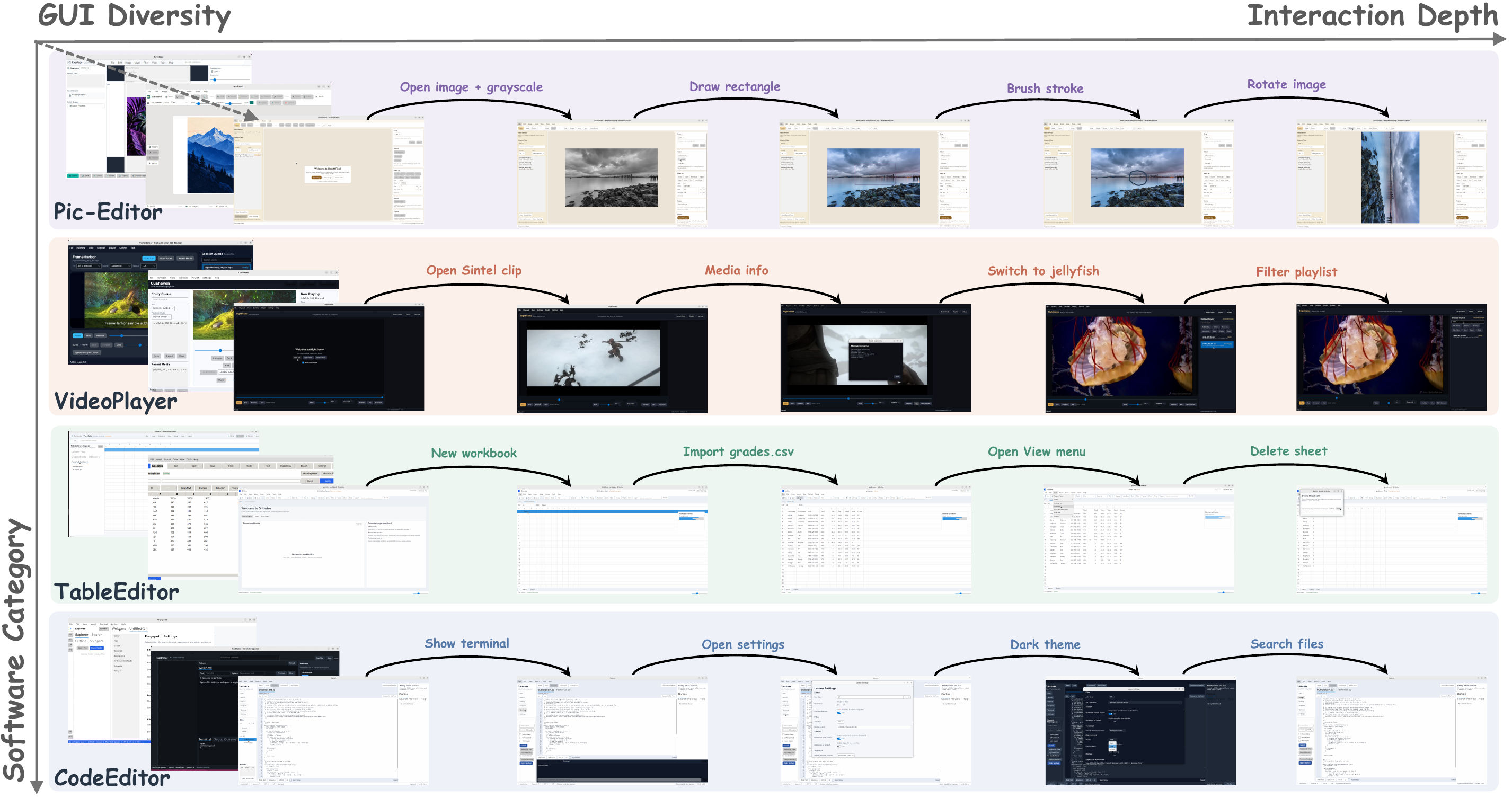}
\vspace{-4mm}
\caption{Representative applications synthesized by SoftwareGym. Pic-Editor,
VideoPlayer, TableEditor, and CodeEditor represent diverse software categories
and GUI designs. Each row traces a representative multi-step workflow,
demonstrating substantial interaction depth.}
\label{fig:launch-screens}
\end{figure*}

\noindent Using this pipeline, we construct a large, diverse, and functionally
reliable software library in substantially less time and at lower monetary cost
than manual development. We synthesize 182 runnable applications across twelve
categories, with a mean GUI task pass rate of 91.5\%.
Figure~\ref{fig:launch-screens} illustrates their functional depth across
representative software categories. We use trajectories collected from these applications to
train a model. Without using any OSWorld applications, tasks, or trajectories
for training, it improves Pass@3 from 41.39\% to 56.97\% on OSWorld,
demonstrating strong out-of-domain generalization. For complementary evaluation,
we evaluate the model on SoftwareGym-Bench, a dedicated 120-task test set
spanning the same twelve categories; it achieves 40.0\% Pass@3 and
outperforms several larger specialized computer-use agents. Our contributions
are threefold:
\begin{list}{\textbf{(\arabic{enumi})}}{%
  \usecounter{enumi}
  \setlength{\leftmargin}{2.4em}
  \setlength{\labelwidth}{1.8em}
  \setlength{\labelsep}{0.4em}
  \setlength{\topsep}{0pt}
  \setlength{\partopsep}{0pt}
  \setlength{\itemsep}{0pt}
  \setlength{\parsep}{0pt}}
\item We introduce SoftwareGym, an automated environment-synthesis
pipeline that produces functionally reliable and structurally diverse software
at scale, with substantially lower development time and monetary cost than
comparable manually developed applications.
\item We release 182 runnable applications across twelve categories and
construct SoftwareGym-Bench, a dedicated 120-task test set spanning the same
twelve categories.
\item Through extensive evaluations and ablations, we
validate the pipeline's key components and show that the model trained only on
SoftwareGym trajectories improves OSWorld Pass@3 by 15.58 points to 56.97\%.
\end{list}
\section{Related Work}
\label{sec:related}

\paragraph{Data Synthesis in Existing Software.}
Benchmarks such as OSWorld, Windows Agent Arena, and AndroidWorld first make
real desktop and mobile applications reproducibly accessible for CUA evaluation
\citep{xie2024osworld,bonatti2024windowsagentarena,rawles2024androidworld}.
Subsequent work synthesizes tasks and trajectories within existing desktop and
mobile applications
\citep{sun2025osgenesis,xie2026agentsynth,jung2026procua,
yang2025ultracua,xue2026evocua}. CUA-Universe and Gym-Anything further adapt
and configure existing software as reproducible agent environments
\citep{shi2026cuauniversescalabledynamicenvironment,aggarwal2026gymanything}.
Complementary systems refine online traces, collect human demonstrations, or
train over decentralized CUA data
\citep{qin2025uitars,wang2025opencua,wang2025fedmabench}.
Although these approaches preserve authentic
software states and complex workflows, their coverage remains constrained by
the availability and integration cost of existing applications. Repeated
training on this fixed software pool may also encourage application-specific
overfitting. SoftwareGym instead synthesizes new executable applications with
controllable functions, interfaces, and interaction logic, directly expanding
the software distribution.

\paragraph{Synthesis of Browser-Hosted Software Environments.}
InfiniteWeb and AutoWebWorld first establish scalable browser-hosted environments
through multi-page websites and state-machine-generated Web worlds
\citep{zhang2026infiniteweb,wu2026autowebworld}. GUI-GENESIS and ScaleWoB further
extend this approach toward software interaction by reconstructing real
applications or synthesizing software interfaces as backend-free webpages
\citep{cao2026guigenesis,liu2026scalewob}. Although these approaches simplify
deployment and state control, their shared browser/HTML/DOM substrate limits
native application structure and persistent workflows. SoftwareGym instead
synthesizes standalone desktop applications.

\section{Method}
\label{sec:method}

\begin{figure*}[t]
\centering
\includegraphics[width=\textwidth]{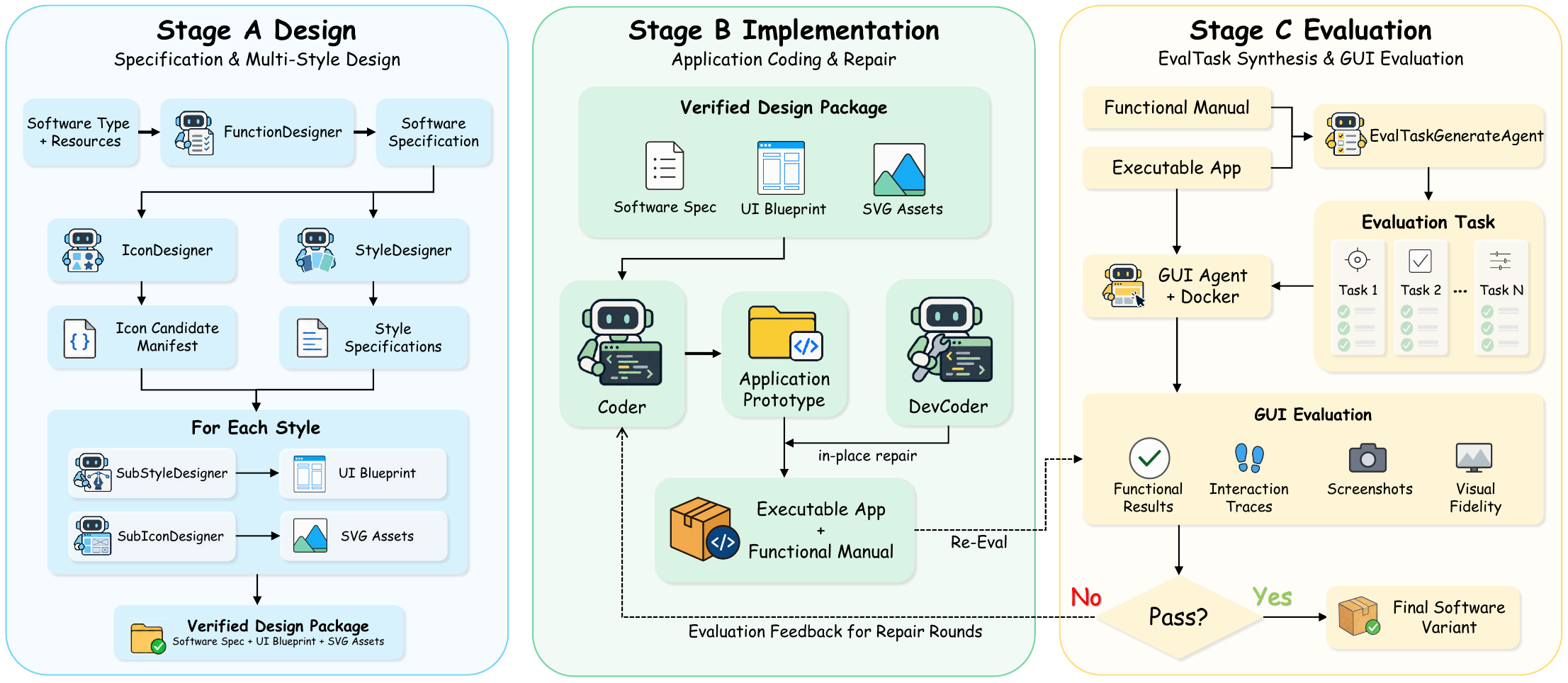}
\caption{SoftwareGym synthesizes diverse, reliable, and executable applications
through design, implementation, and evaluation. Design produces rich functional
specifications and distinct interfaces. Implementation realizes and verifies
the applications. Evaluation tests their GUI behavior and returns failures for
subsequent implementation repair.}
\label{fig:pipeline}
\end{figure*}

Figure~\ref{fig:pipeline} summarizes software synthesis in SoftwareGym.
\emph{Software Design}~(\S\ref{sec:design}) specifies application functions and
interfaces. \emph{Software Implementation and Evaluation}~
(\S\ref{sec:implementation}) then realizes each application and iteratively
repairs it using failures observed through GUI interaction. Applications that
complete this verification loop enter the software library. \emph{Task Synthesis}~
(\S\ref{sec:tasks}; Figure~\ref{fig:task-synthesis-example}) converts their
verified functions into executable GUI tasks.

\subsection{Software Design}
\label{sec:synth}
\label{sec:design}

The design stage produces rich, realistic functional specifications and
specifies diverse GUI structures and interaction patterns for each software
variant. Given a software
type and supporting resources, it produces a shared Software Specification and
multiple style-specific Verified Design Packages. Each package contains the
Software Specification, a UI Blueprint, and SVG Assets.

\label{sec:style}

\paragraph{Function Design.}
Function design specifies the purpose of an application and the detailed
functions it should provide. Given the Software Type and Resources,
\agent{FunctionDesigner} expands the initial description through a
four-step workflow. First, it establishes the functional baseline of a mature
product by identifying its target users, core objects, and expected operations,
such as creating, opening, editing, organizing, searching, recovering, and
saving content. Second, it extends each module with domain-specific professional
functions, including multi-condition retrieval, batch operations, history
management, persistent view settings, and input validation. Third, it connects
these functions into complete user workflows, from creating or opening content
to editing, organization, retrieval, and final save or export. The workflows
also cover cross-module state changes, dependent operations, failure handling,
and recovery. Finally, it records each function
in the Software Specification as an implementation and evaluation contract.
The contract specifies the user entry, user action, expected state update,
visible GUI response, and acceptance checkpoint. These steps provide functional
breadth and operational depth while making every function implementable and
directly verifiable through GUI interaction.

\paragraph{Style Diversification.}
Given a shared functional specification from \agent{FunctionDesigner}, this
stage generates multiple diverse, implementation-ready Avalonia UI blueprints.
These blueprints realize the same software functionality with different
interface and interaction designs. \agent{StyleDesigner} creates distinct
style specifications that define different product positions, information
architectures, interaction patterns, and visual identities. Each specification
follows a layout archetype that determines navigation, spatial organization,
object browsing, and action placement. To prevent variants from converging to
similar interfaces, every pair of variants must differ along at least two
layout dimensions.
Appendix~\ref{app:metric-definitions} formally defines this distance. Each style
specification is assigned to an independent \agent{SubStyleDesigner}. Each
\agent{SubStyleDesigner} converts its specification into concrete screen
structures and interaction flows. We compile every generated XAML file to
verify that it is valid and executable. To further ensure that the generated
variants are structurally different rather than merely cosmetically different,
we represent each blueprint as a normalized layout tree over windows, regions,
panels, and controls. We then compute pairwise tree-edit distance between these
trees. If the minimum pairwise distance falls below a predefined threshold, we
regenerate the least distinct blueprint.
Appendix~\ref{app:metric-definitions} defines the metric and its normalization.
These checks ensure that the retained blueprints are executable and structurally
distinct.

\paragraph{Icon and SVG Asset Generation.}
\agent{IconDesigner} derives a style-independent Icon Candidate Manifest from
the Software Specification, covering application, action, navigation, status,
content-type, entity, avatar, and illustration icons. For each interface
variant, \agent{SubStyleDesigner} assigns icon candidates to GUI slots, and
\agent{SubIconDesigner} renders them as style-specific SVG assets. Icons with
the same semantic role across variants are required to differ structurally
rather than only by color. A platform-side validator checks asset completeness,
SVG validity, and safety constraints. The resulting icon assets and slot
mappings are included in the Verified Design Package. Together, these
constraints preserve functional consistency while introducing visual diversity
across synthesized software variants.

\subsection{Software Implementation and Evaluation}
\label{sec:implementation}
\label{sec:evaluation}
\label{sec:function}
Software Implementation and Evaluation turns each Verified Design Package into
a runnable application and verifies that it satisfies the specified functional
and interface requirements. For each software variant, we implement the
application, evaluate its GUI behavior, and use observed failures to guide
iterative repair.

\paragraph{Application Implementation and Source-Code Verification.}
\agent{Coder} uses the Verified Design Package as the implementation
specification and realizes the complete Avalonia application, including its
structure, visual system, icons, and functionality. Each specified function is
implemented as a complete GUI workflow, covering user entry points, command
bindings, state updates, persistence, and visible feedback. \agent{Coder} also
records each function and its acceptance check in a functional manual.
\agent{DevCoder} then audits the implementation against this manual. It checks
command bindings, application logic, persistence, startup behavior, and file
interfaces, and repairs missing or incomplete implementations. In later
refinement rounds, \agent{Coder} and \agent{DevCoder} use GUI failure traces
and runtime evidence to perform localized repairs while preserving already
correct behavior. This process ensures that every specified function is backed
by executable logic and produces observable GUI behavior.

\paragraph{GUI Test Generation and Evaluation.}
\label{sec:verify}
\label{sec:gui-verification}
Source-code verification alone cannot confirm how an application behaves under
direct interaction. Evaluation therefore tests the running application
independently, covering both functional GUI behavior and visual fidelity to the
design package.
\agent{EvalTaskGenerateAgent} generates cases from the source-verified functional manual
produced during implementation. The cases cover control
reachability, navigation, object creation and editing, persistence, retrieval,
settings, dialogs, and file workflows. Each case begins from an explicit state,
uses only documented functions, and defines visible checkpoints. The case set is
fixed after the first implementation and reused across repair rounds.
\agent{GUIAgent} executes the cases with mouse and keyboard actions and records
the interaction trace, checkpoint evidence, final state, failure location, and
repair suggestion. Functional pass rate is
$P_{\mathrm{func}}=N_{\mathrm{pass}}/N_{\mathrm{all}}$, where
$N_{\mathrm{pass}}$ is the number of passed tasks and $N_{\mathrm{all}}$ is the
total number of generated cases. The evaluation report and failed trajectories
are returned to implementation for the next repair round. Evaluation stops when
all cases pass or the version budget is exhausted. This process grounds
implementation repair in reproducible GUI failures and progressively improves
functional reliability.

\paragraph{Design Fidelity.}
Functional success alone does not guarantee that a repaired application retains
its intended interface. Design fidelity therefore ensures that implementation
and source-code repair preserve the
interface identity of each software variant. After each GUI-evaluation round,
SoftwareGym computes a normalized visual-fidelity score between the implemented
application and its UI blueprint; its calculation is defined in
Appendix~\ref{app:metric-definitions}. When the score falls below a
pre-calibrated threshold, the
evaluator provides targeted feedback for repair. Structural checks also compare
the implemented layout and interface elements with the blueprint. This ensures
that implementation repair improves functionality without deviating from the
intended interface structure or visual design.

\begin{figure}[t]
\centering
\includegraphics[width=\textwidth,trim=26.71bp 65.30bp 26.71bp 26.22bp,clip]{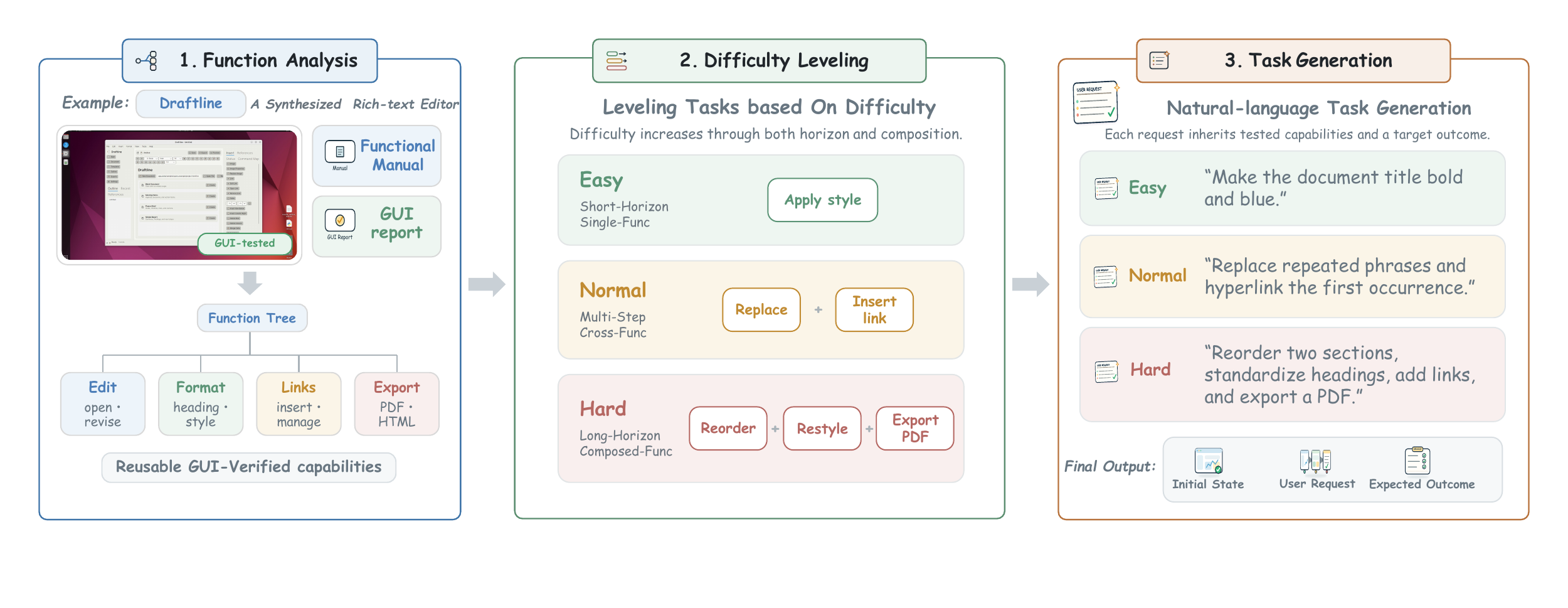}
\caption{SoftwareGym grounds every synthesized task in capabilities
verified through GUI interaction. It first extracts evidence-supported
capabilities, then assigns coherent combinations to difficulty- and
state-controlled slots, and finally realizes each slot as a natural request
with an explicit initial state and expected outcome.}
\label{fig:task-synthesis-example}
\end{figure}

\subsection{Task Synthesis}
\label{sec:tasks}

Task synthesis constructs GUI tasks using only application capabilities verified
through GUI evaluation. As shown in Figure~\ref{fig:task-synthesis-example}, it
consists of three steps: function analysis, difficulty leveling, and task
generation.

\paragraph{Function Analysis.}
Function analysis identifies the application capabilities that have been
verified through GUI interaction and can be used for task synthesis. It
constructs an evidence-grounded function tree for each retained software
variant. To build this tree, the pipeline extracts a functional manual
produced during implementation. It then
uses the GUI evaluation report to determine which functions are supported. Each leaf
records a task-level capability, its support status, GUI entry points,
constraints, observable state, and evaluation evidence. A capability is marked
\emph{supported} only when it cites a passed case, checkpoint, or interaction
step. Untested functions, failed functions, and blocked paths are excluded. This
evidence gate prevents task generation from relying on functions that have not
been demonstrated in the running application.

\paragraph{Difficulty Leveling.}
Difficulty leveling controls task complexity through dependencies among
verified capabilities. Tasks are assigned to three difficulty levels: easy,
normal, and hard.
Supported capabilities are selected according to the assigned difficulty. An
easy task has one primary goal with limited functional
composition. A normal task combines related objects or operations with a
necessary dependency between them. A hard task forms a multi-stage workflow in
which each stage establishes state required by the next stage. Each difficulty
level contains two task types. A file-based task starts with an existing input
file. A blank-state task starts without an input file and operates from the
application's default GUI state. All operations in a task must contribute to
the same user goal; unrelated or repeated actions cannot be added merely to
increase difficulty. These constraints produce
comparable difficulty levels while preserving natural, goal-directed workflows.

\paragraph{Task Generation.}
Finally, task generation converts each planned workflow into a natural user request and
an executable evaluation specification. It checks the selected capabilities
against the manual, passed GUI evidence, entry points,
state dependencies, persistence behavior, and completion feedback. The request
describes one user goal and the constraints required for evaluation. The
user-facing request does not expose coordinates, detailed operation paths,
capability identifiers, or evaluation mechanisms. The internal task specification records the initial
state, outcome criteria, process criteria, and expected final state.
Deterministic validation rejects unsupported capabilities, inconsistent initial
states, incompatible assets, and unreachable final states. This stage produces
natural-language tasks whose requirements remain feasible, observable, and
deterministically verifiable.

\FloatBarrier
\section{Experiments}
\label{sec:exp}

\newcommand{\mainresulttables}{%
\begin{table*}[t!]
\caption{Pass@3 (\%) on OSWorld. Within each model category, bold and
underlined scores denote the best and second-best results, respectively.}
\label{tab:main-osworld}
\centering
\fontsize{6.8}{7.8}\selectfont
\setlength{\tabcolsep}{1pt}
\setlength{\aboverulesep}{0pt}
\setlength{\belowrulesep}{0pt}
\begin{tabular*}{\textwidth}{@{\extracolsep{\fill}}l*{8}{c}>{\columncolor{overallcolumnbg}[1pt][0pt]}c@{}}
\toprule
\textbf{Model} & \multicolumn{1}{c}{Chrome} & \multicolumn{1}{c}{GIMP} &
\multicolumn{1}{c}{Calc} & \multicolumn{1}{c}{Impress} &
\multicolumn{1}{c}{Writer} & \multicolumn{1}{c}{Thunderbird} &
\multicolumn{1}{c}{VLC} & \multicolumn{1}{c}{VS Code} &
\multicolumn{1}{>{\columncolor{overallcolumnbg}[1pt][0pt]}c}{\textbf{Overall}} \\
\midrule
\multicolumn{10}{c}{\textbf{\textit{API Foundation Agents (Parameters Undisclosed)}}} \\
\midrule
\makebox[1.30em][c]{\raisebox{-0.18em}{\includegraphics[height=1.18em,width=1.18em,keepaspectratio]{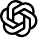}}}\hspace{0.25em}GPT-5.5 & 34.78 & 65.38 & 85.11 & \textbf{85.11} & 73.91 & 80.00 & \textbf{82.35} & 78.26 & 71.31 \\
\makebox[1.30em][c]{\raisebox{-0.18em}{\includegraphics[height=1.18em,width=1.18em,keepaspectratio]{Figures/model_logos/openai.pdf}}}\hspace{0.25em}GPT-5.4 & 60.87 & 76.92 & 82.98 & 70.21 & 73.91 & \underline{86.67} & \textbf{82.35} & 69.57 & 73.77 \\
\makebox[1.30em][c]{\raisebox{-0.18em}{\includegraphics[height=1.18em,width=1.18em,keepaspectratio]{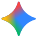}}}\hspace{0.25em}Gemini 3.1 Pro & 52.17 & \underline{84.62} & \underline{97.87} & 70.21 & \textbf{86.96} & 80.00 & \textbf{82.35} & \textbf{95.65} & \underline{79.10} \\
\makebox[1.30em][c]{\raisebox{-0.18em}{\includegraphics[height=1.18em,width=1.18em,keepaspectratio]{Figures/model_logos/gemini.pdf}}}\hspace{0.25em}Gemini 3 Flash & 50.00 & 57.69 & \textbf{100.00} & 68.09 & 65.22 & \underline{86.67} & \textbf{82.35} & 78.26 & 72.54 \\
\makebox[1.30em][c]{\raisebox{-0.18em}{\includegraphics[height=1.18em,width=1.18em,keepaspectratio]{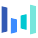}}}\hspace{0.25em}Seed 2.1 Pro & \textbf{71.74} & 73.08 & 93.62 & \underline{74.47} & 78.26 & \underline{86.67} & \underline{76.47} & \underline{82.61} & \textbf{79.51} \\
\makebox[1.30em][c]{\raisebox{-0.18em}{\includegraphics[height=1.18em,width=1.18em,keepaspectratio]{Figures/model_logos/bytedance.pdf}}}\hspace{0.25em}Seed 2.0 Pro & \textbf{71.74} & \textbf{92.31} & 91.49 & 65.96 & 69.57 & \textbf{93.33} & \textbf{82.35} & 78.26 & \underline{79.10} \\
\makebox[1.30em][c]{\raisebox{-0.18em}{\includegraphics[height=1.18em,width=1.18em,keepaspectratio]{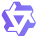}}}\hspace{0.25em}Qwen3.8-Flash & 60.87 & 80.77 & 89.36 & \underline{74.47} & 78.26 & 73.33 & \underline{76.47} & 60.87 & 74.59 \\
\makebox[1.30em][c]{\raisebox{-0.18em}{\includegraphics[height=1.18em,width=1.18em,keepaspectratio]{Figures/model_logos/qwen.pdf}}}\hspace{0.25em}Qwen3.7-Plus & \underline{67.39} & 69.23 & 63.83 & 68.09 & \underline{82.61} & 60.00 & \textbf{82.35} & 52.17 & 67.62 \\
\midrule
\multicolumn{10}{c}{\textbf{\textit{Large Open-Weight Agents ($\geq$32B Total)}}} \\
\midrule
\makebox[1.30em][c]{\raisebox{-0.18em}{\includegraphics[height=1.18em,width=1.18em,keepaspectratio]{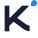}}}\hspace{0.25em}Kimi K2.5 & \textbf{58.70} & \textbf{92.31} & \underline{55.32} & \underline{65.96} & 47.83 & 73.33 & \underline{58.82} & \textbf{65.22} & \underline{63.52} \\
\makebox[1.30em][c]{\raisebox{-0.18em}{\includegraphics[height=1.18em,width=1.18em,keepaspectratio]{Figures/model_logos/qwen.pdf}}}\hspace{0.25em}Qwen3-VL-235B-A22B-Thinking & \textbf{58.70} & 50.00 & 42.55 & 63.83 & 65.22 & \underline{86.67} & \underline{58.82} & 39.13 & 56.15 \\
\makebox[1.30em][c]{\raisebox{-0.18em}{\includegraphics[height=1.18em,width=1.18em,keepaspectratio]{Figures/model_logos/qwen.pdf}}}\hspace{0.25em}Qwen3-VL-235B-A22B-Instruct & 47.83 & 50.00 & 36.17 & 44.68 & 56.52 & \underline{86.67} & 47.06 & 34.78 & 47.13 \\
\makebox[1.30em][c]{\raisebox{-0.18em}{\includegraphics[height=1.18em,width=1.18em,keepaspectratio]{Figures/model_logos/qwen.pdf}}}\hspace{0.25em}Qwen3-VL-32B-Instruct & 41.30 & 53.85 & 23.40 & 40.43 & \underline{69.57} & \underline{86.67} & 41.18 & 39.13 & 44.26 \\
\makebox[1.30em][c]{\raisebox{-0.18em}{\includegraphics[height=1.18em,width=1.18em,keepaspectratio]{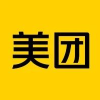}}}\hspace{0.25em}EvoCUA-32B & 50.00 & \underline{84.62} & \textbf{61.70} & \textbf{70.21} & \textbf{73.91} & \textbf{93.33} & \textbf{76.47} & \underline{60.87} & \textbf{67.62} \\
\makebox[1.30em][c]{\raisebox{-0.18em}{\includegraphics[height=1.18em,width=1.18em,keepaspectratio]{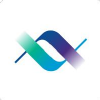}}}\hspace{0.25em}OpenCUA-32B & \underline{52.17} & \underline{84.62} & 27.66 & 44.68 & 43.48 & 60.00 & 52.94 & 47.83 & 48.77 \\
\midrule
\multicolumn{10}{c}{\textbf{\textit{Compact Open-Weight Agents (7--9B)}}} \\
\midrule
\makebox[1.30em][c]{\raisebox{-0.18em}{\includegraphics[height=1.18em,width=1.18em,keepaspectratio]{Figures/model_logos/longcat.pdf}}}\hspace{0.25em}EvoCUA-8B & \underline{54.35} & \textbf{88.46} & \textbf{44.68} & 42.55 & 52.17 & \textbf{80.00} & \textbf{58.82} & \textbf{65.22} & \underline{56.56} \\
\makebox[1.30em][c]{\raisebox{-0.18em}{\includegraphics[height=1.18em,width=1.18em,keepaspectratio]{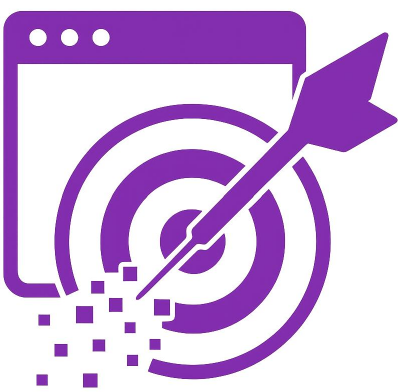}}}\hspace{0.25em}DART-GUI-7B & 43.48 & 76.92 & 23.40 & \textbf{55.32} & \underline{56.52} & \underline{73.33} & \underline{47.06} & \underline{60.87} & 50.41 \\
\makebox[1.30em][c]{\raisebox{-0.18em}{\includegraphics[height=1.18em,width=1.18em,keepaspectratio]{Figures/model_logos/opencua.pdf}}}\hspace{0.25em}OpenCUA-7B & 39.13 & 61.54 & 8.51 & 36.17 & 30.43 & 60.00 & 35.29 & 39.13 & 35.25 \\
\makebox[1.30em][c]{\raisebox{-0.18em}{\includegraphics[height=1.18em,width=1.18em,keepaspectratio]{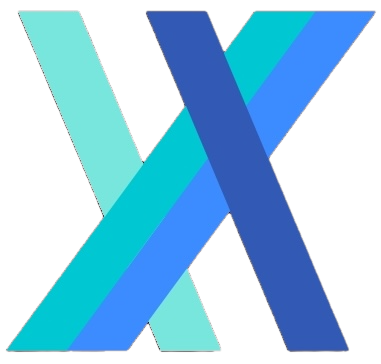}}}\hspace{0.25em}UI-TARS-1.5-7B & 26.09 & 42.31 & 12.77 & 36.17 & 43.48 & 40.00 & 41.18 & 52.17 & 33.20 \\
\addlinespace[1pt]
\makebox[1.30em][c]{\raisebox{-0.18em}{\includegraphics[height=1.18em,width=1.18em,keepaspectratio]{Figures/model_logos/qwen.pdf}}}\hspace{0.25em}Qwen3.5-9B Base & 28.26 & 53.85 & 27.66 & 48.94 & \underline{56.52} & 66.67 & \underline{47.06} & 30.43 & 41.39 \\
\makebox[1.30em][c]{\raisebox{-0.18em}{\includegraphics[height=1.18em,width=1.18em,keepaspectratio]{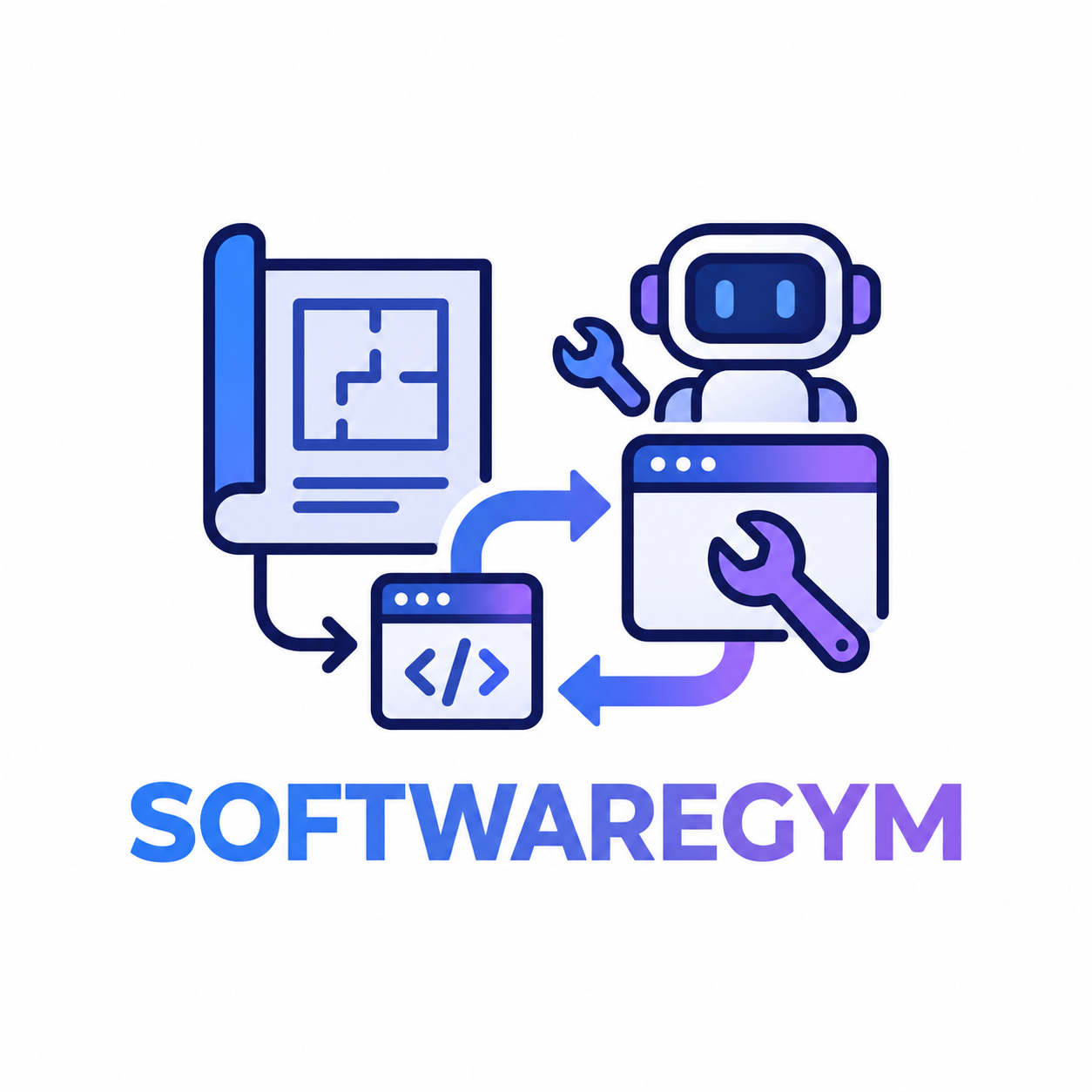}}}\hspace{0.25em}\textbf{Qwen3.5-9B SFT (Ours)} & \textbf{58.70} & \underline{84.62} & \underline{31.91} & \underline{53.19} & \textbf{65.22} & \underline{73.33} & \textbf{58.82} & \underline{60.87} & \textbf{56.97} \\
\bottomrule
\end{tabular*}
\end{table*}

\begin{table*}[t!]
\caption{Pass@3 (\%) on SoftwareGym-Bench. Within each model category, bold
and underlined scores denote the best and second-best results, respectively.}
\label{tab:main-softwaregym}
\centering
\fontsize{6.8}{7.8}\selectfont
\setlength{\tabcolsep}{1pt}
\setlength{\aboverulesep}{0pt}
\setlength{\belowrulesep}{0pt}
\begin{tabular*}{\textwidth}{@{\extracolsep{\fill}}l*{12}{c}>{\columncolor{overallcolumnbg}[1pt][0pt]}c@{}}
\toprule
\textbf{Model} & \multicolumn{1}{c}{Pic} & \multicolumn{1}{c}{Mail} &
\multicolumn{1}{c}{PDF} & \multicolumn{1}{c}{Music} &
\multicolumn{1}{c}{Code} & \multicolumn{1}{c}{Forum} &
\multicolumn{1}{c}{Writer} & \multicolumn{1}{c}{Browser} &
\multicolumn{1}{c}{Video} & \multicolumn{1}{c}{Slides} &
\multicolumn{1}{c}{Table} & \multicolumn{1}{c}{Chat} &
\multicolumn{1}{>{\columncolor{overallcolumnbg}[1pt][0pt]}c}{\textbf{Overall}} \\
\midrule
\multicolumn{14}{c}{\textbf{\textit{API Foundation Agents (Parameters Undisclosed)}}} \\
\midrule
\makebox[1.30em][c]{\raisebox{-0.18em}{\includegraphics[height=1.18em,width=1.18em,keepaspectratio]{Figures/model_logos/openai.pdf}}}\hspace{0.25em}GPT-5.5 & 70.0 & \underline{90.0} & \textbf{100.0} & \underline{50.0} & \textbf{90.0} & 70.0 & \textbf{60.0} & \underline{90.0} & \textbf{80.0} & 40.0 & \textbf{70.0} & \underline{70.0} & \underline{73.3} \\
\makebox[1.30em][c]{\raisebox{-0.18em}{\includegraphics[height=1.18em,width=1.18em,keepaspectratio]{Figures/model_logos/openai.pdf}}}\hspace{0.25em}GPT-5.4 & \underline{80.0} & \underline{90.0} & \underline{90.0} & \underline{50.0} & \underline{80.0} & \underline{80.0} & \underline{50.0} & \underline{90.0} & \underline{70.0} & \textbf{60.0} & \textbf{70.0} & \textbf{80.0} & \textbf{74.2} \\
\makebox[1.30em][c]{\raisebox{-0.18em}{\includegraphics[height=1.18em,width=1.18em,keepaspectratio]{Figures/model_logos/gemini.pdf}}}\hspace{0.25em}Gemini 3.1 Pro & \underline{80.0} & \textbf{100.0} & \underline{90.0} & \underline{50.0} & \underline{80.0} & \textbf{90.0} & \textbf{60.0} & 70.0 & \underline{70.0} & 40.0 & 50.0 & \underline{70.0} & 70.8 \\
\makebox[1.30em][c]{\raisebox{-0.18em}{\includegraphics[height=1.18em,width=1.18em,keepaspectratio]{Figures/model_logos/gemini.pdf}}}\hspace{0.25em}Gemini 3 Flash & \textbf{90.0} & \underline{90.0} & 70.0 & 40.0 & \underline{80.0} & \underline{80.0} & \textbf{60.0} & \textbf{100.0} & 60.0 & 40.0 & 50.0 & \underline{70.0} & 69.2 \\
\makebox[1.30em][c]{\raisebox{-0.18em}{\includegraphics[height=1.18em,width=1.18em,keepaspectratio]{Figures/model_logos/bytedance.pdf}}}\hspace{0.25em}Seed 2.1 Pro & 50.0 & 80.0 & 80.0 & \underline{50.0} & 70.0 & 70.0 & \underline{50.0} & 70.0 & 50.0 & \underline{50.0} & 30.0 & \underline{70.0} & 60.0 \\
\makebox[1.30em][c]{\raisebox{-0.18em}{\includegraphics[height=1.18em,width=1.18em,keepaspectratio]{Figures/model_logos/bytedance.pdf}}}\hspace{0.25em}Seed 2.0 Pro & 60.0 & \underline{90.0} & 80.0 & \underline{50.0} & \underline{80.0} & 50.0 & \textbf{60.0} & 70.0 & 50.0 & 30.0 & 30.0 & \textbf{80.0} & 60.8 \\
\makebox[1.30em][c]{\raisebox{-0.18em}{\includegraphics[height=1.18em,width=1.18em,keepaspectratio]{Figures/model_logos/qwen.pdf}}}\hspace{0.25em}Qwen3.8-Flash & 60.0 & 80.0 & 70.0 & \textbf{60.0} & \underline{80.0} & 50.0 & \textbf{60.0} & \underline{90.0} & \underline{70.0} & 30.0 & \underline{60.0} & \underline{70.0} & 65.0 \\
\makebox[1.30em][c]{\raisebox{-0.18em}{\includegraphics[height=1.18em,width=1.18em,keepaspectratio]{Figures/model_logos/qwen.pdf}}}\hspace{0.25em}Qwen3.7-Plus & 30.0 & 40.0 & 50.0 & 40.0 & 60.0 & 40.0 & 40.0 & 30.0 & 30.0 & 30.0 & 20.0 & 50.0 & 38.3 \\
\midrule
\multicolumn{14}{c}{\textbf{\textit{Large Open-Weight Agents ($\geq$32B Total)}}} \\
\midrule
\makebox[1.30em][c]{\raisebox{-0.18em}{\includegraphics[height=1.18em,width=1.18em,keepaspectratio]{Figures/model_logos/kimi.pdf}}}\hspace{0.25em}Kimi K2.5 & \textbf{70.0} & \textbf{90.0} & \textbf{90.0} & \textbf{40.0} & \textbf{80.0} & \textbf{60.0} & \textbf{50.0} & \textbf{70.0} & \textbf{60.0} & \textbf{30.0} & \textbf{40.0} & \textbf{70.0} & \textbf{62.5} \\
\makebox[1.30em][c]{\raisebox{-0.18em}{\includegraphics[height=1.18em,width=1.18em,keepaspectratio]{Figures/model_logos/qwen.pdf}}}\hspace{0.25em}Qwen3-VL-235B-A22B-Thinking & 30.0 & \underline{40.0} & 40.0 & \underline{30.0} & 40.0 & \underline{50.0} & \underline{20.0} & 20.0 & 30.0 & \underline{10.0} & \underline{30.0} & \underline{40.0} & 31.7 \\
\makebox[1.30em][c]{\raisebox{-0.18em}{\includegraphics[height=1.18em,width=1.18em,keepaspectratio]{Figures/model_logos/qwen.pdf}}}\hspace{0.25em}Qwen3-VL-235B-A22B-Instruct & 20.0 & 10.0 & 30.0 & 20.0 & 10.0 & 30.0 & 0.0 & 10.0 & 0.0 & \underline{10.0} & 0.0 & 0.0 & 11.7 \\
\makebox[1.30em][c]{\raisebox{-0.18em}{\includegraphics[height=1.18em,width=1.18em,keepaspectratio]{Figures/model_logos/qwen.pdf}}}\hspace{0.25em}Qwen3-VL-32B-Instruct & 10.0 & 10.0 & 10.0 & 0.0 & 20.0 & 0.0 & 0.0 & 10.0 & 20.0 & \underline{10.0} & 0.0 & \underline{40.0} & 10.8 \\
\makebox[1.30em][c]{\raisebox{-0.18em}{\includegraphics[height=1.18em,width=1.18em,keepaspectratio]{Figures/model_logos/longcat.pdf}}}\hspace{0.25em}EvoCUA-32B & \underline{40.0} & 20.0 & \underline{60.0} & \textbf{40.0} & \underline{60.0} & 30.0 & 10.0 & \underline{40.0} & \underline{40.0} & \textbf{30.0} & 10.0 & \underline{40.0} & \underline{35.0} \\
\makebox[1.30em][c]{\raisebox{-0.18em}{\includegraphics[height=1.18em,width=1.18em,keepaspectratio]{Figures/model_logos/opencua.pdf}}}\hspace{0.25em}OpenCUA-32B & 20.0 & 20.0 & 40.0 & 10.0 & 20.0 & 10.0 & 0.0 & 10.0 & 20.0 & \underline{10.0} & 10.0 & \underline{40.0} & 17.5 \\
\midrule
\multicolumn{14}{c}{\textbf{\textit{Compact Open-Weight Agents (7--9B)}}} \\
\midrule
\makebox[1.30em][c]{\raisebox{-0.18em}{\includegraphics[height=1.18em,width=1.18em,keepaspectratio]{Figures/model_logos/longcat.pdf}}}\hspace{0.25em}EvoCUA-8B & \underline{20.0} & \underline{30.0} & 20.0 & \underline{30.0} & \underline{20.0} & 20.0 & 0.0 & \underline{10.0} & \textbf{20.0} & 10.0 & \underline{10.0} & \textbf{40.0} & 19.2 \\
\makebox[1.30em][c]{\raisebox{-0.18em}{\includegraphics[height=1.18em,width=1.18em,keepaspectratio]{Figures/model_logos/dartgui.pdf}}}\hspace{0.25em}DART-GUI-7B & \underline{20.0} & 10.0 & 30.0 & \underline{30.0} & \underline{20.0} & 10.0 & 0.0 & 0.0 & \underline{10.0} & 10.0 & 0.0 & 10.0 & 12.5 \\
\makebox[1.30em][c]{\raisebox{-0.18em}{\includegraphics[height=1.18em,width=1.18em,keepaspectratio]{Figures/model_logos/opencua.pdf}}}\hspace{0.25em}OpenCUA-7B & 10.0 & 20.0 & 20.0 & \underline{30.0} & 10.0 & \underline{30.0} & 0.0 & 0.0 & 0.0 & \underline{20.0} & 0.0 & \underline{30.0} & 14.2 \\
\makebox[1.30em][c]{\raisebox{-0.18em}{\includegraphics[height=1.18em,width=1.18em,keepaspectratio]{Figures/model_logos/uitars.png}}}\hspace{0.25em}UI-TARS-1.5-7B & \underline{20.0} & 10.0 & 20.0 & 20.0 & 10.0 & 0.0 & 0.0 & 0.0 & \textbf{20.0} & 0.0 & 0.0 & 10.0 & 9.2 \\
\addlinespace[1pt]
\makebox[1.30em][c]{\raisebox{-0.18em}{\includegraphics[height=1.18em,width=1.18em,keepaspectratio]{Figures/model_logos/qwen.pdf}}}\hspace{0.25em}Qwen3.5-9B Base & \underline{20.0} & \underline{30.0} & \underline{40.0} & \underline{30.0} & 10.0 & \textbf{40.0} & \underline{20.0} & \underline{10.0} & \textbf{20.0} & \textbf{30.0} & \textbf{20.0} & \textbf{40.0} & \underline{25.8} \\
\makebox[1.30em][c]{\raisebox{-0.18em}{\includegraphics[height=1.18em,width=1.18em,keepaspectratio]{Figures/model_logos/softwaregym_new.png}}}\hspace{0.25em}\textbf{Qwen3.5-9B SFT (Ours)} & \textbf{40.0} & \textbf{40.0} & \textbf{60.0} & \textbf{50.0} & \textbf{70.0} & \textbf{40.0} & \textbf{30.0} & \textbf{60.0} & \textbf{20.0} & \underline{20.0} & \underline{10.0} & \textbf{40.0} & \textbf{40.0} \\
\bottomrule
\end{tabular*}
\end{table*}
}

\subsection{Experimental Setup}

\paragraph{Software Synthesis.}
We synthesize applications across twelve categories using Codex CLI agents with
GPT-5.5 for design, implementation, and evaluation-case generation. Each
application is refined for up to three versions ($v_1$--$v_3$), with the
$v_1$ GUI test suite held fixed across rounds and early stopping when all cases
pass. Seed 2.1 Pro evaluates each version through mouse-and-keyboard interaction
for up to 50 steps per case in isolated virtual machines running Ubuntu
22.04.3 LTS (x86\_64).

\paragraph{Training.}
After constructing the software library, we perform multimodal SFT on
Qwen3.5-9B~\citep{qwenteam2026qwen35} using trajectories collected only from
SoftwareGym applications. Appendix~\ref{app:trajectory-quality} describes
trajectory scoring and filtering, and Appendix~\ref{app:training-configuration}
provides the full training configuration.
\paragraph{SoftwareGym-Bench.}
SoftwareGym-Bench is a dedicated evaluation suite built on the SoftwareGym
application library. We select one application from each of twelve software
categories and ten manually verified tasks per application, yielding 120 tasks
in total. Each task is constructed to ensure that its target state is reachable
and programmatically verifiable. Tasks also include state-grounded distractors
and protected objects, requiring agents to reason about which objects to modify,
ignore, or preserve rather than simply follow a fixed sequence of operations.
Further construction details are provided in
Appendix~\ref{app:softwaregym-bench}.

\paragraph{Evaluation and Baselines.}
We evaluate the trained model and baselines on OSWorld and SoftwareGym-Bench.
OSWorld uses 244 tasks from eight application domains, whereas
SoftwareGym-Bench contains the 120 tasks described above. We evaluate GPT-5.5~\citep{openai2026gpt55},
GPT-5.4~\citep{openai2026gpt54}, Gemini 3.1 Pro
and Gemini 3 Flash~\citep{googledeepmind2026gemini31pro,googledeepmind2025gemini3flash},
Seed 2.0 Pro~\citep{bytedanceseed2026seed20}, Seed 2.1
Pro~\citep{bytedanceseed2026seed21}, Kimi
K2.5~\citep{kimiteam2026kimik25visualagentic},
Qwen3.8-Flash~\citep{alibabacloud2026qwen38flash},
Qwen3.7-Plus~\citep{alibabacloud2026qwen37plus},
Qwen3.5~\citep{qwenteam2026qwen35}, Qwen3-VL~\citep{bai2025qwen3vl},
EvoCUA~\citep{xue2026evocua}, DART-GUI~\citep{li2025dart},
OpenCUA~\citep{wang2025opencua}, and
UI-TARS~\citep{qin2025uitars}. We use the native context structure and tool protocol from
the OSWorld repository~\citep{xie2024osworld}. We set the temperature to 1 and the interaction budget
to 100 steps. Each task is evaluated with three independent rollouts. Complete
per-run results are provided in Appendix~\ref{app:main-per-run-results}.

\mainresulttables

\subsection{Main Result Analysis}
\paragraph{OSWorld Result Analysis.}
SoftwareGym-only supervision delivers strong out-of-domain transfer on the eight
OSWorld application domains included in our evaluation
(Table~\ref{tab:main-osworld}).
Without any OSWorld training data, it improves every evaluated domain and raises overall Pass@3 from
41.39\% to 56.97\%. It remains competitive with much larger open-source
baselines across the expanded comparison. This gain shows that reliable and structurally diverse synthesized
environments teach transferable GUI interaction rather than
application-specific procedures, complementing native software by broadening
the training distribution.

\paragraph{SoftwareGym-Bench Result Analysis.}
Complementing this out-of-domain result, SoftwareGym supervision supports
competitive in-domain learning at a compact model scale
(Table~\ref{tab:main-softwaregym}). It substantially improves the same 9B base
model, raising Pass@3 by 14.2 points from 25.8\% to 40.0\%. The resulting model
exceeds Qwen3.7-Plus and several substantially larger specialized CUAs.

\begin{figure*}[t!]
\centering
\setlength{\tabcolsep}{0pt}
\begin{tabular}{@{}p{0.56\textwidth}@{\hspace{0.02\textwidth}}p{0.42\textwidth}@{}}
\begin{minipage}[t]{\linewidth}
  \vspace{0pt}
  \centering
  \begin{minipage}[b]{0.56\linewidth}
    \centering
    \includegraphics[width=\linewidth,height=3.35cm,keepaspectratio]{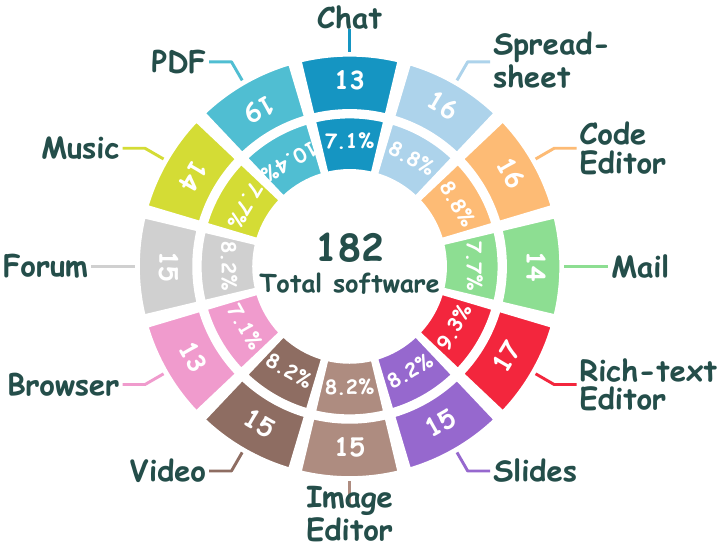}\\[-1pt]
    \textbf{(a)} Software distribution
  \end{minipage}\hfill
  \begin{minipage}[b]{0.43\linewidth}
    \centering
    \includegraphics[width=\linewidth,height=3.35cm,keepaspectratio]{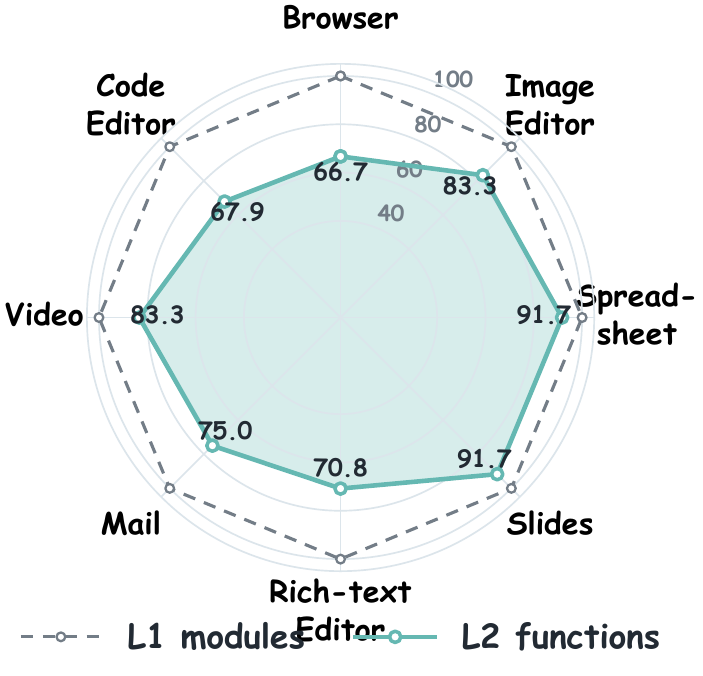}\\[-1pt]
    \textbf{(b)} Function coverage
  \end{minipage}
\end{minipage}
&
\begin{minipage}[t]{\linewidth}
  \vspace{0pt}
  \centering
  \includegraphics[width=\linewidth]{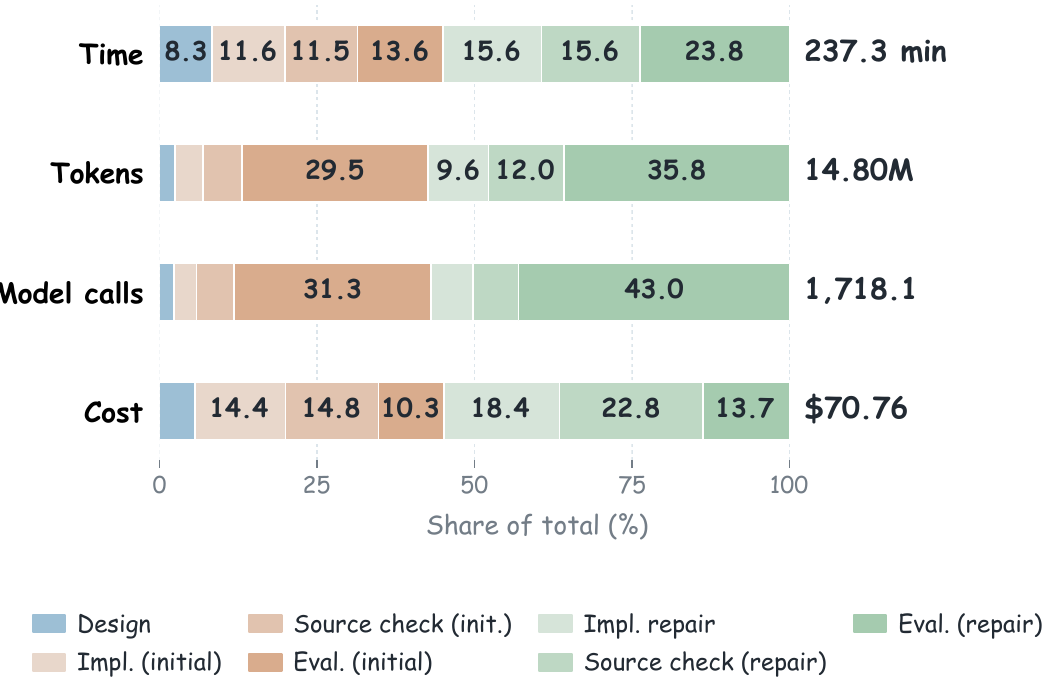}
\end{minipage}
\\
\begin{minipage}[t]{\linewidth}
  \vspace{0pt}
  \captionof{figure}{SoftwareGym scales breadth while preserving runtime-verified
  functionality. (a) 182 applications span twelve categories. (b) Across 121
  applications in eight aligned categories, GUI evidence covers all 49 major
  modules and 78.8\% of concrete native functions.}
  \label{fig:software-library}
\end{minipage}
&
\begin{minipage}[t]{\linewidth}
  \vspace{0pt}
  \captionsetup{justification=raggedright,singlelinecheck=false}
  \captionof{figure}{SoftwareGym makes quality-controlled synthesis practical
  at scale. Bars decompose per-application resource use across design,
  implementation, and evaluation.}
  \label{fig:gen-cost}
\end{minipage}
\end{tabular}
\par\vspace{2pt}
\includegraphics[width=\textwidth]{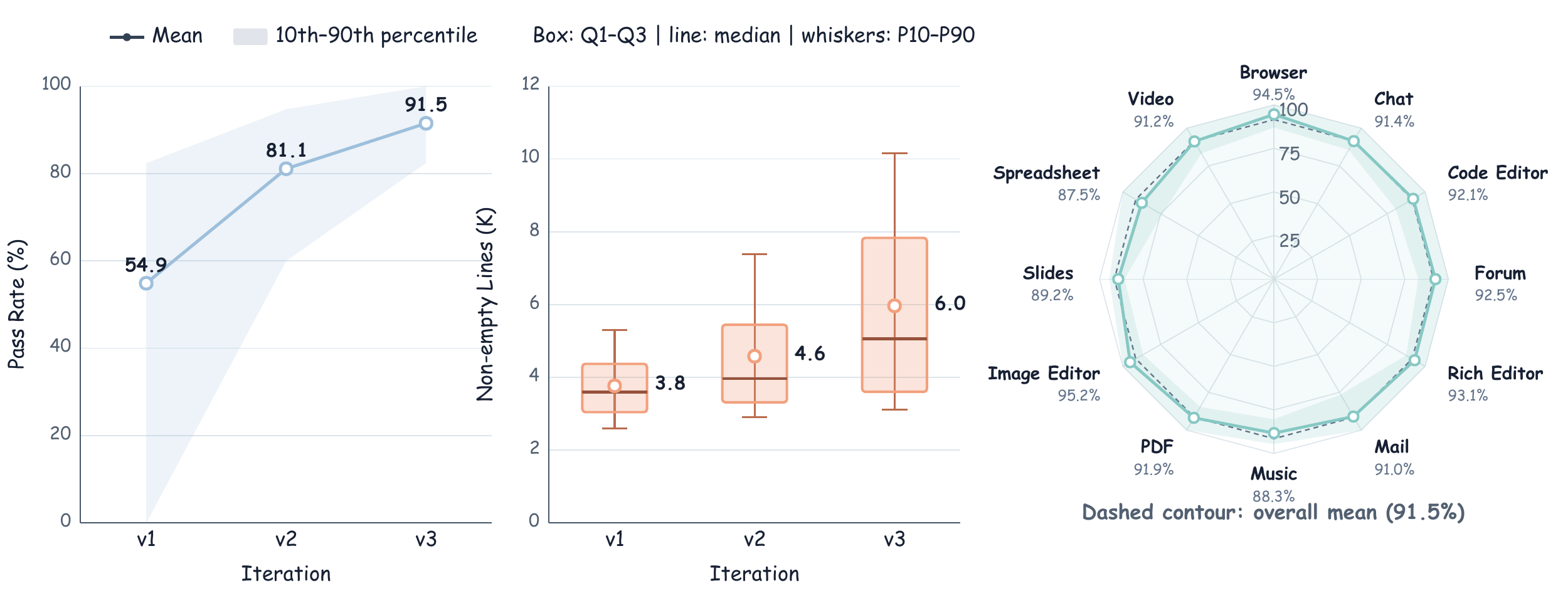}
\begin{minipage}{\textwidth}
\begin{tabular*}{\linewidth}{@{\extracolsep{\fill}}>{\centering\arraybackslash}p{0.31\linewidth}>{\centering\arraybackslash}p{0.31\linewidth}>{\centering\arraybackslash}p{0.31\linewidth}@{}}
\textbf{(a)} GUI pass-rate gains &
\textbf{(b)} Source-code growth &
\textbf{(c)} Category-level pass rates
\end{tabular*}
\end{minipage}
\captionof{figure}{Iterative evaluation and implementation repair consistently improve
software reliability across the synthesized application library.}
\label{fig:quality-iteration}
\end{figure*}
\FloatBarrier
\vspace{-10pt}

\paragraph{Software Synthesis Analysis.}
SoftwareGym scales the software library while preserving broad,
runtime-verified functionality. Its 182 runnable applications span twelve
categories (Figure~\ref{fig:software-library}(a)). Among 121 applications from
eight categories represented in OSWorld, GUI evaluations cover all 49 major
functional modules and 78.8\% of concrete functions on average
(Figure~\ref{fig:software-library}(b)). Appendix~\ref{app:metric-definitions}
defines the evidence criterion. This coverage demonstrates that SoftwareGym
expands software breadth while retaining substantial verified functionality.
We further characterize synthesis cost by sampling 20 applications. Each application
requires 237.3 minutes, 14.8 million tokens, 1,718 model calls, and an estimated
\$70.76 on average. Figure~\ref{fig:gen-cost} decomposes these resources across
pipeline stages, showing that automated synthesis substantially reduces the
cost of constructing software environments at scale.
Finally, iterative evaluation and repair improve software reliability across
the library. The mean GUI task pass rate increases from 54.9\% at $v_1$ to
81.1\% at $v_2$ and 91.5\% at $v_3$, with consistent gains across categories
(Figure~\ref{fig:quality-iteration}). This improvement demonstrates that
evaluation-guided repair progressively enhances implementation reliability.

\vspace{-6pt}
\subsection{Ablation Studies}
We next examine the contribution of three key components: structured style
design, GUI-guided implementation repair, and the source of training
trajectories.

\begin{figure}[t!]
\centering
\setlength{\tabcolsep}{1pt}
\begin{tabular}{@{}cc@{}}
\raisebox{-\height}{\includegraphics[width=0.495\textwidth]{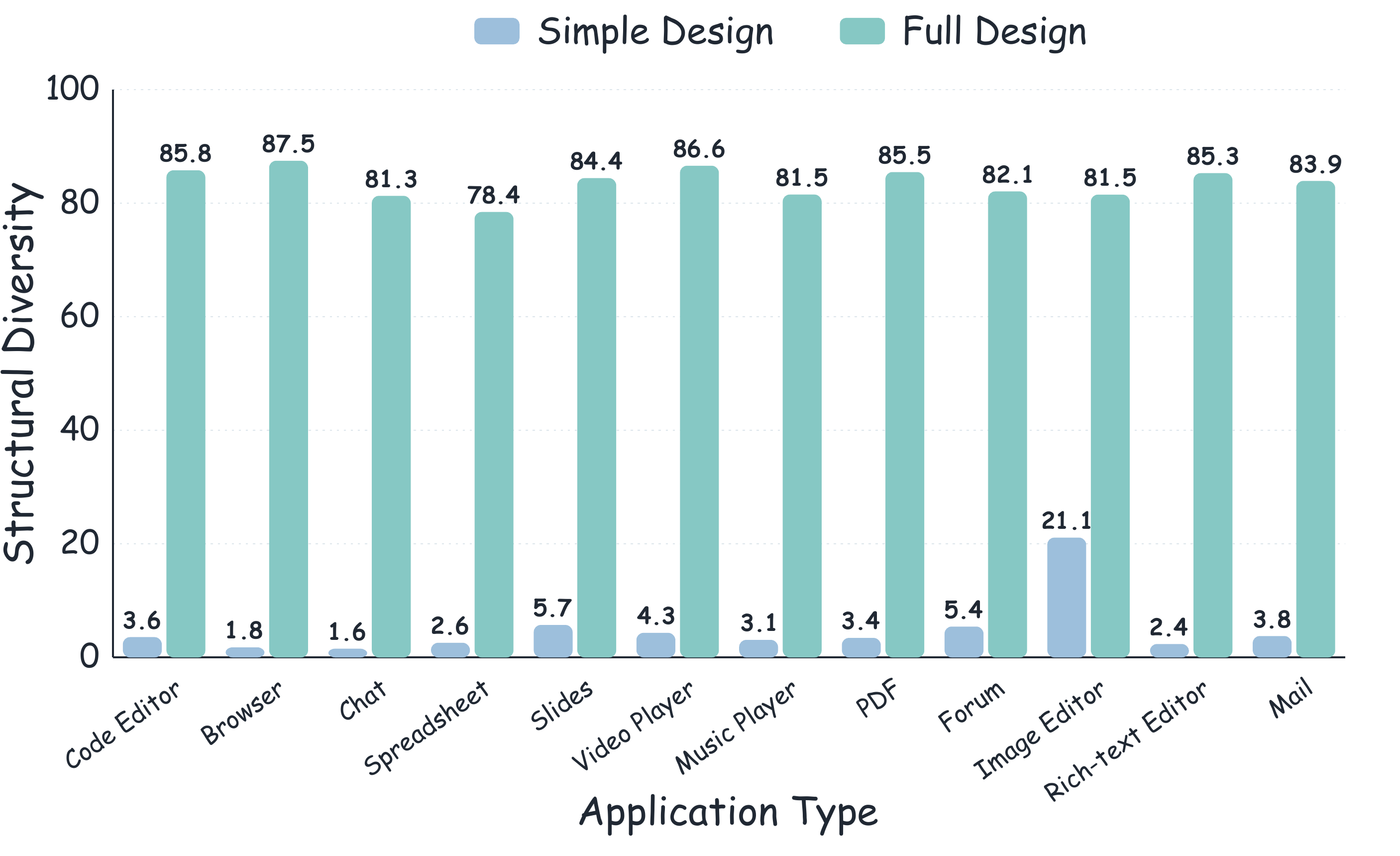}} &
\raisebox{-\height}{\includegraphics[width=0.495\textwidth]{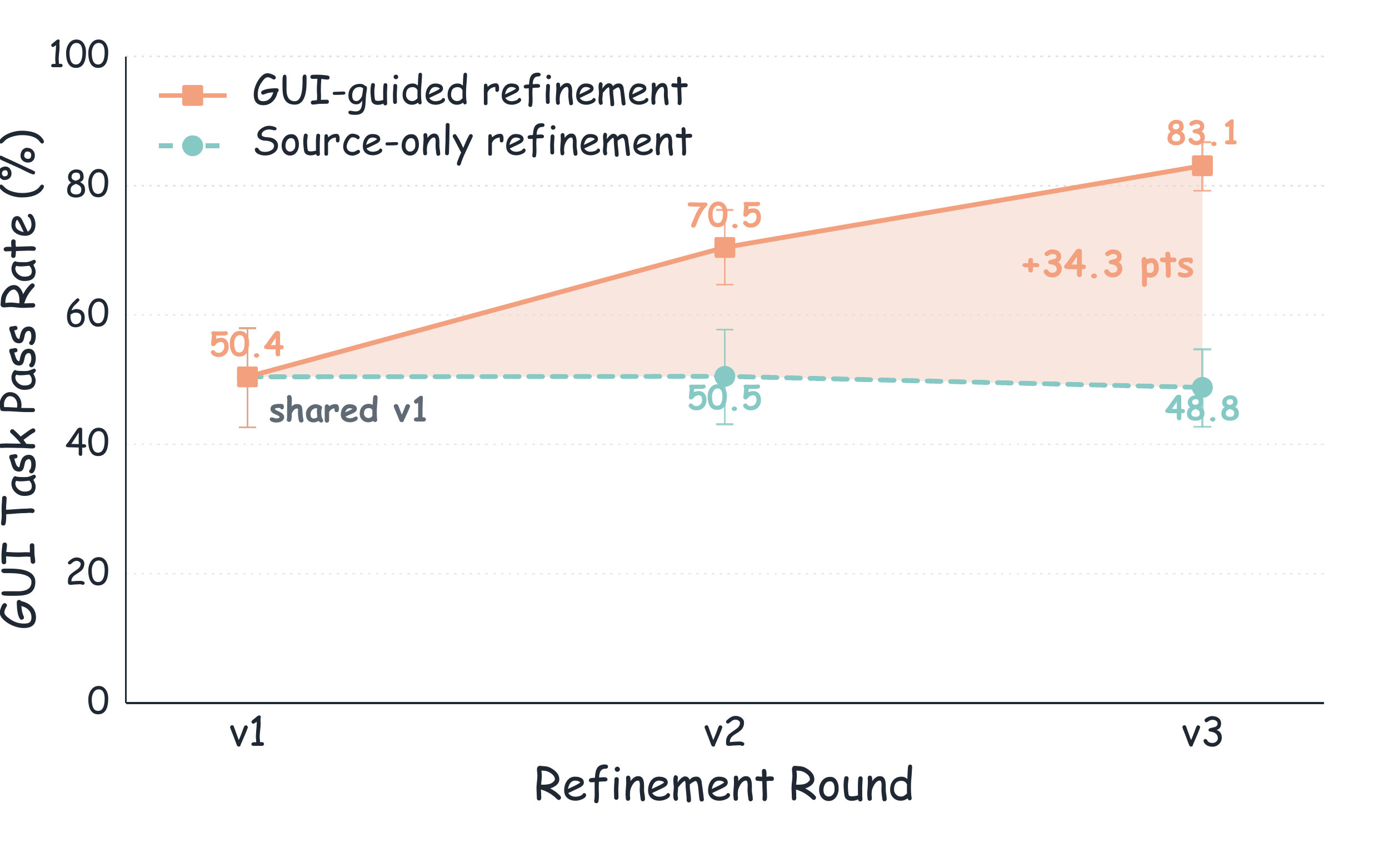}} \\
\parbox[t]{0.49\textwidth}{\centering\textbf{(a)} Full vs. simplified design} &
\parbox[t]{0.49\textwidth}{\centering\textbf{(b)} GUI-guided vs. source-only repair}
\end{tabular}
\caption{Structured style design improves blueprint diversity, while the GUI task pass
rate under GUI-guided repair rises across both rounds and remains near its start under
source-only repair.}
\label{fig:component-ablations}
\vspace{-4pt}
\end{figure}

\noindent\textbf{Style Design.}\quad
We compare Full Design with a simplified baseline under the same functional
requirements and measure structural diversity using normalized tree-edit
distance. Full Design improves the macro-average diversity score from 4.9 to
83.6 and achieves higher diversity across all categories
(Figure~\ref{fig:component-ablations}(a)). This result shows that explicit
style control prevents independently generated designs from collapsing to
similar interface structures.
Appendix~\ref{app:style-design-ablation} provides details.

\noindent\textbf{Evaluation-Guided Implementation Repair.}\quad
We compare GUI-guided repair with source-only repair under the same repair
process. Starting from the same initial pass rate of 50.4\%, GUI-guided repair
reaches 83.1\%, whereas source-only repair ends at 48.8\%
(Figure~\ref{fig:component-ablations}(b)). This result demonstrates that GUI
feedback provides essential runtime evidence for reliable implementation
repair. Appendix~\ref{app:gui-repair-ablation} provides details.

\noindent\textbf{Training Software Source.}\quad
We compare models trained with trajectories collected from SoftwareGym
applications, OSWorld applications, and their combination, while keeping the
model and evaluation protocol fixed. The model trained only on SoftwareGym
trajectories achieves 56.97\% on OSWorld and 40.0\% on SoftwareGym-Bench, while
combining SoftwareGym and OSWorld trajectories further improves
SoftwareGym-Bench performance to 47.5\%
(Table~\ref{tab:data-composition-ablation}). These results show that synthesized
software provides transferable supervision and complements native-software
data. Appendix~\ref{app:training-configuration} provides details.

\begin{table}[t!]
\caption{Pass@3 (\%) under different training-software sources on OSWorld and
SoftwareGym-Bench. Bold and underlined values denote the best and second-best
results in each column.}
\label{tab:data-composition-ablation}
\centering
\fontsize{6.5}{7.4}\selectfont
\setlength{\tabcolsep}{0.8pt}
\setlength{\aboverulesep}{0pt}
\setlength{\belowrulesep}{0pt}
\textbf{(a) OSWorld}\\[1pt]
\begin{tabular*}{\textwidth}{@{\extracolsep{\fill}}l*{8}{c}>{\columncolor{overallcolumnbg}[7pt][0pt]}c@{}}
\toprule
\textbf{Training data} & \multicolumn{1}{c}{Chrome} & \multicolumn{1}{c}{GIMP} &
\multicolumn{1}{c}{Calc} & \multicolumn{1}{c}{Impress} &
\multicolumn{1}{c}{Writer} & \multicolumn{1}{c}{Thunderbird} &
\multicolumn{1}{c}{VLC} & \multicolumn{1}{c}{VS Code} &
\multicolumn{1}{>{\columncolor{overallcolumnbg}[7pt][0pt]}c}{\textbf{Overall}} \\
\midrule
OSWorld & \underline{54.35} & \textbf{84.62} & \underline{44.68} & \textbf{68.09} & \textbf{69.57} & \textbf{73.33} & \textbf{64.71} & 52.17 & \textbf{61.48} \\
SoftwareGym & \textbf{58.70} & \textbf{84.62} & 31.91 & 53.19 & \underline{65.22} & \textbf{73.33} & \underline{58.82} & \textbf{60.87} & 56.97 \\
OSWorld + SoftwareGym & \textbf{58.70} & \underline{65.38} & \textbf{46.81} & \underline{59.57} & \underline{65.22} & \underline{66.67} & \textbf{64.71} & \underline{56.52} & \underline{58.61} \\
\bottomrule
\end{tabular*}\\[2pt]
\textbf{(b) SoftwareGym-Bench}\\[1pt]
\begin{tabular*}{\textwidth}{@{\extracolsep{\fill}}l*{12}{c}>{\columncolor{overallcolumnbg}[7pt][0pt]}c@{}}
\toprule
\textbf{Training data} & \multicolumn{1}{c}{Pic} & \multicolumn{1}{c}{Mail} &
\multicolumn{1}{c}{PDF} & \multicolumn{1}{c}{Music} &
\multicolumn{1}{c}{Code} & \multicolumn{1}{c}{Forum} &
\multicolumn{1}{c}{Writer} & \multicolumn{1}{c}{Browser} &
\multicolumn{1}{c}{Video} & \multicolumn{1}{c}{Slides} &
\multicolumn{1}{c}{Table} & \multicolumn{1}{c}{Chat} &
\multicolumn{1}{>{\columncolor{overallcolumnbg}[7pt][0pt]}c}{\textbf{Overall}} \\
\midrule
OSWorld & 20.0 & \underline{30.0} & \underline{60.0} & \underline{30.0} & 50.0 & 30.0 & \underline{20.0} & \underline{30.0} & \textbf{40.0} & \underline{30.0} & \underline{20.0} & \textbf{40.0} & 33.3 \\
SoftwareGym & \underline{40.0} & \textbf{40.0} & \underline{60.0} & \textbf{50.0} & \underline{70.0} & \underline{40.0} & \textbf{30.0} & \textbf{60.0} & \underline{20.0} & 20.0 & 10.0 & \textbf{40.0} & \underline{40.0} \\
OSWorld + SoftwareGym & \textbf{50.0} & \textbf{40.0} & \textbf{80.0} & \textbf{50.0} & \textbf{80.0} & \textbf{50.0} & 10.0 & \textbf{60.0} & \underline{20.0} & \textbf{40.0} & \textbf{50.0} & \textbf{40.0} & \textbf{47.5} \\
\bottomrule
\end{tabular*}
\end{table}

\vspace{-6pt}
\section{Conclusion}
\label{sec:conclusion}
We introduced SoftwareGym, an automated design--implementation--evaluation
pipeline that synthesizes diverse and functionally reliable desktop applications
at scale. Its library contains 182 applications across twelve categories and
achieves a mean GUI task pass rate of 91.5\%; ablations show the value of
structured design and evaluation-guided repair.
Trajectories collected from these environments provide transferable supervision
beyond synthesized software. Without any OSWorld application data for training,
it improves Pass@3 from 41.39\% to 56.97\% on OSWorld and reaches 40.0\% on
SoftwareGym-Bench, outperforming several larger specialized agents. SoftwareGym
therefore provides a rich, scalable complement to native software, broadening
the interfaces, interaction patterns, and workflows available for CUA training.

\clearpage
\begingroup
\renewcommand{\baselinestretch}{0.99}\selectfont
\setlength{\bibsep}{0pt}
\bibliography{softwaregym_references}
\bibliographystyle{iclr2027_conference}
\endgroup

\clearpage
\appendix
\raggedbottom
\setlength{\parskip}{6pt}
\setlength{\baselineskip}{11pt}
\newcommand{\metricdefinitionsappendix}{%
\section{Definitions and Computation of Reported Measures}
\label{app:metric-definitions}

\paragraph{Layout-Archetype Distance.}
Layout-archetype distance quantifies how differently two style variants
organize interaction at a high level. For layout archetypes $L_i$ and $L_j$,
we define their distance as the Hamming distance~\citep{hamming1950error} over
navigation, spatial organization, object browsing, and action placement:
\[
d_{\mathrm{layout}}(L_i,L_j)=
\sum_{k\in\{\mathrm{nav},\mathrm{space},\mathrm{browse},\mathrm{action}\}}
\mathbf{1}[\ell_{i,k}\neq\ell_{j,k}].
\]
Thus, $d_{\mathrm{layout}}(L_i,L_j)\geq2$ requires the two archetypes to differ
in at least two layout dimensions. Enforcing this constraint in SoftwareGym
separates style variants before detailed blueprint generation and reduces
convergence toward the same interface organization.

\paragraph{Structural Blueprint Diversity.}
Structural blueprint diversity tests whether generated interfaces differ in
their component hierarchies rather than only in wording or decoration. We
convert each blueprint into a normalized layout tree whose nodes represent
windows, regions, panels, and interactive controls and whose edges represent
containment. Text, colors, fonts, and other visual attributes are removed.
Let $T_i$ and $T_j$ be two such trees, and let $|T|$ denote the number of nodes
in a tree. Using unit costs for node
insertion, deletion, and substitution~\citep{zhang1989treeedit}, we define
\[
d_{ij}=\min\!\left\{1,
\frac{\operatorname{TED}(T_i,T_j)}
{\max(|T_i|,|T_j|)}\right\}.
\]
The resulting distance lies in $[0,1]$, with larger values indicating greater
structural difference. For a set of $K$ blueprints, we summarize overall
structural diversity on a 0--100 scale as
\[
D_{\mathrm{struct}}=\frac{200}{K(K-1)}
\sum_{1\leq i<j\leq K}d_{ij}.
\]
We use $0.60$ as the pipeline's structural-diversity
gate: if $\min_{i\ne j}d_{ij}<0.60$, we regenerate the blueprint with the
smallest average distance to the remaining blueprints. This gate removes
near-duplicate structures and ensures that the retained variants provide
meaningfully different interface organizations for implementation.

\paragraph{Visual Fidelity.}
Visual fidelity measures whether an implemented application preserves the
design tokens specified by its UI blueprint during iterative repair. Let $C_b$
and $C_a$ denote the sets of color values extracted from the UI blueprint and
implemented application, and let $R_b$ and $R_a$ denote their corner-radius
sets. We use the Jaccard similarity
$J(A,B)=|A\cap B|/|A\cup B|$, with $J(\emptyset,\emptyset)=1$, and define
$F_{\mathrm{pal}}=J(C_b,C_a)$ and $F_{\mathrm{rad}}=J(R_b,R_a)$.
$F_{\mathrm{pri}}$ and $F_{\mathrm{font}}$ are binary indicators of whether the
blueprint's primary color and main font are retained in the application.
$F_{\mathrm{bind}}=N_{\mathrm{res}}/(N_{\mathrm{res}}+N_{\mathrm{inline}})$,
where $N_{\mathrm{res}}$ counts resource-based visual references and
$N_{\mathrm{inline}}$ counts inline color values; the ratio is set to 1 when its
denominator is zero. The primary-color, corner-radius, and font components are
set to 1 when the corresponding blueprint property is unspecified. The overall
score is
\[
F_{\mathrm{vis}}=0.40F_{\mathrm{pal}}+0.15F_{\mathrm{pri}}+
0.20F_{\mathrm{rad}}+0.10F_{\mathrm{font}}+0.15F_{\mathrm{bind}}.
\]
This is a design-token implementation gate, not a perceptual image-similarity
measure. We use a pre-calibrated threshold of $0.80$, so the gate passes when
$F_{\mathrm{vis}}\geq0.80$ and the primary color is
retained. When an implementation does not satisfy this gate, the evaluator
identifies the visual tokens that deviate from the blueprint and returns those
mismatches as concrete repair targets. Subsequent repair can therefore correct
functional issues without altering the application's intended visual identity.

\paragraph{Functional Coverage.}
Functional coverage measures how comprehensively synthesized applications
reproduce the functions available in their corresponding real-world applications. For
each synthesized application $a$, $R_a^{(2)}$ contains frozen-taxonomy
functions verified by at least one passed GUI evaluation case with a directly
relevant successful checkpoint; claims in the functional manual alone do not count.
$R_a^{(1)}$ contains the parent modules, so a level-1 module is covered when at
least one child function has such evidence. For category $c$ and level $l$,
coverage is
$C_c^{(l)}=|F_{c,\mathrm{native}}^{(l)}\cap
(\bigcup_{a\in\mathcal{A}_c}R_a^{(l)})|/
|F_{c,\mathrm{native}}^{(l)}|$, where $\mathcal{A}_c$ contains the
release-selected applications in category $c$. Related but insufficient
evidence is excluded. We report the unweighted category macro-average as the
primary measure and the pooled micro-average as a supplementary statistic. We
use this measure to verify that expanding the synthesized library preserves
broad, directly exercised functionality from corresponding real-world software.
}

\onecolumn
\section{SoftwareGym-Bench: Construction and Task Examples}
\label{app:softwaregym-bench}

\subsection{Benchmark Construction}
\label{sec:benchmark}

\paragraph{Outcome-Driven Task Construction.}
We construct each task backward from a desired final state. We first define the
target outcome and then identify the key semantic steps required to reach it.
The initial state is built to support these steps and their dependencies. This
construction ensures that every task has a reachable solution path and a
well-defined terminal state.

\paragraph{State-Grounded Task Difficulty.}
We introduce distractors and protected objects into the initial state to make
task completion require state-grounded decisions. Some objects must be
modified, some ignored, and some preserved. The correct action therefore
depends on the current application state rather than on a fixed sequence of
interface operations. This design increases task difficulty while preserving a
coherent user goal.

\paragraph{Programmatic Evaluation.}
Each task is paired with a reproducible initializer and a programmatic
evaluator. The initializer restores the task to the same initial state before
each attempt. The evaluator checks the resulting application state and any
produced artifacts against the target outcome. This makes task success
deterministically verifiable and results comparable across models and runs.

To make the benchmark and its evaluation protocol concrete, we present
representative successful trajectories across software categories.

\noindent\textbf{Instruction:} Please replace the ``Draft total'' entry in the
active workbook's \texttt{Sheet1} cell \texttt{B2} with ``Final total,'' and
leave the Find and Replace panel open.

\begin{figure}[H]
\centering
\begin{minipage}[t]{0.235\textwidth}
  \centering
  \includegraphics[width=\linewidth]{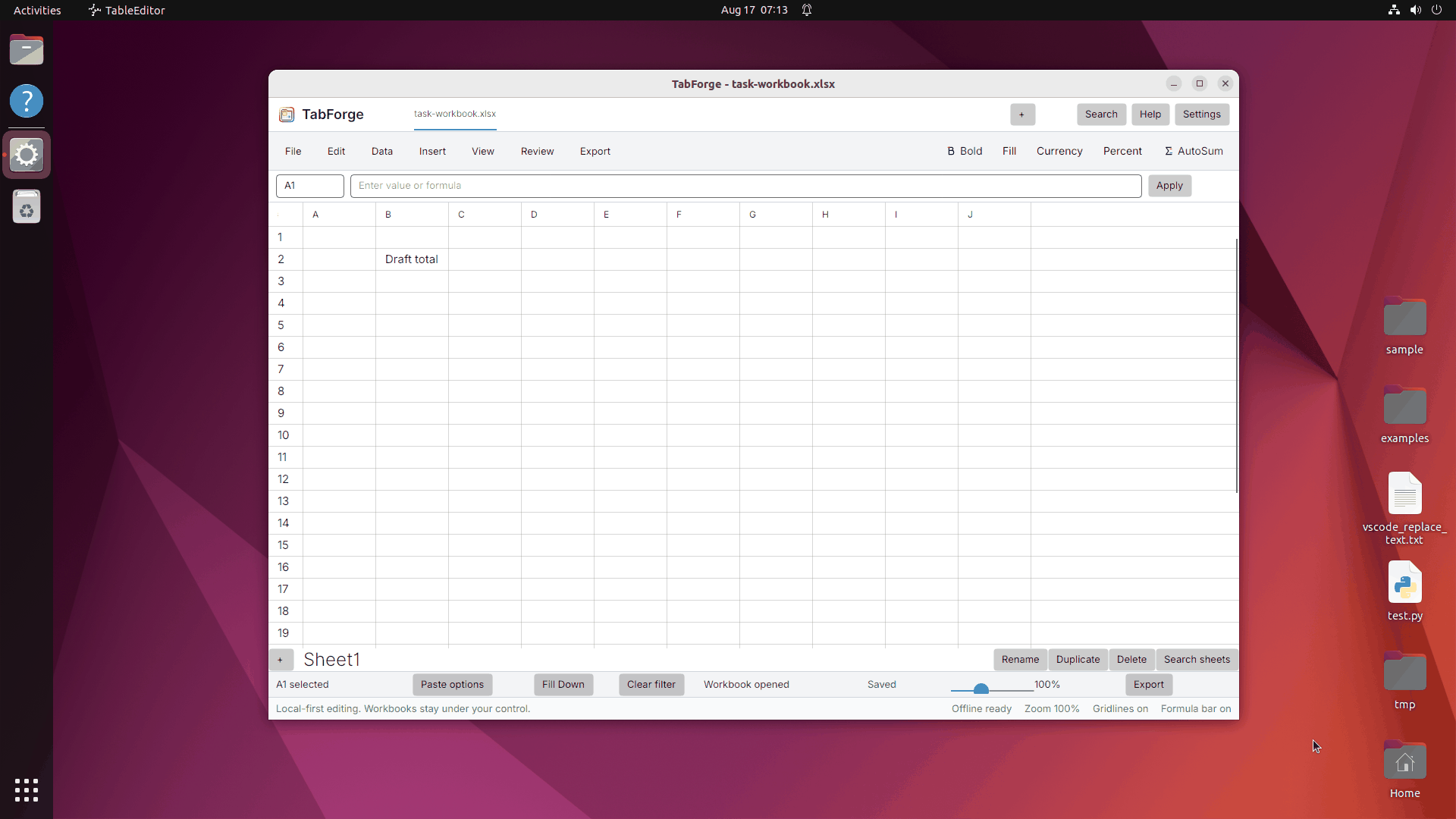}
  \par\vspace{0.6ex}

  {\scriptsize\textbf{Step 1:} Locate ``Draft total''.\par}
\end{minipage}\hspace{0.012\textwidth}
\begin{minipage}[t]{0.235\textwidth}
  \centering
  \includegraphics[width=\linewidth]{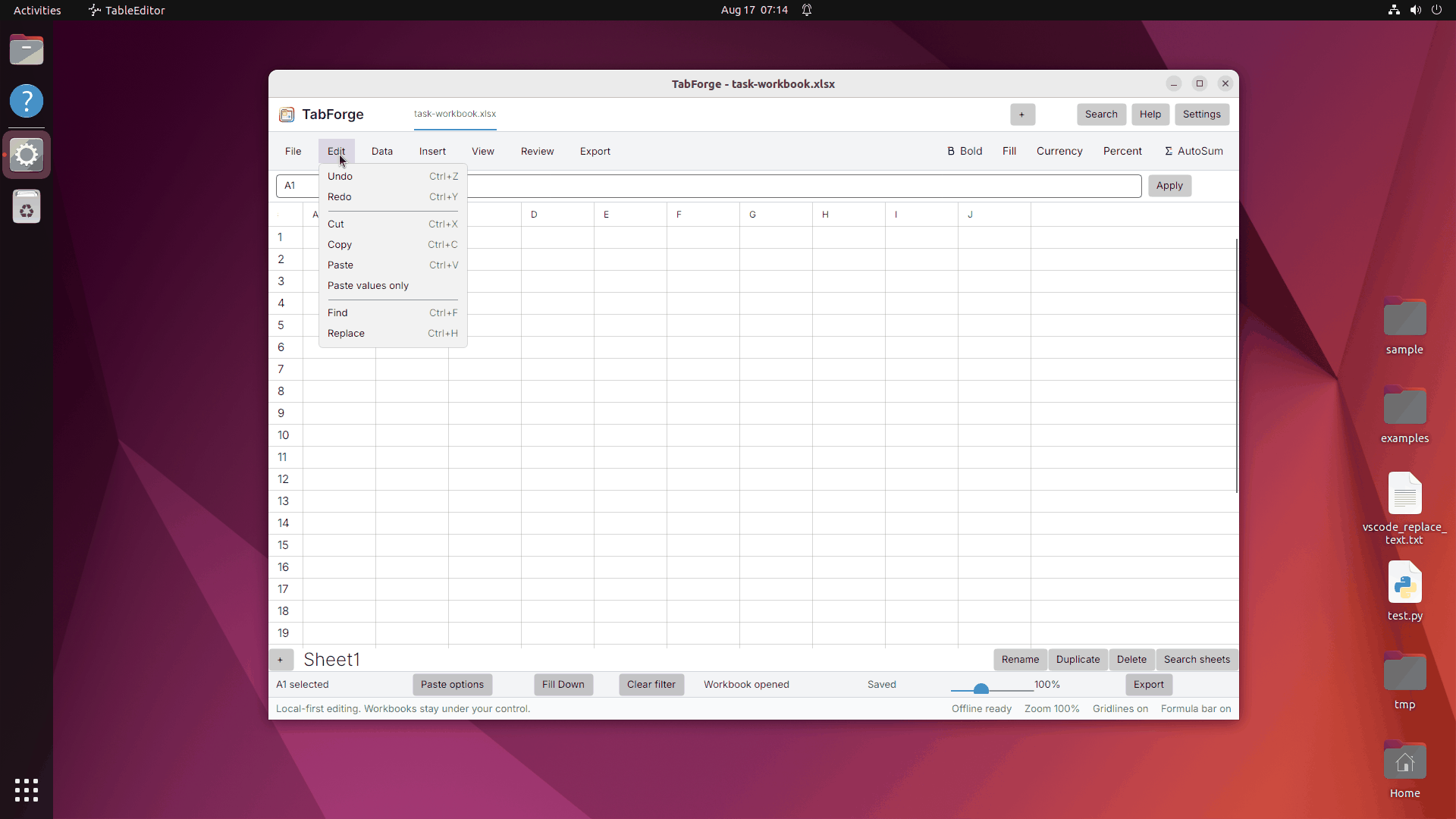}
  \par\vspace{0.6ex}

  {\scriptsize\textbf{Step 2:} Open Edit.\par}
\end{minipage}\hspace{0.012\textwidth}
\begin{minipage}[t]{0.235\textwidth}
  \centering
  \includegraphics[width=\linewidth]{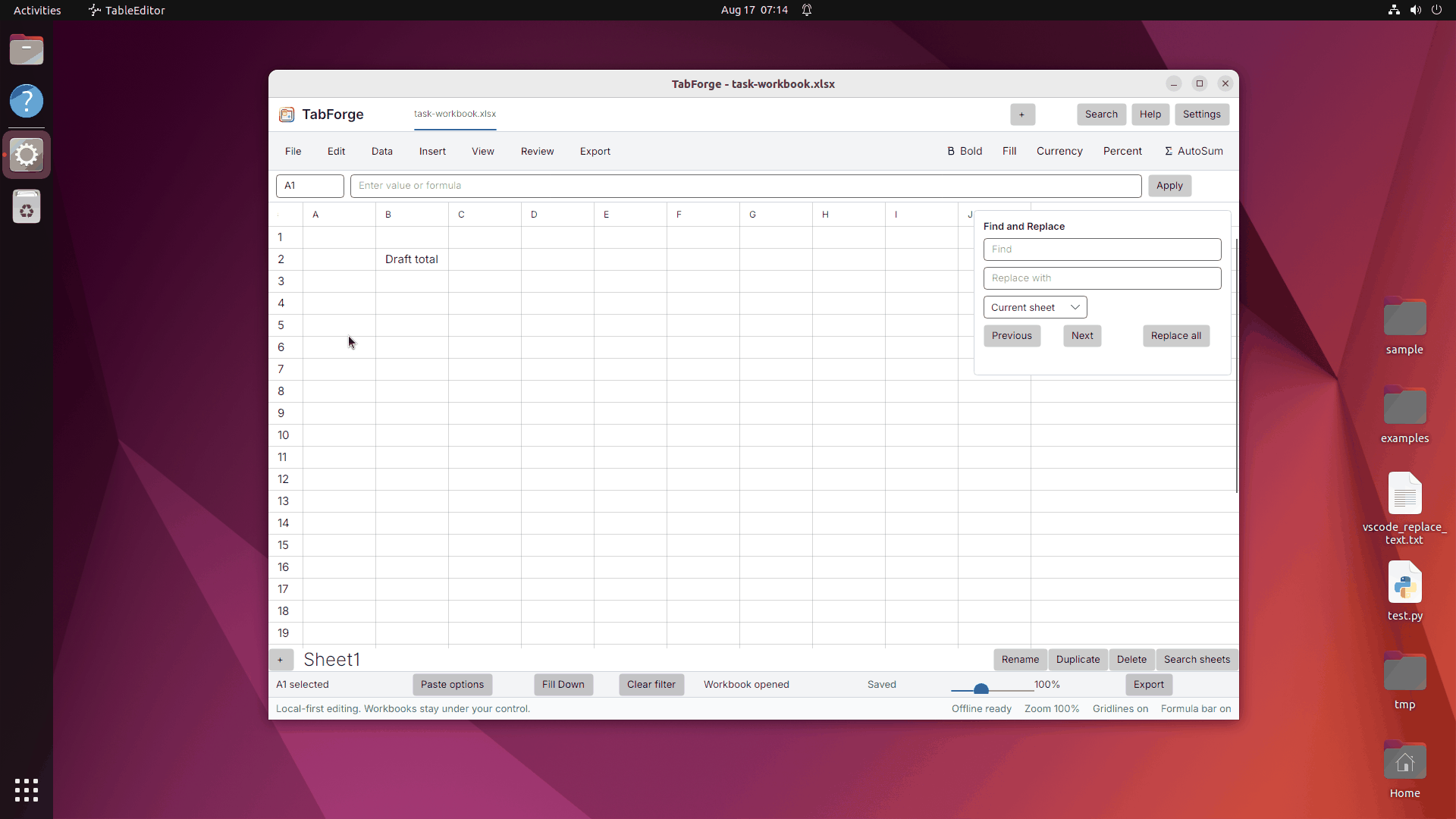}
  \par\vspace{0.6ex}

  {\scriptsize\textbf{Step 3:} Open Find and Replace.\par}
\end{minipage}\hspace{0.012\textwidth}
\begin{minipage}[t]{0.235\textwidth}
  \centering
  \includegraphics[width=\linewidth]{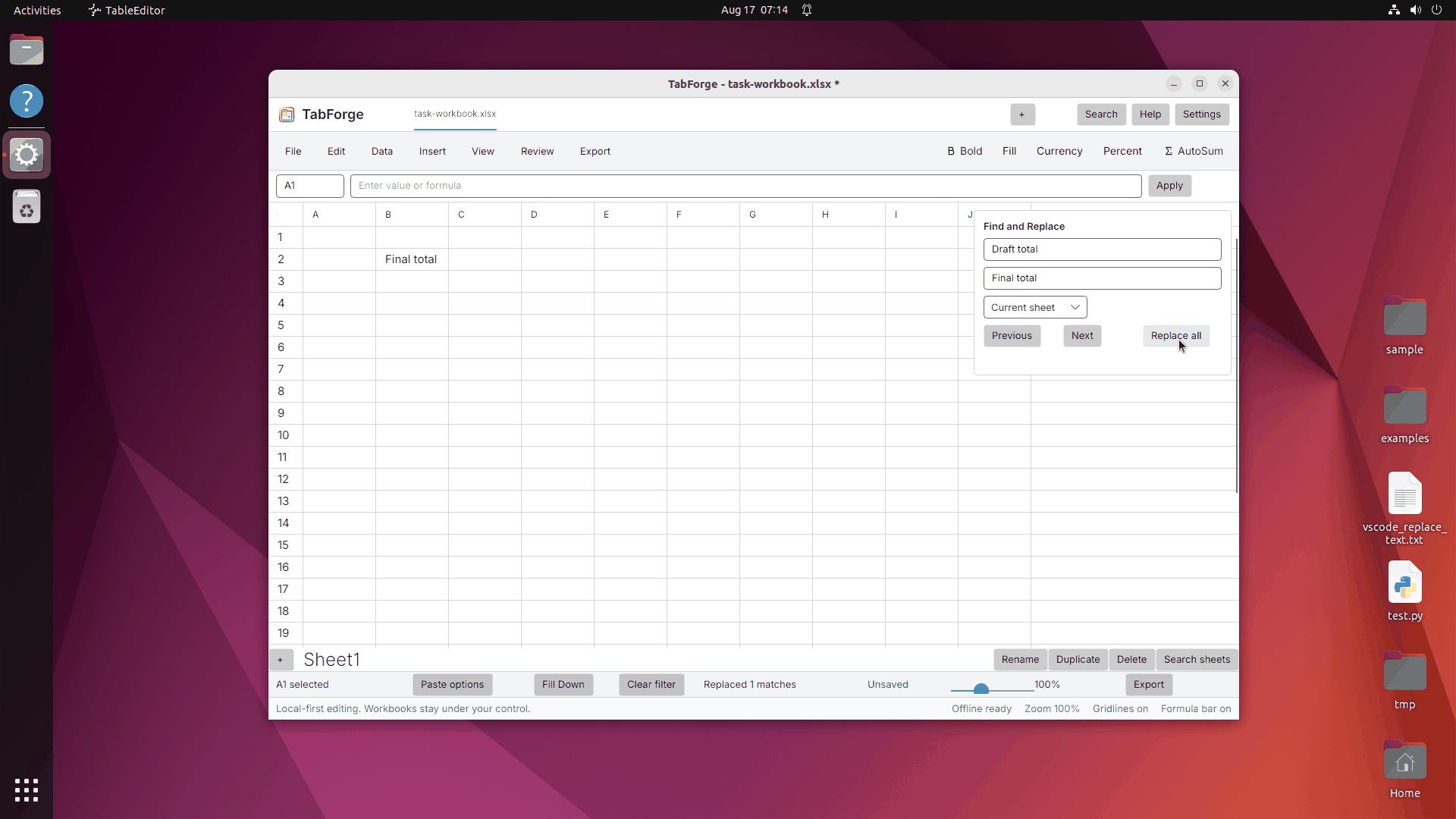}
  \par\vspace{0.6ex}

  {\scriptsize\textbf{Step 4:} Enter ``Final total''.\par}
\end{minipage}

\vspace{0.3em}

\makebox[\textwidth][l]{%
\begin{minipage}[t]{0.235\textwidth}
  \centering
  \includegraphics[width=\linewidth]{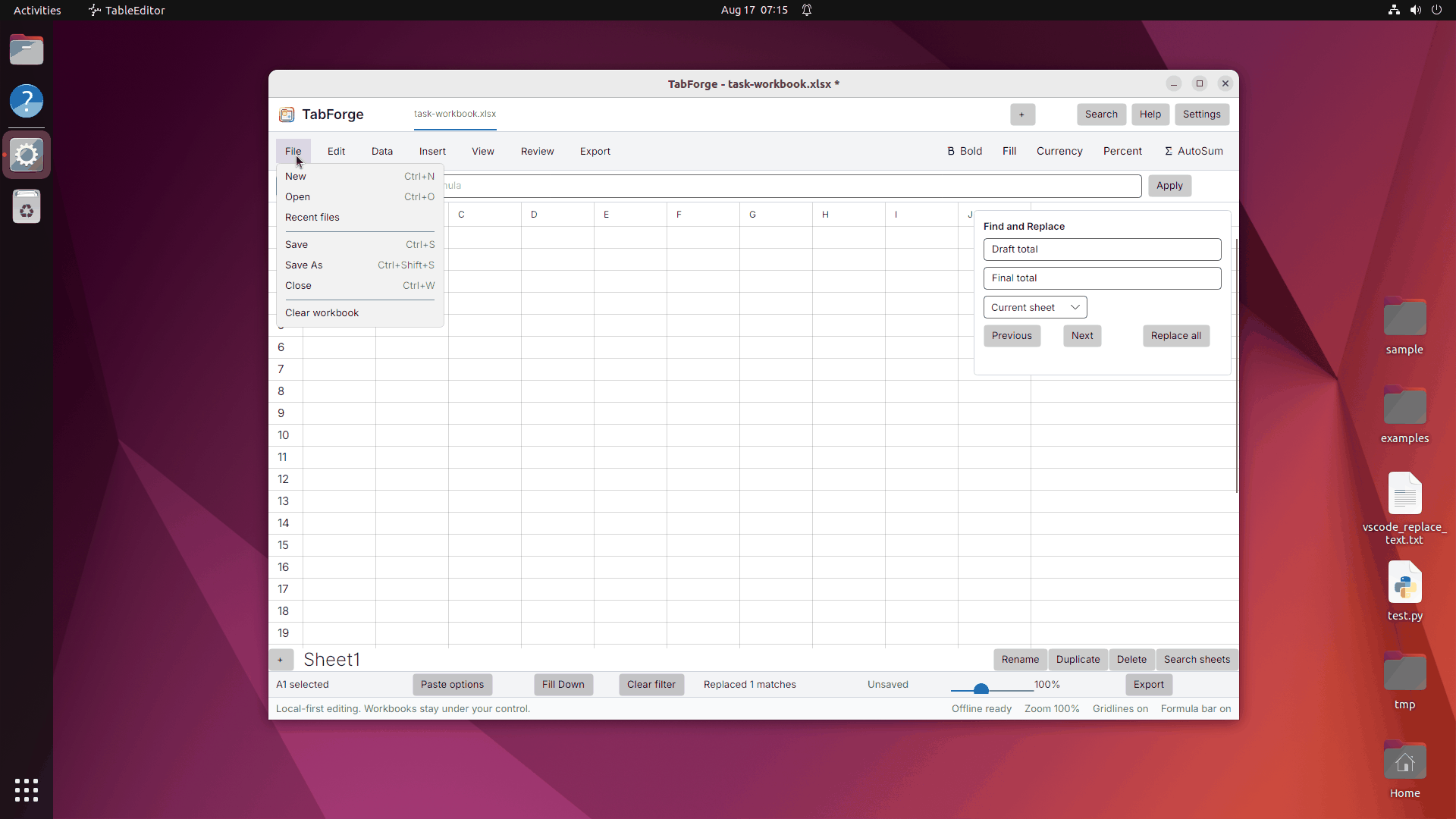}
  \par\vspace{0.6ex}

  {\scriptsize\textbf{Step 5:} Open File.\par}
\end{minipage}\hspace{0.012\textwidth}
\begin{minipage}[t]{0.235\textwidth}
  \centering
  \includegraphics[width=\linewidth]{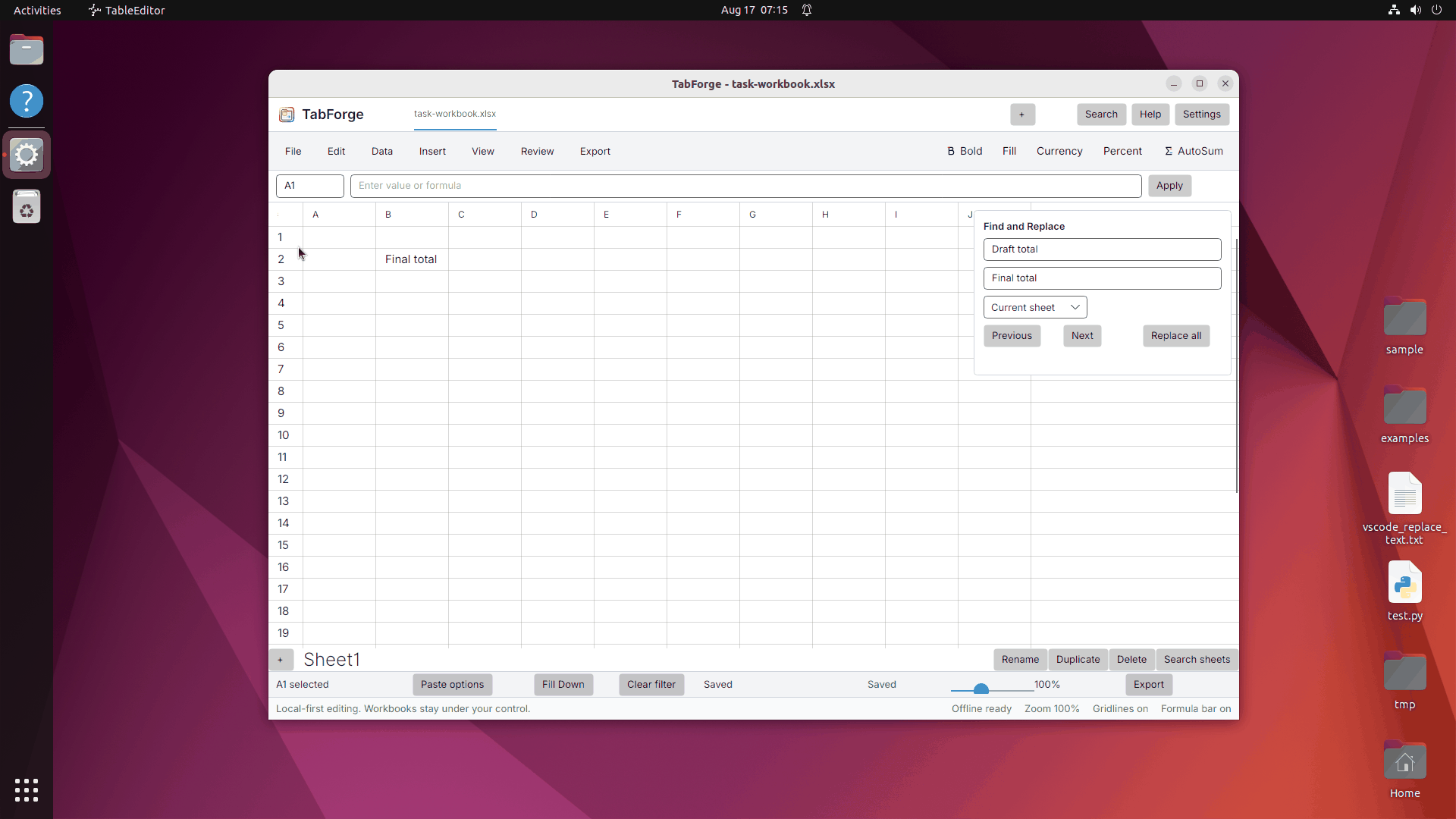}
  \par\vspace{0.6ex}

  {\scriptsize\textbf{Step 6:} Save with panel open.\par}
\end{minipage}%
}
\caption{A successful TableEditor interaction that replaces ``Draft total'' in
\texttt{Sheet1!B2} with ``Final total'' while leaving the Find and Replace
panel open.}
\label{fig:table-editor-replace-trajectory}
\end{figure}

\begingroup
\footnotesize
\urlstyle{same}
\setlength{\emergencystretch}{3em}
\noindent\textbf{Instruction:} I need a web proof batch finished. Please leave
\nolinkurl{/home/user/Desktop/examples/import/web-proof-batch/hero-card.png} and
\nolinkurl{/home/user/Desktop/examples/import/web-proof-batch/detail-card.jpg}
processed as JPEG files at
\nolinkurl{/home/user/Desktop/examples/export/web-proof-batch/hero-card.jpg} and
\nolinkurl{/home/user/Desktop/examples/export/web-proof-batch/detail-card.jpg},
with 640 pixel resize width, quality 72, both rows marked Done, and the Batch
Queue page open.\par
\endgroup

\begin{figure}[H]
\centering
\makebox[\textwidth][l]{%
\begin{minipage}[t]{0.158\textwidth}
  \centering
  \includegraphics[width=\linewidth]{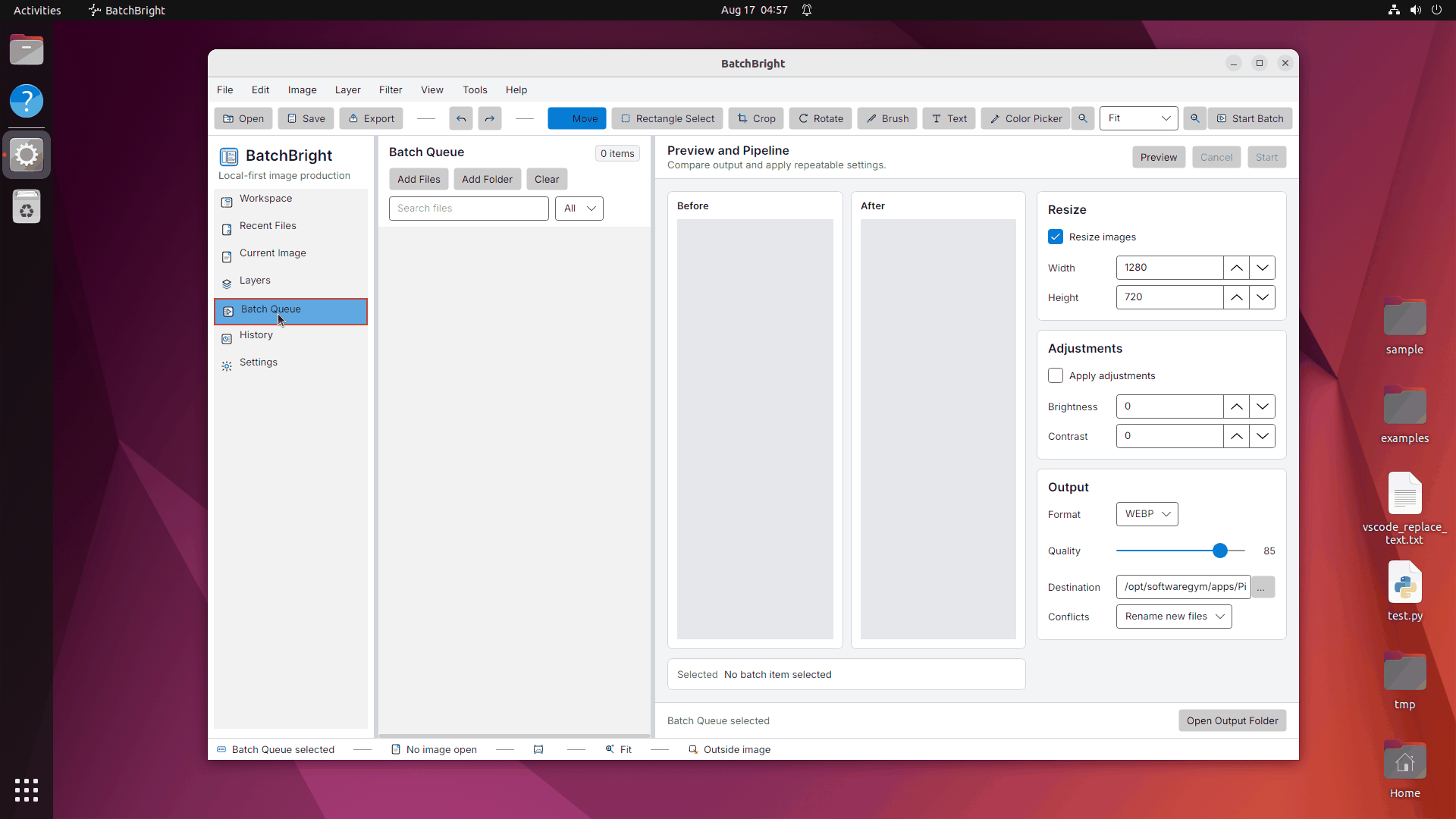}
  \par\vspace{0.6ex}

  {\scriptsize\textbf{Step 1:} Open Batch Queue.\par}
\end{minipage}\hspace{0.008\textwidth}
\begin{minipage}[t]{0.158\textwidth}
  \centering
  \includegraphics[width=\linewidth]{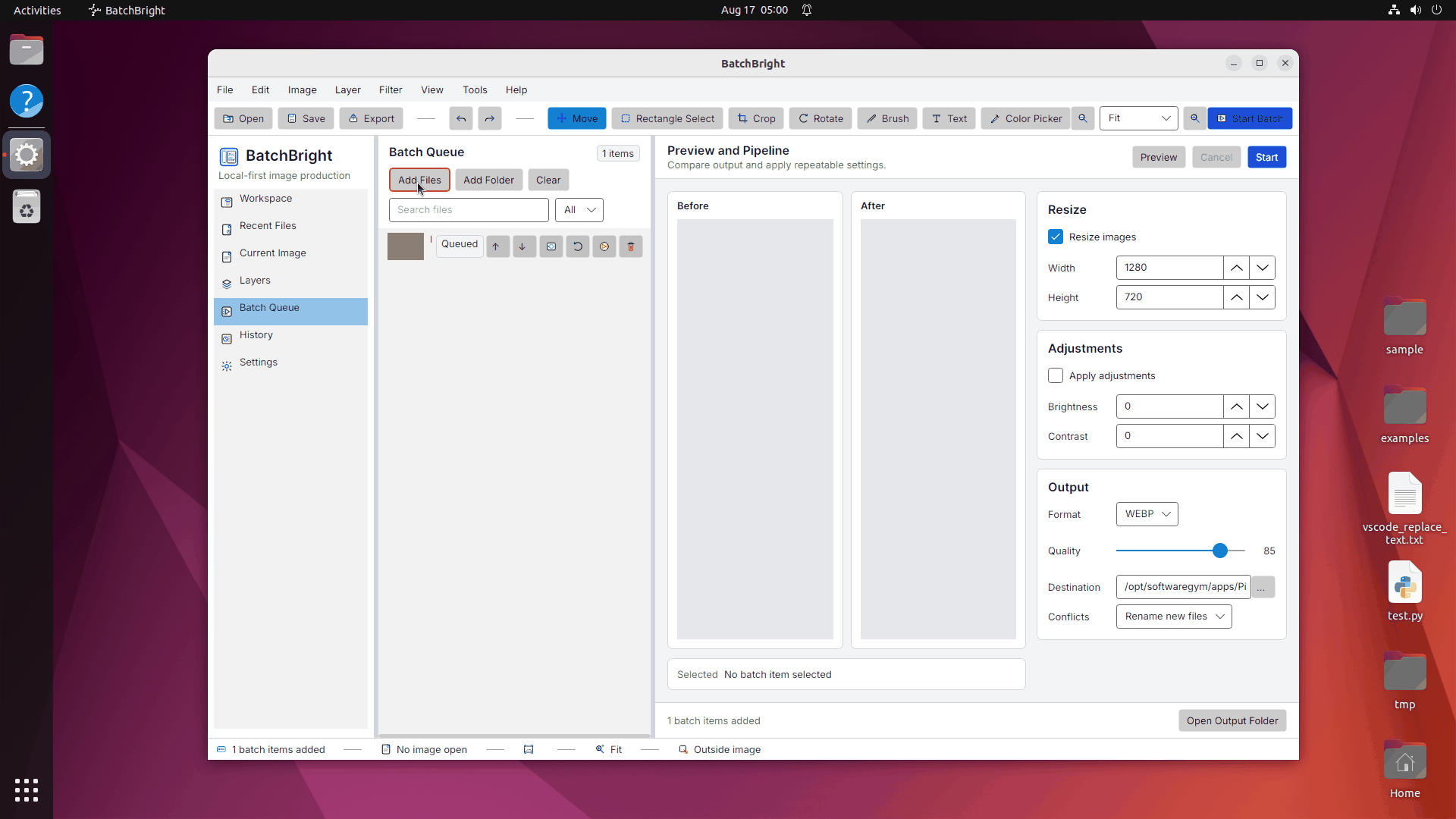}
  \par\vspace{0.6ex}

  {\scriptsize\textbf{Step 2:} Add ``hero-card.png''.\par}
\end{minipage}\hspace{0.008\textwidth}
\begin{minipage}[t]{0.158\textwidth}
  \centering
  \includegraphics[width=\linewidth]{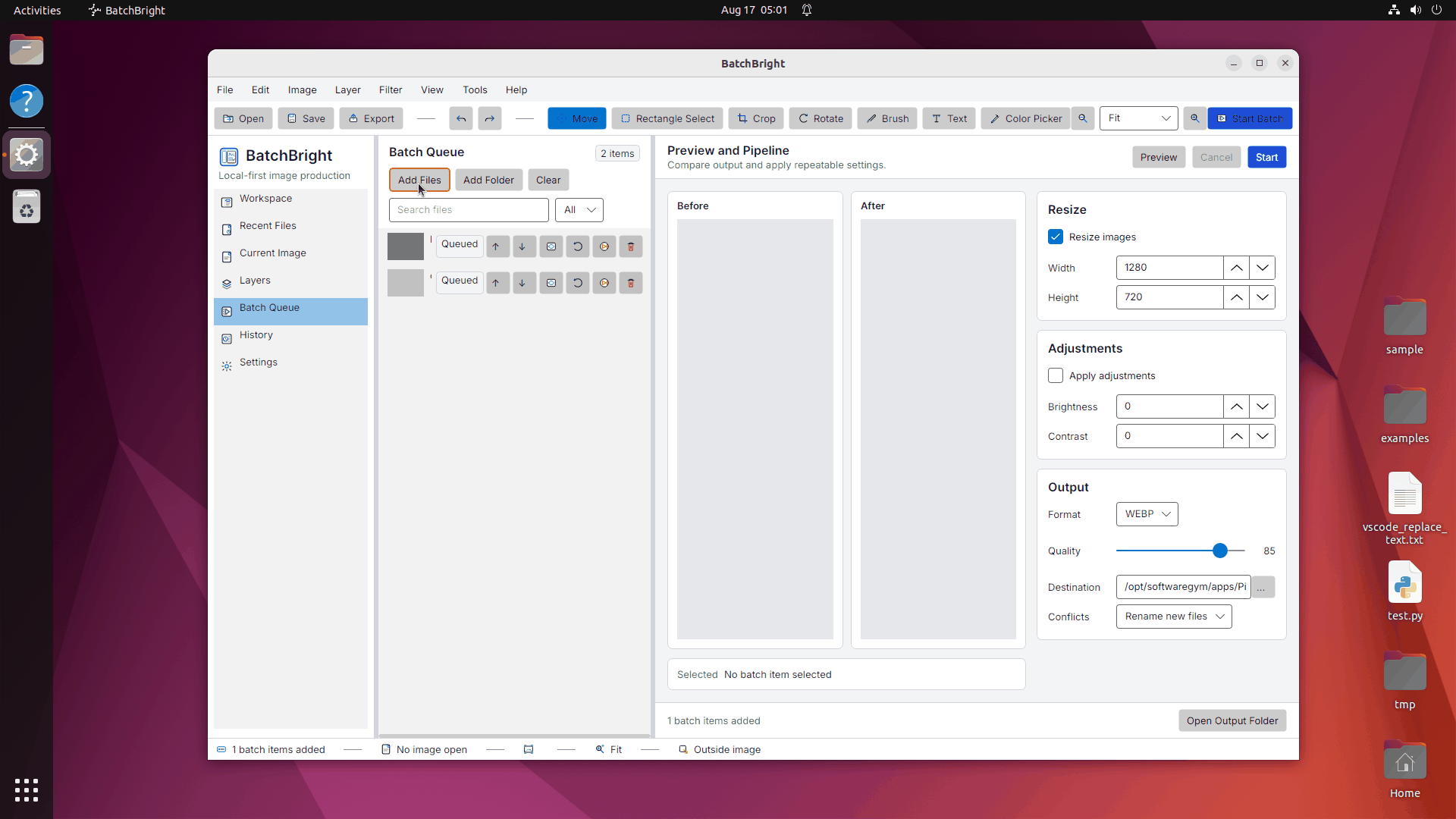}
  \par\vspace{0.6ex}

  {\scriptsize\textbf{Step 3:} Add ``detail-card.jpg''.\par}
\end{minipage}\hspace{0.008\textwidth}
\begin{minipage}[t]{0.158\textwidth}
  \centering
  \includegraphics[width=\linewidth]{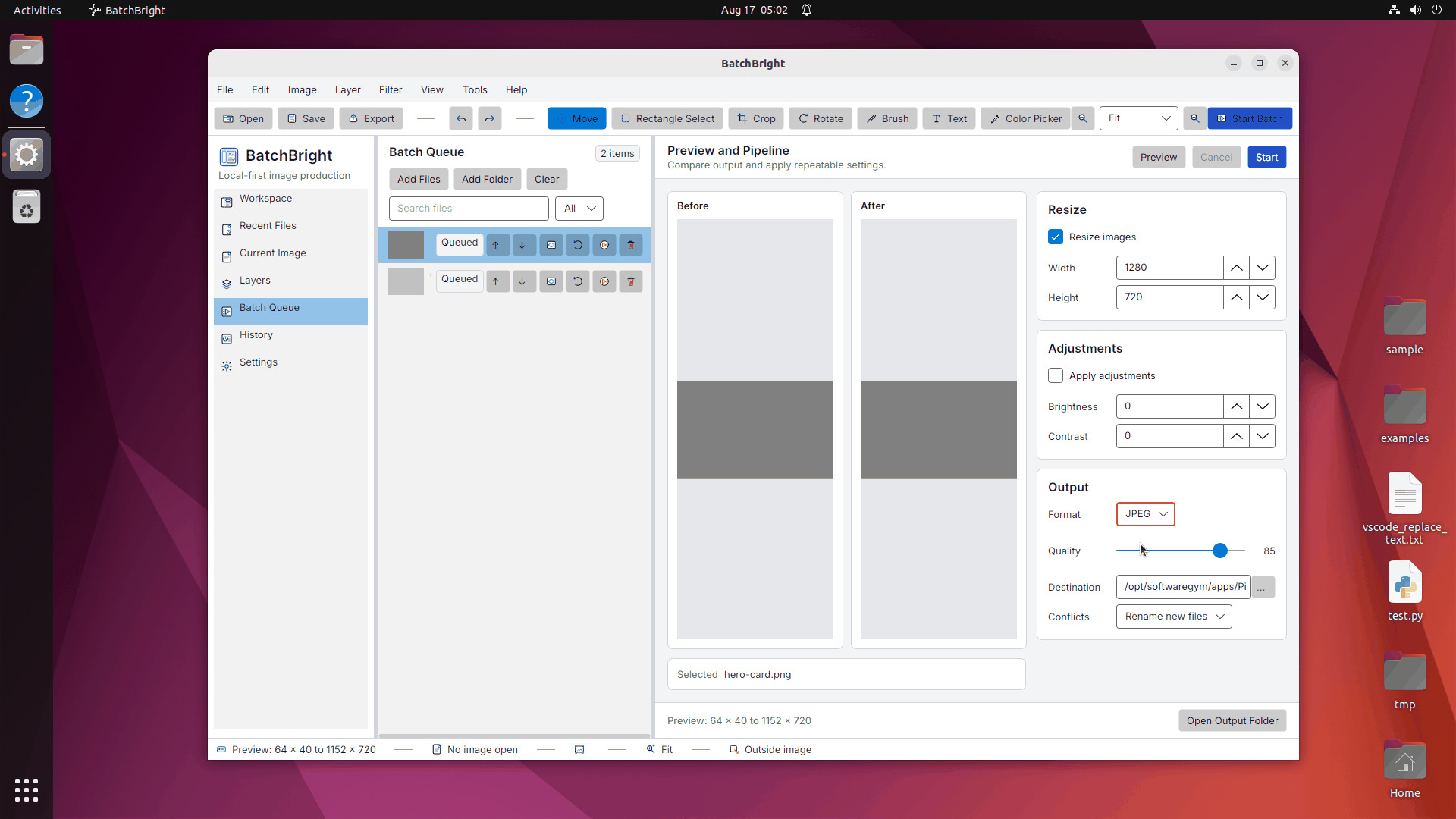}
  \par\vspace{0.6ex}

  {\scriptsize\textbf{Step 4:} Set format to JPEG.\par}
\end{minipage}\hspace{0.008\textwidth}
\begin{minipage}[t]{0.158\textwidth}
  \centering
  \includegraphics[width=\linewidth]{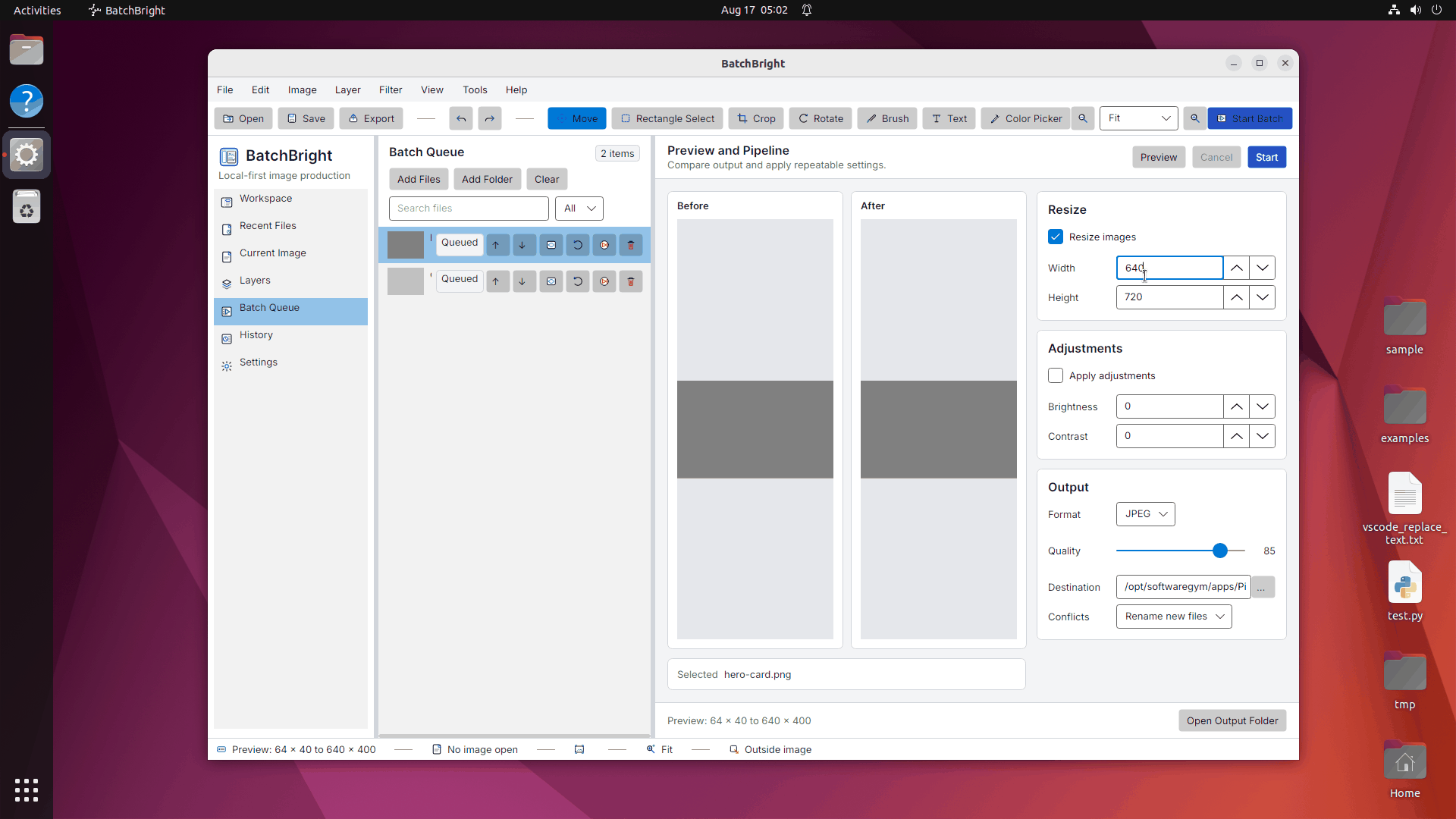}
  \par\vspace{0.6ex}

  {\scriptsize\textbf{Step 5:} Set width to 640 px.\par}
\end{minipage}\hspace{0.008\textwidth}
\begin{minipage}[t]{0.158\textwidth}
  \centering
  \includegraphics[width=\linewidth]{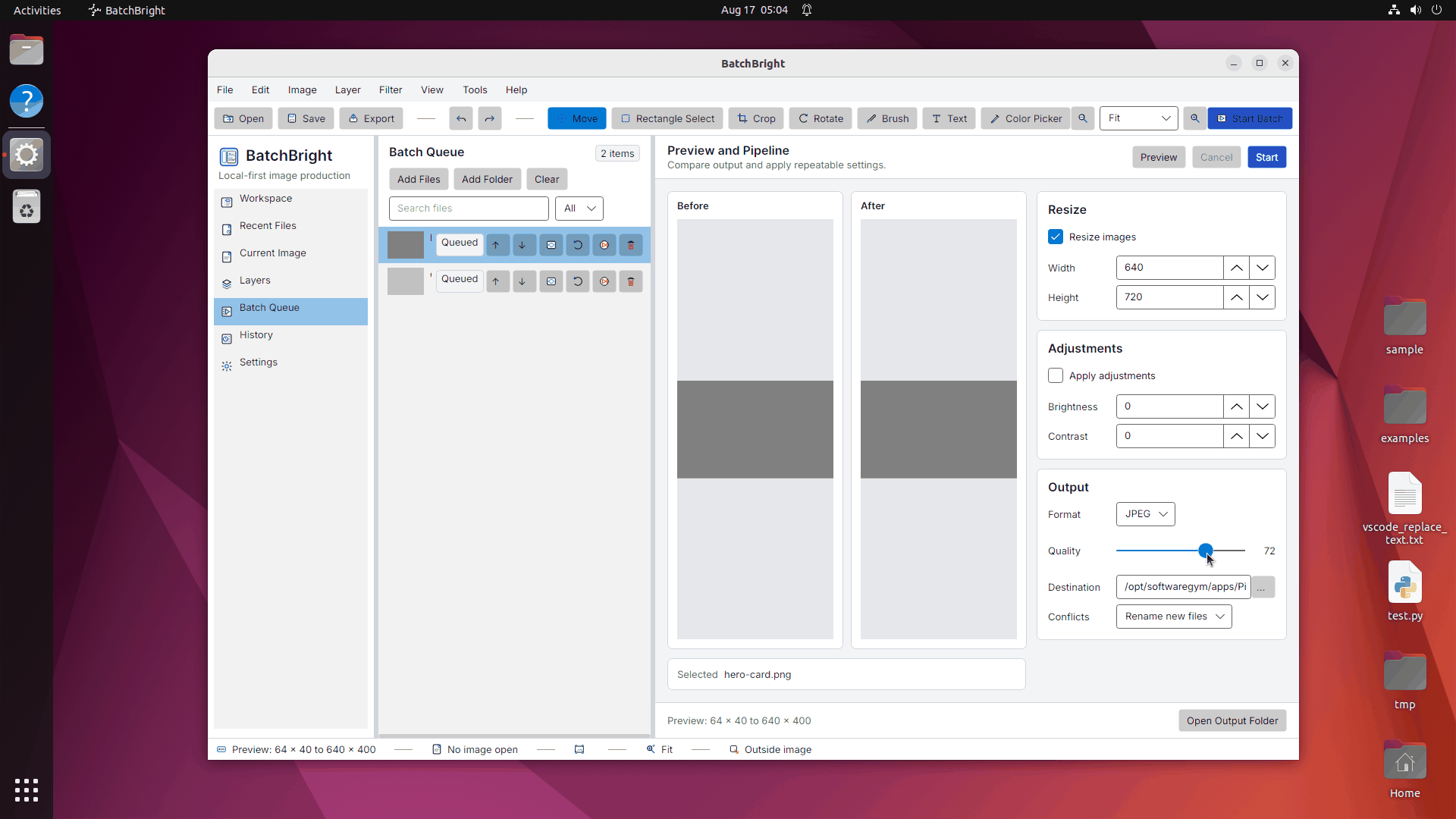}
  \par\vspace{0.6ex}

  {\scriptsize\textbf{Step 6:} Set quality to 72.\par}
\end{minipage}%
}

\vspace{0.3em}

\makebox[\textwidth][l]{%
\begin{minipage}[t]{0.158\textwidth}
  \centering
  \includegraphics[width=\linewidth]{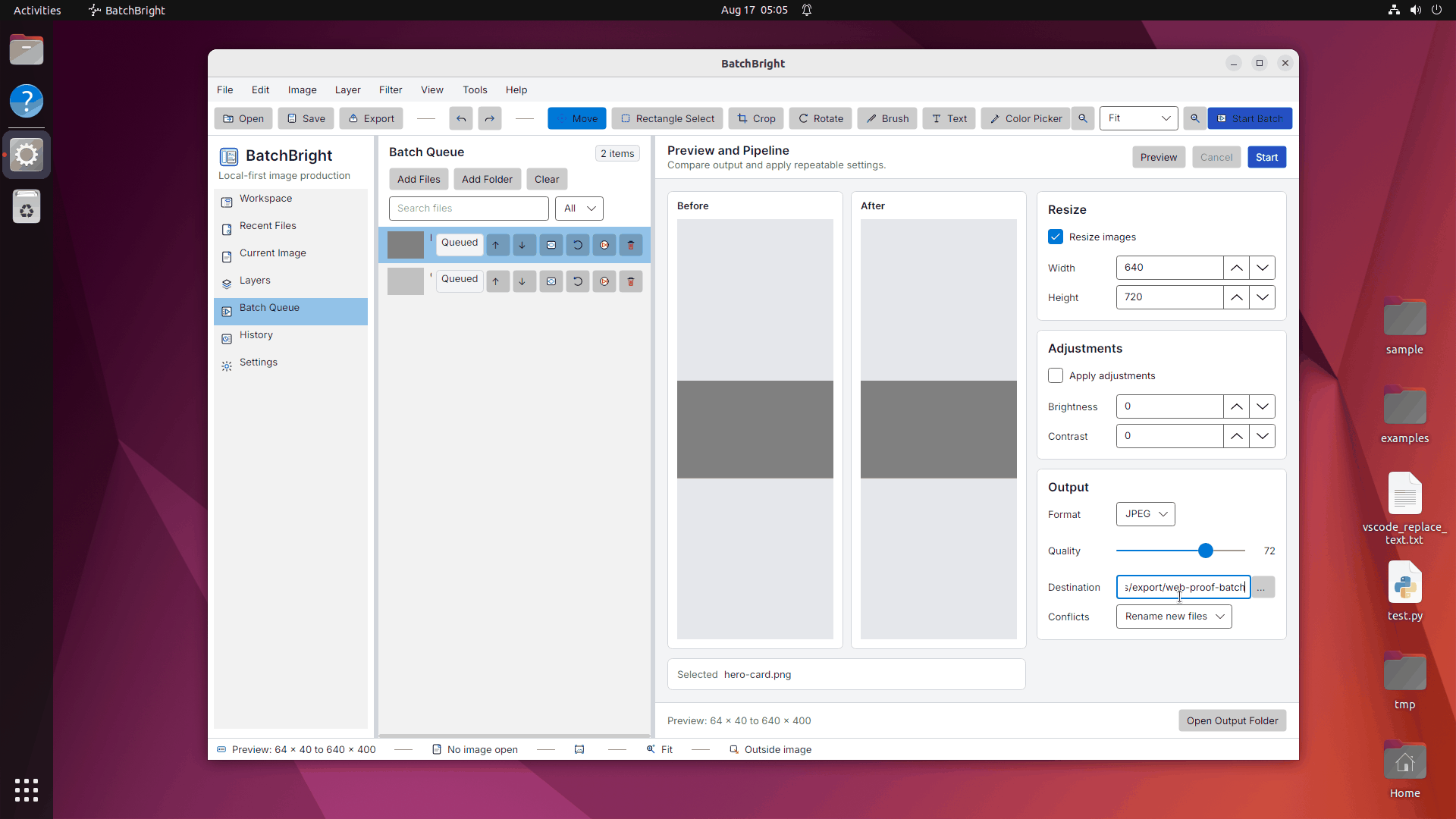}
  \par\vspace{0.6ex}

  {\scriptsize\textbf{Step 7:} Set export destination.\par}
\end{minipage}\hspace{0.008\textwidth}
\begin{minipage}[t]{0.158\textwidth}
  \centering
  \includegraphics[width=\linewidth]{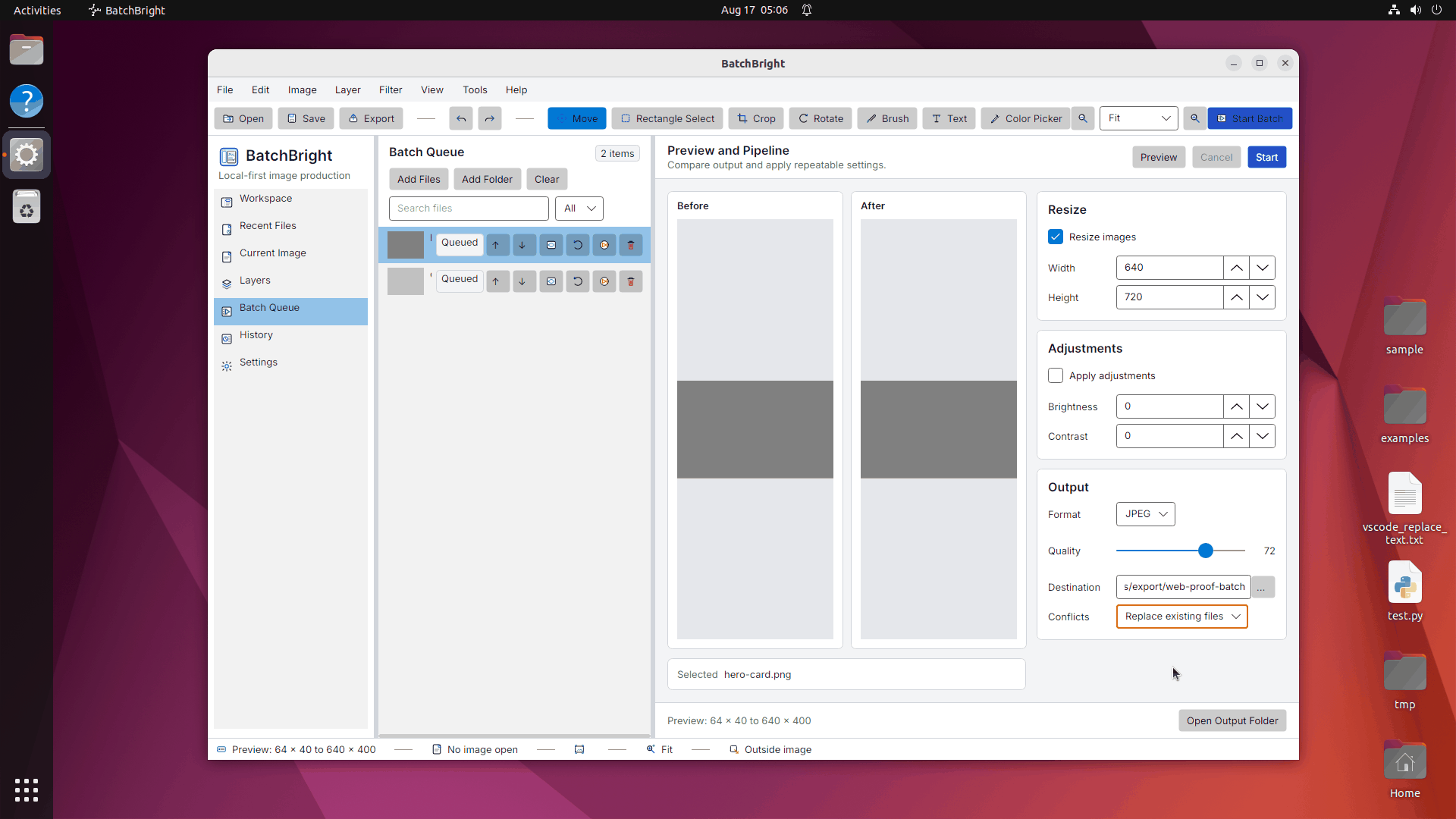}
  \par\vspace{0.6ex}

  {\scriptsize\textbf{Step 8:} Replace existing files.\par}
\end{minipage}\hspace{0.008\textwidth}
\begin{minipage}[t]{0.158\textwidth}
  \centering
  \includegraphics[width=\linewidth]{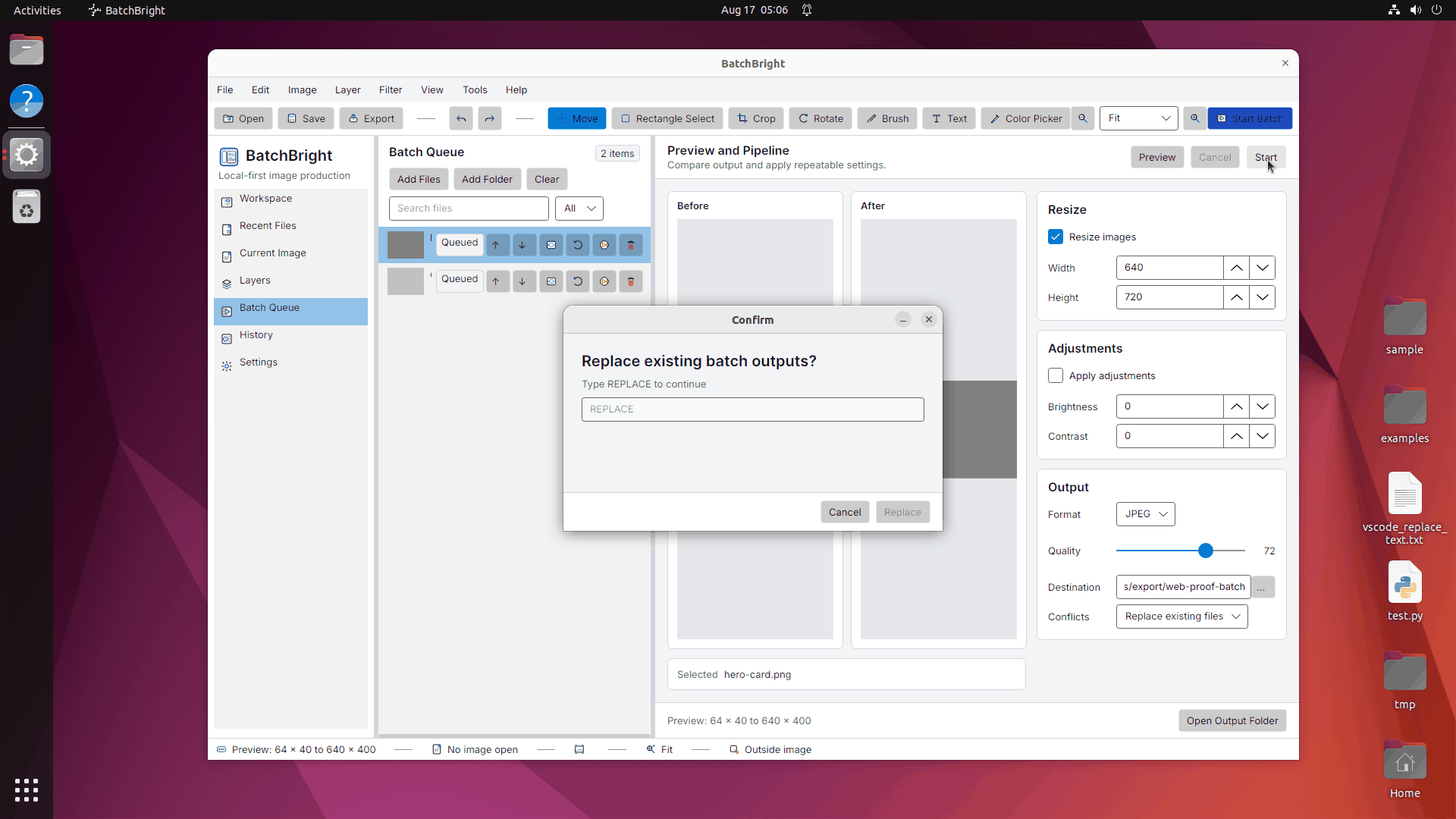}
  \par\vspace{0.6ex}

  {\scriptsize\textbf{Step 9:} Start the batch.\par}
\end{minipage}\hspace{0.008\textwidth}
\begin{minipage}[t]{0.158\textwidth}
  \centering
  \includegraphics[width=\linewidth]{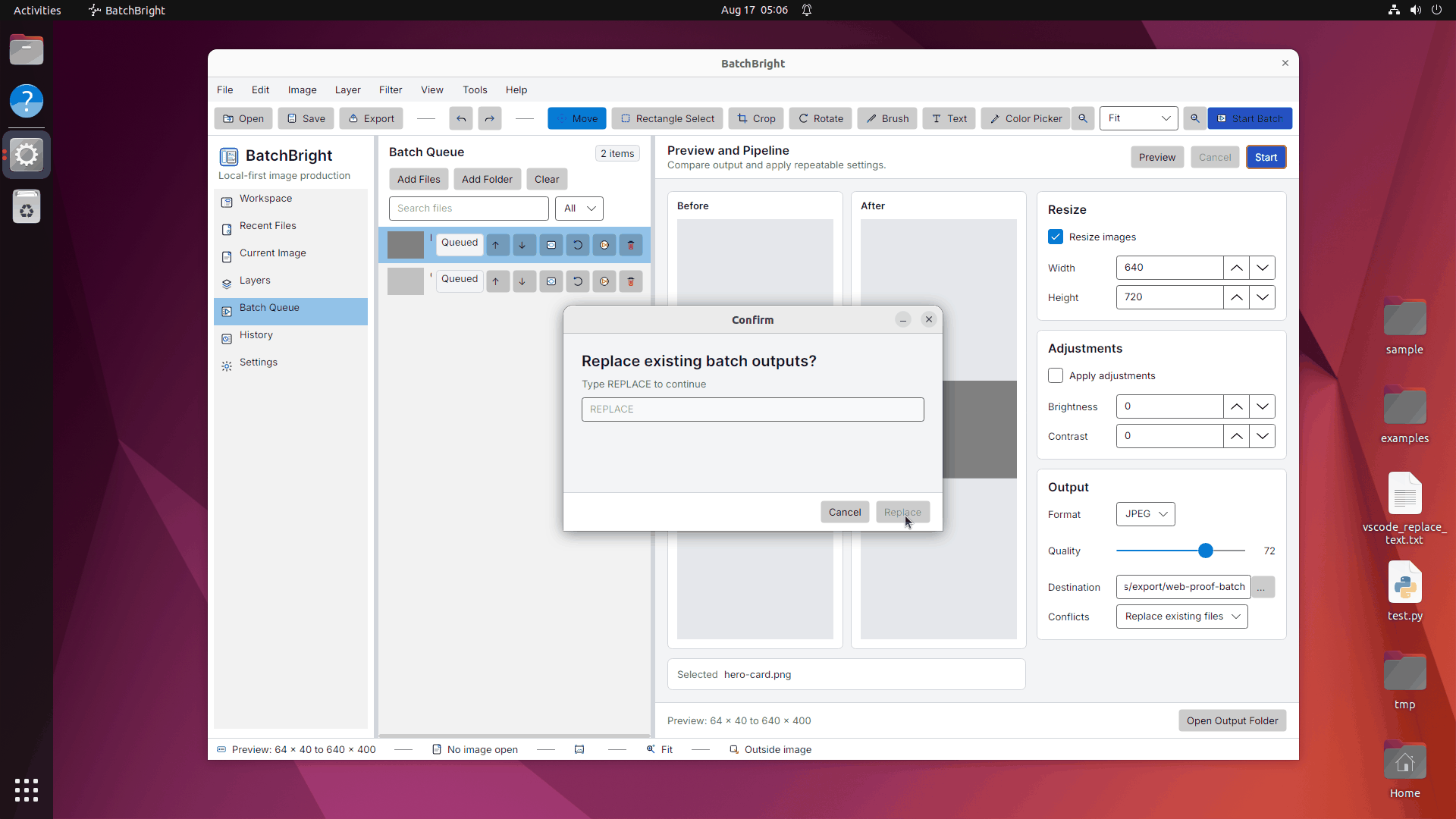}
  \par\vspace{0.6ex}

  {\scriptsize\textbf{Step 10:} Confirm replacement.\par}
\end{minipage}\hspace{0.008\textwidth}
\begin{minipage}[t]{0.158\textwidth}
  \centering
  \includegraphics[width=\linewidth]{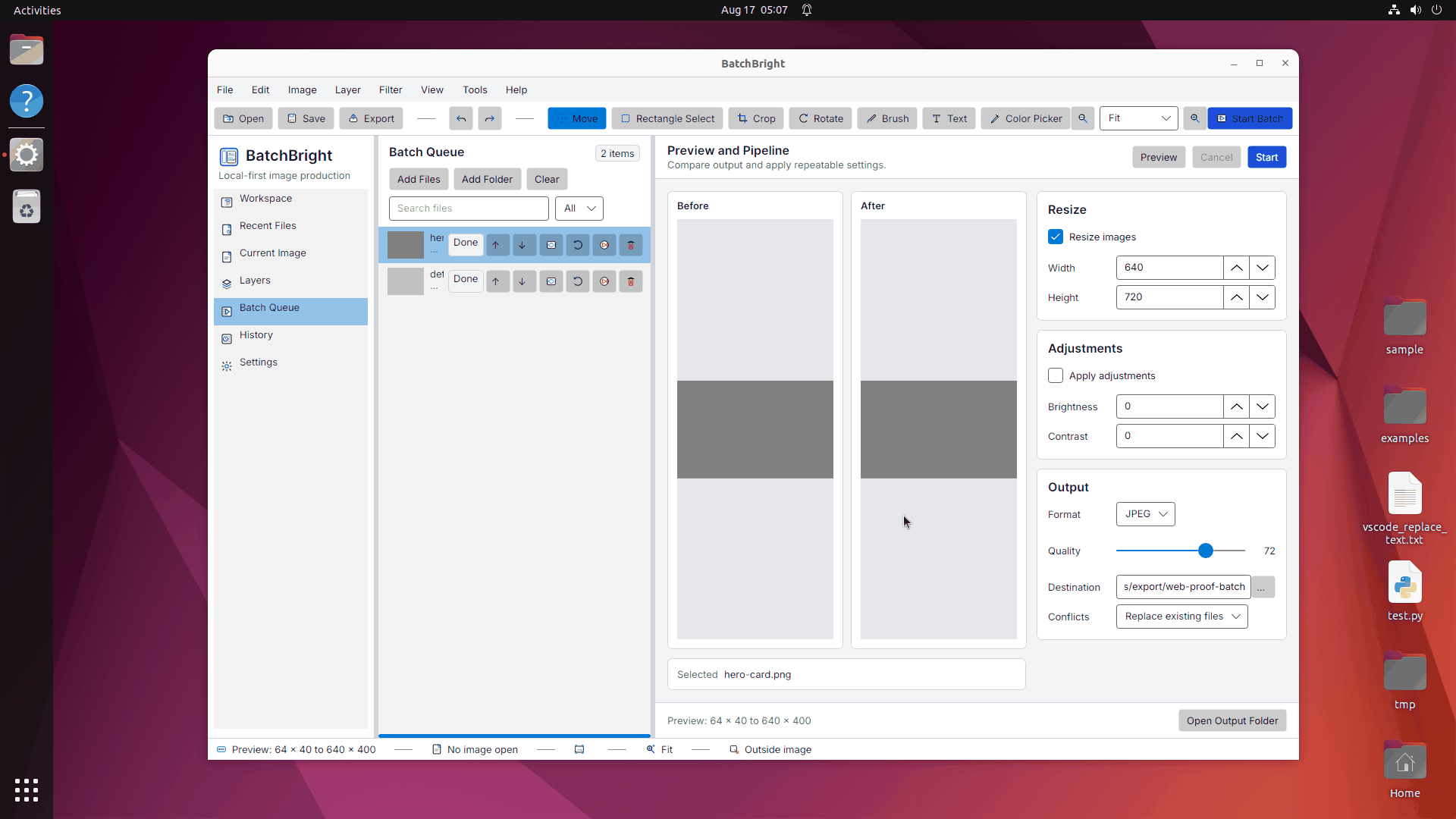}
  \par\vspace{0.6ex}

  {\scriptsize\textbf{Step 11:} Confirm again; finish.\par}
\end{minipage}%
}
\caption{A successful PicEditor interaction that processes two images as JPEG
files at 640-pixel width and quality 72, leaves both queue rows marked Done,
and keeps the Batch Queue page open.}
\label{fig:pic-editor-batch-trajectory}
\end{figure}

\noindent\textbf{Instruction:} Please turn on the favorites-only library
filter.

\begin{figure}[H]
\centering
\makebox[\textwidth][l]{%
\begin{minipage}[t]{0.235\textwidth}
  \centering
  \includegraphics[width=\linewidth]{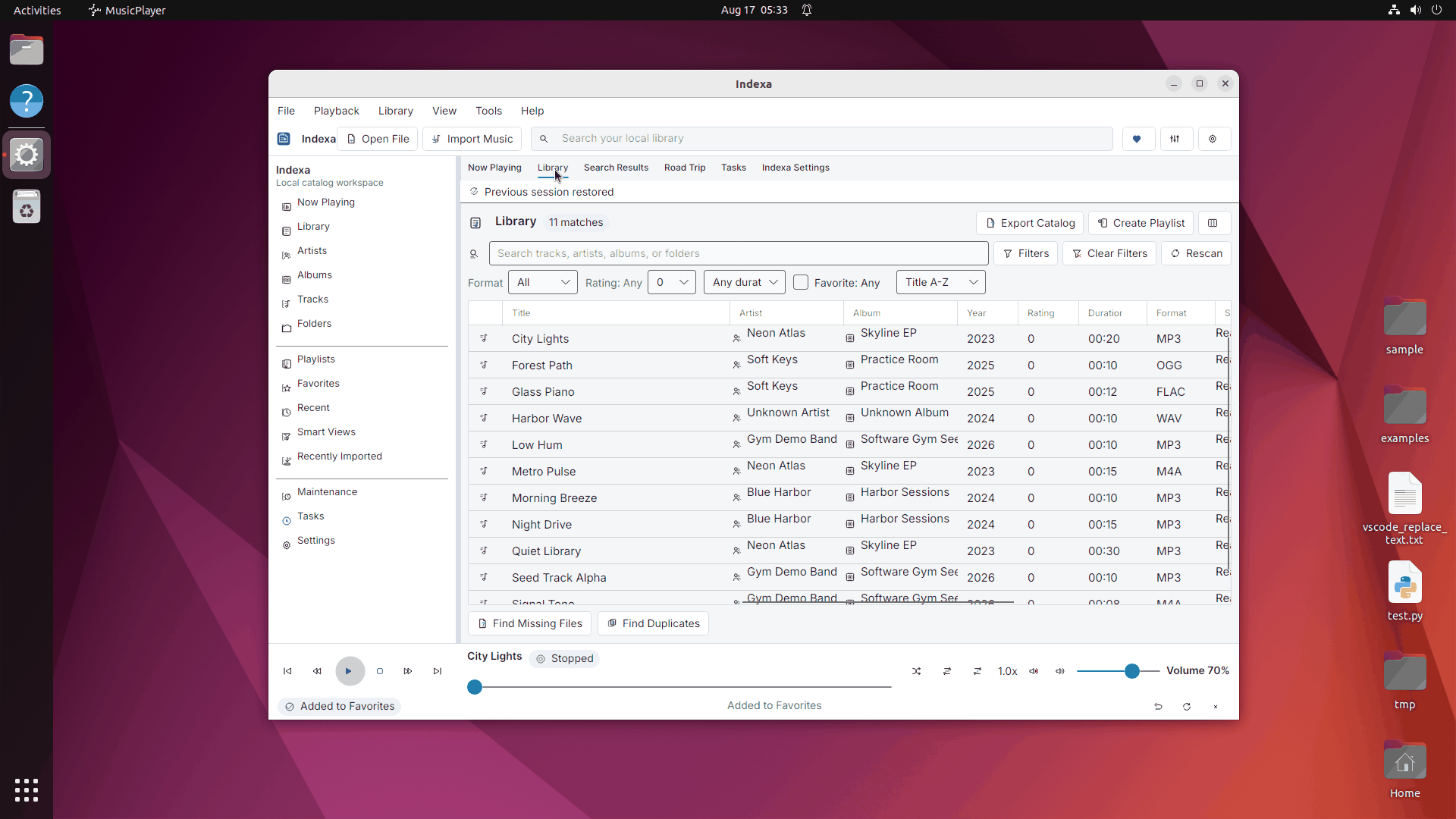}
  \par\vspace{0.6ex}

  {\scriptsize\textbf{Step 1:} Open Library.\par}
\end{minipage}\hspace{0.012\textwidth}
\begin{minipage}[t]{0.235\textwidth}
  \centering
  \includegraphics[width=\linewidth]{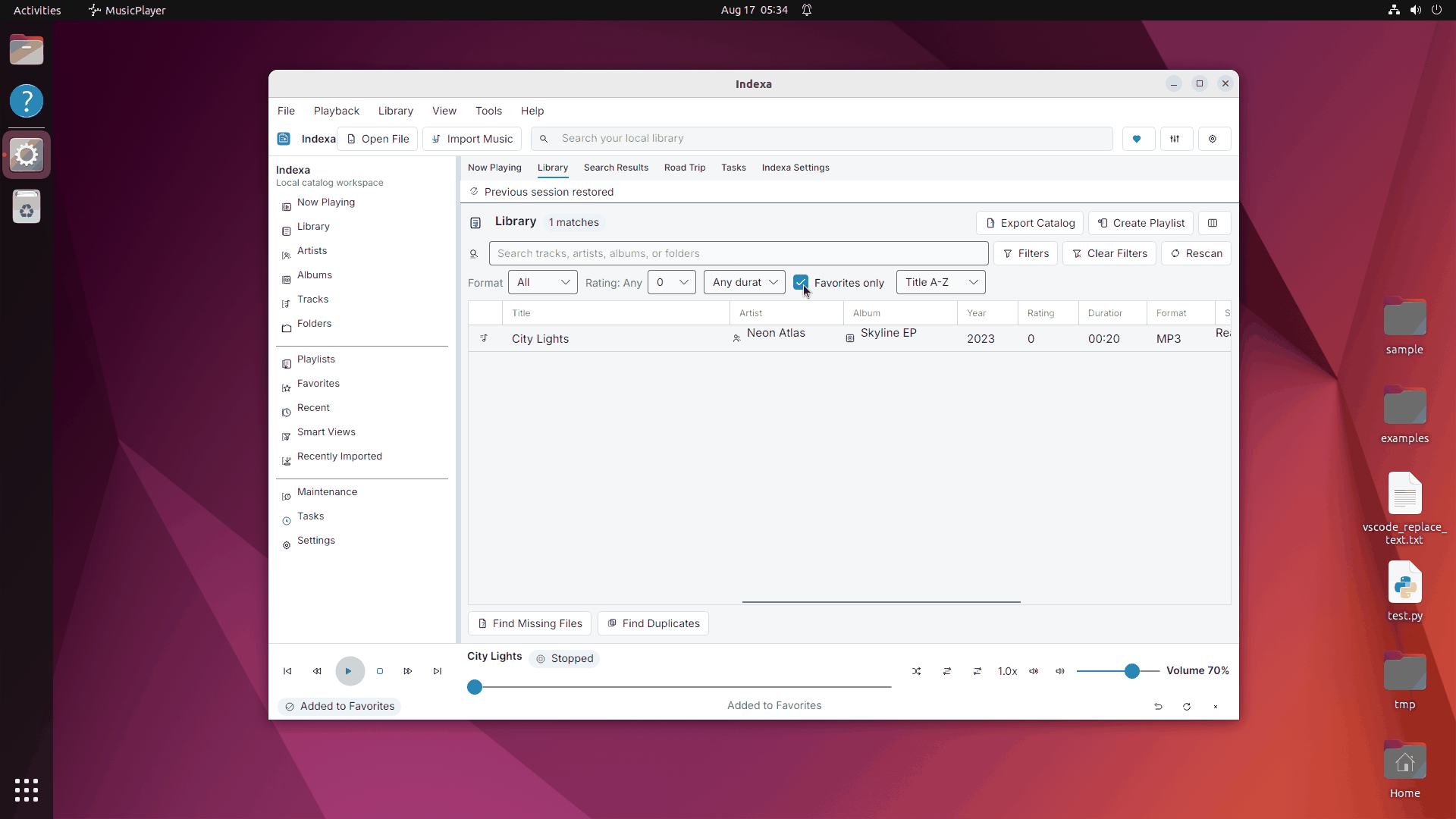}
  \par\vspace{0.6ex}

  {\scriptsize\textbf{Step 2:} Enable Favorites only.\par}
\end{minipage}\hspace{0.012\textwidth}
\begin{minipage}[t]{0.235\textwidth}
  \centering
  \includegraphics[width=\linewidth]{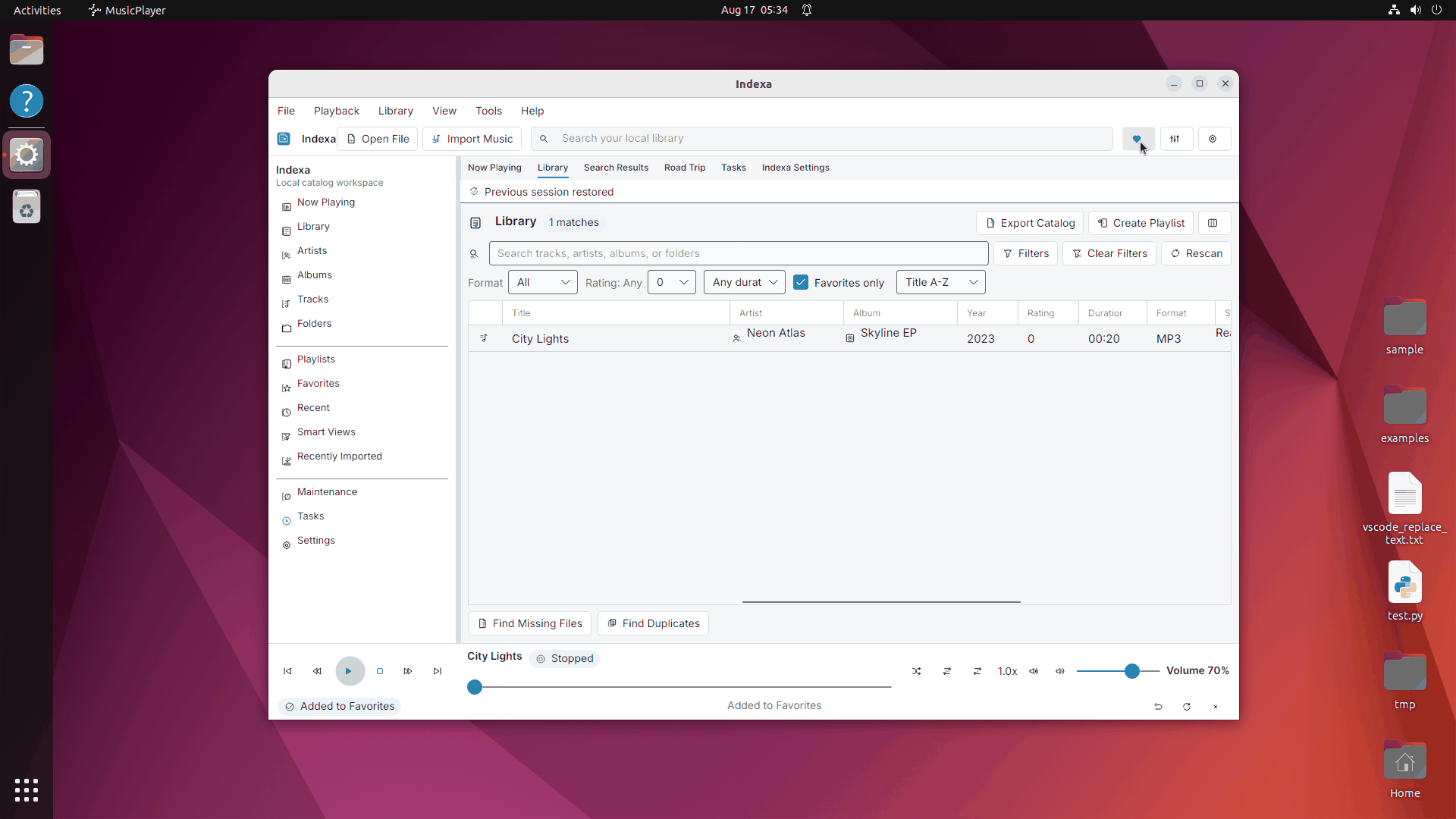}
  \par\vspace{0.6ex}

  {\scriptsize\textbf{Step 3:} Verify filtered results.\par}
\end{minipage}%
}
\caption{A successful MusicPlayer interaction that enables the favorites-only
library filter and leaves the filtered Library view open.}
\label{fig:music-player-favorites-filter-trajectory}
\end{figure}

\noindent\textbf{Instruction:} Clean up the Extensions drawer by removing the
leftover rainbow-indent extension while keeping every built-in extension
enabled.

\begin{figure}[H]
\centering
\makebox[\textwidth][l]{%
\begin{minipage}[t]{0.235\textwidth}
  \centering
  \includegraphics[width=\linewidth]{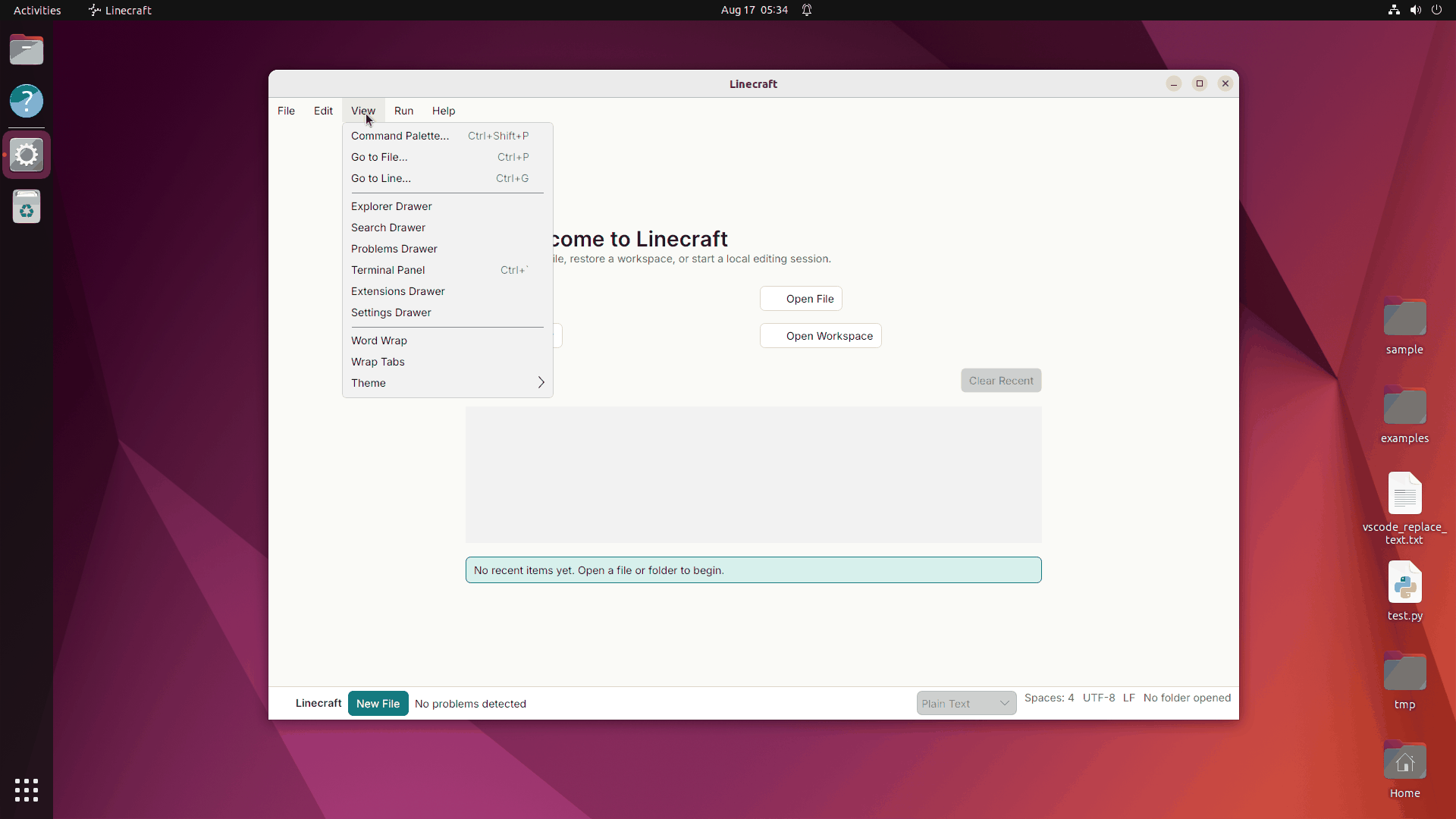}
  \par\vspace{0.6ex}

  {\scriptsize\textbf{Step 1:} Open the View menu.\par}
\end{minipage}\hspace{0.012\textwidth}
\begin{minipage}[t]{0.235\textwidth}
  \centering
  \includegraphics[width=\linewidth]{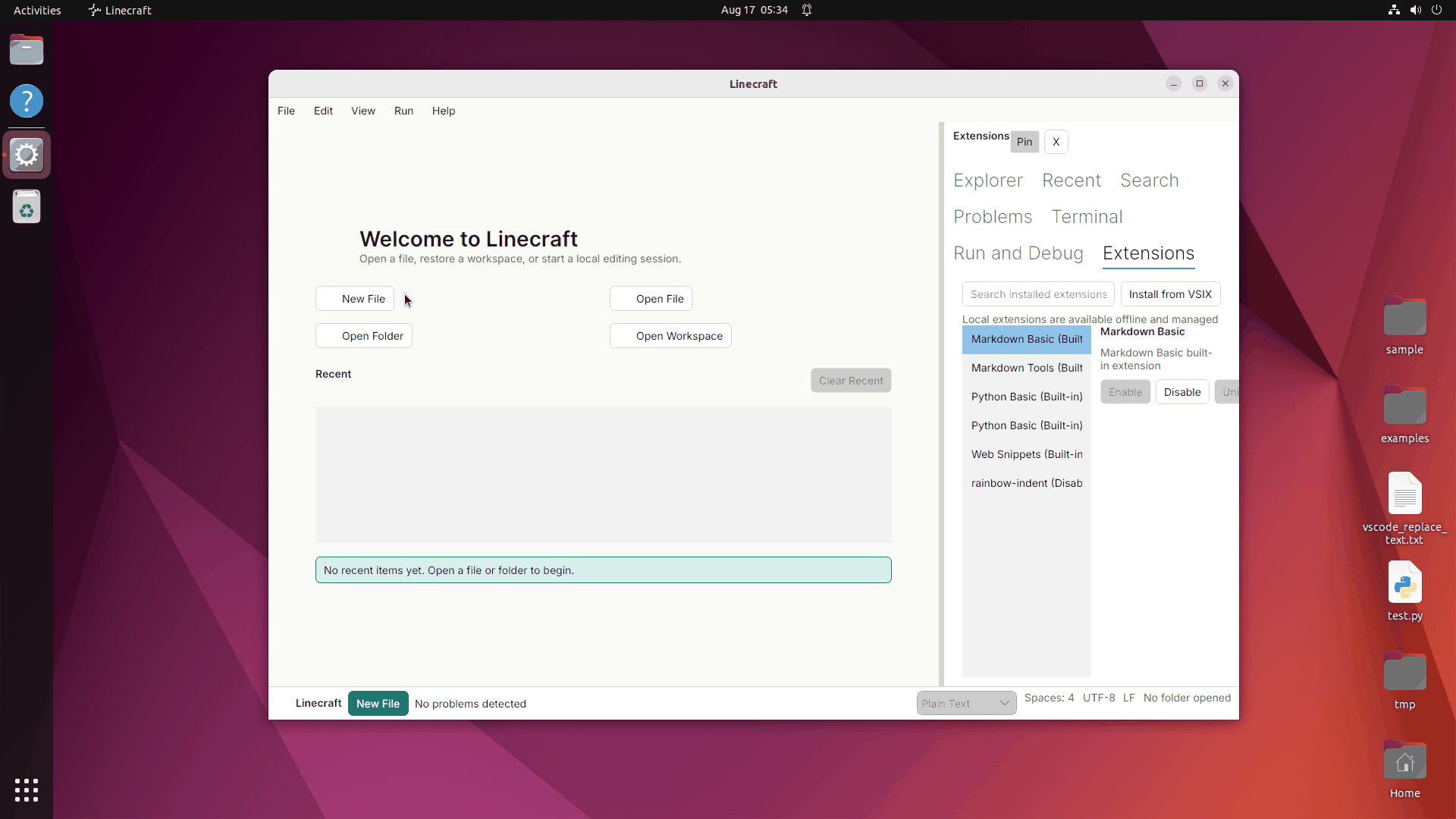}
  \par\vspace{0.6ex}

  {\scriptsize\textbf{Step 2:} Open Extensions.\par}
\end{minipage}\hspace{0.012\textwidth}
\begin{minipage}[t]{0.235\textwidth}
  \centering
  \includegraphics[width=\linewidth]{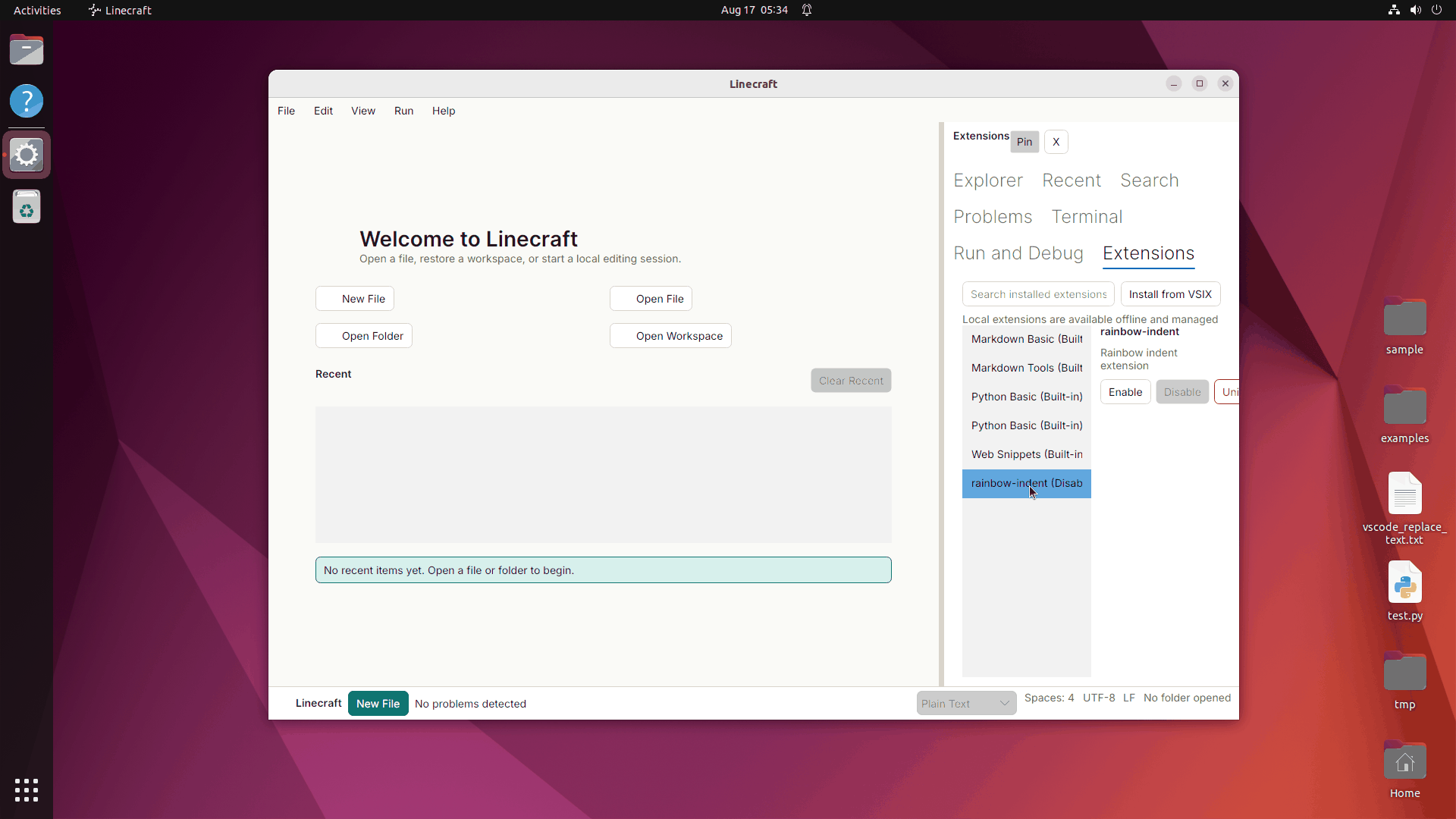}
  \par\vspace{0.6ex}

  {\scriptsize\textbf{Step 3:} Select rainbow-indent.\par}
\end{minipage}\hspace{0.012\textwidth}
\begin{minipage}[t]{0.235\textwidth}
  \centering
  \includegraphics[width=\linewidth]{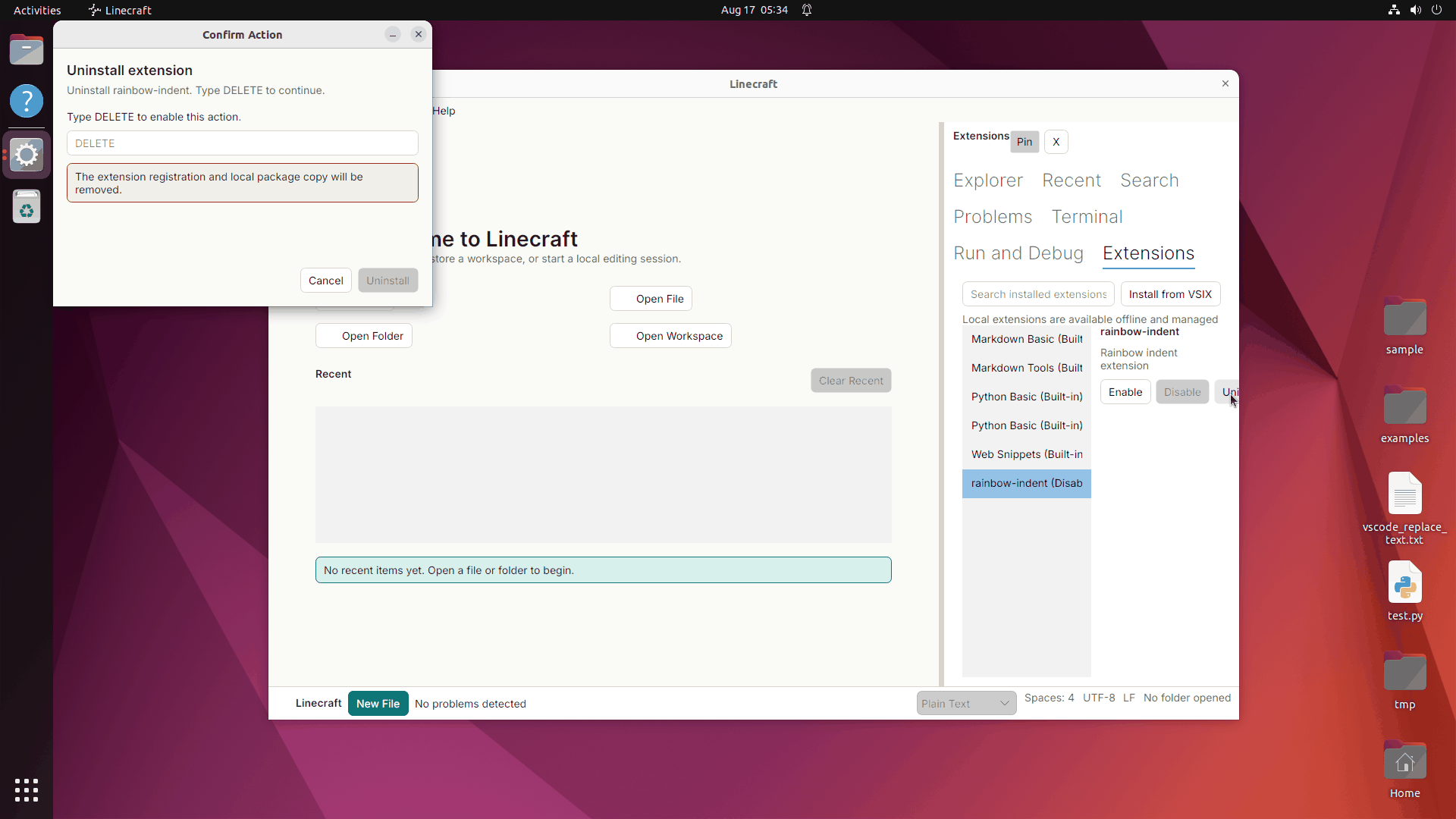}
  \par\vspace{0.6ex}

  {\scriptsize\textbf{Step 4:} Start uninstall.\par}
\end{minipage}%
}

\vspace{0.3em}

\makebox[\textwidth][l]{%
\begin{minipage}[t]{0.235\textwidth}
  \centering
  \includegraphics[width=\linewidth]{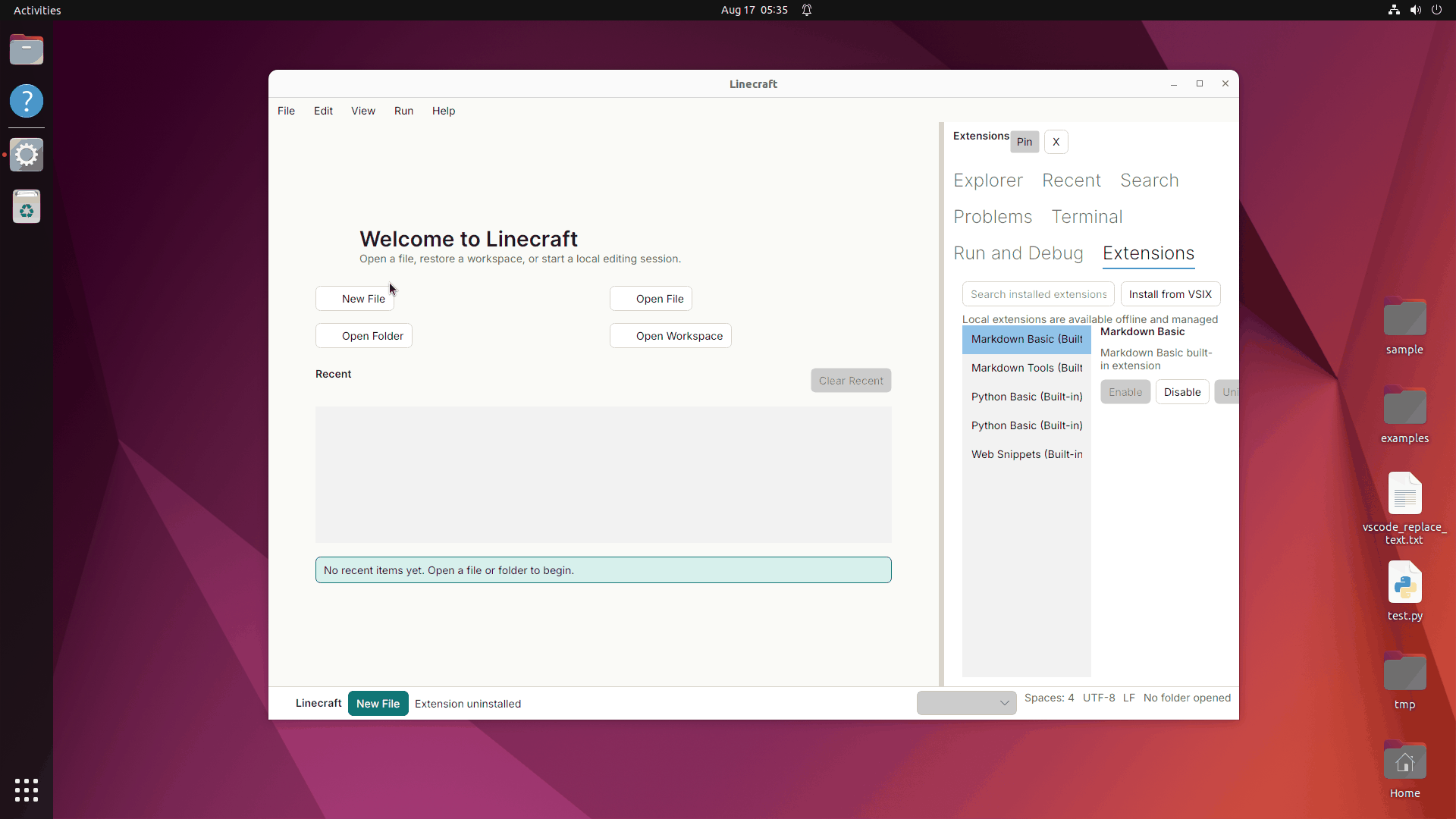}
  \par\vspace{0.6ex}

  {\scriptsize\textbf{Step 5:} Confirm removal.\par}
\end{minipage}%
}
\caption{A successful CodeEditor interaction that removes the leftover
rainbow-indent extension while preserving the enabled built-in extensions.}
\label{fig:code-editor-extension-cleanup-trajectory}
\end{figure}

\noindent\textbf{Instruction:} Please create a public community named Civic
Desk, open for joining, with the summary exactly ``Planning notes for
neighborhood projects.''

\begin{figure}[H]
\centering
\makebox[\textwidth][l]{%
\begin{minipage}[t]{0.235\textwidth}
  \centering
  \includegraphics[width=\linewidth]{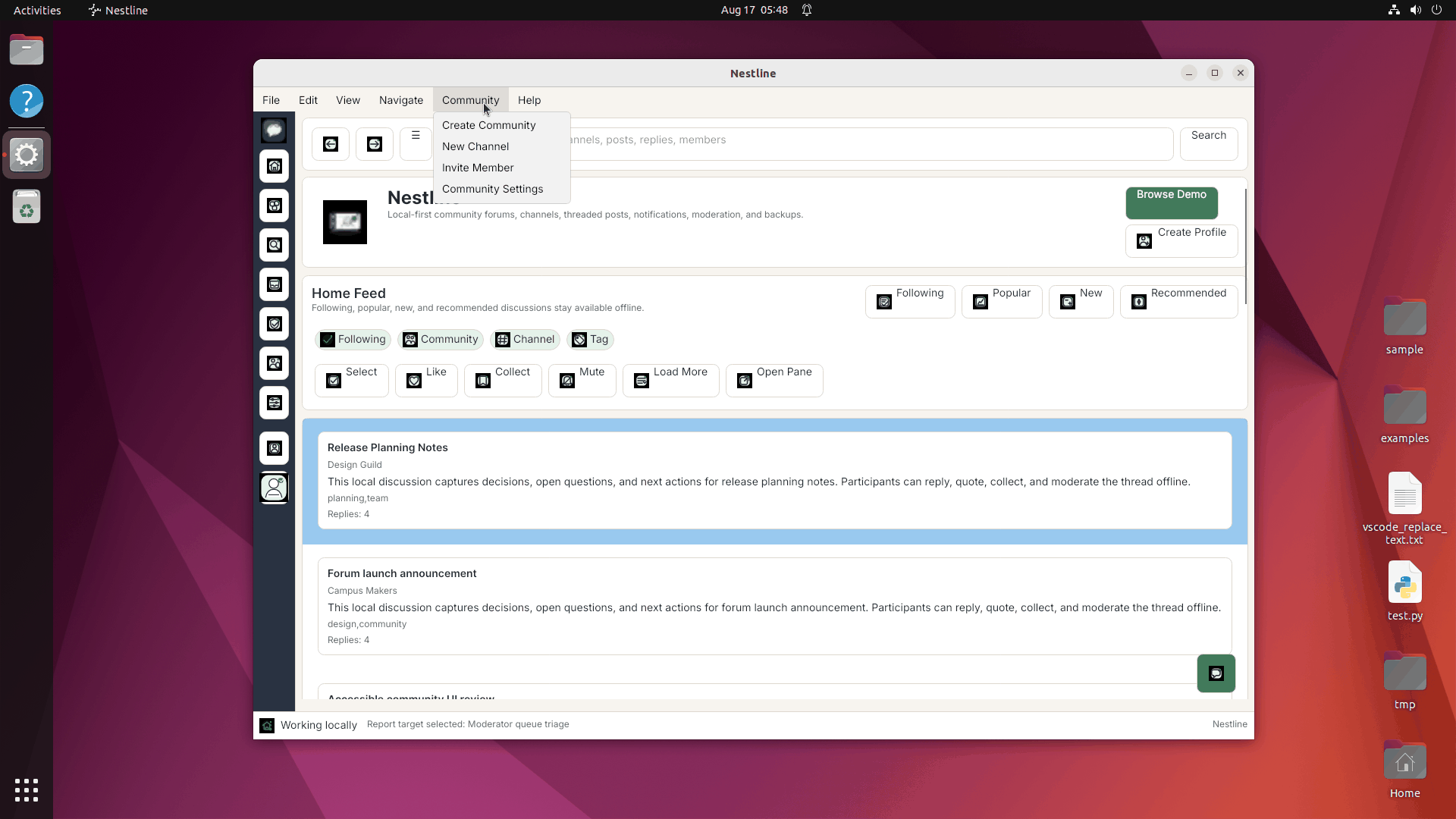}
  \par\vspace{0.6ex}

  {\scriptsize\textbf{Step 1:} Open Community menu.\par}
\end{minipage}\hspace{0.012\textwidth}
\begin{minipage}[t]{0.235\textwidth}
  \centering
  \includegraphics[width=\linewidth]{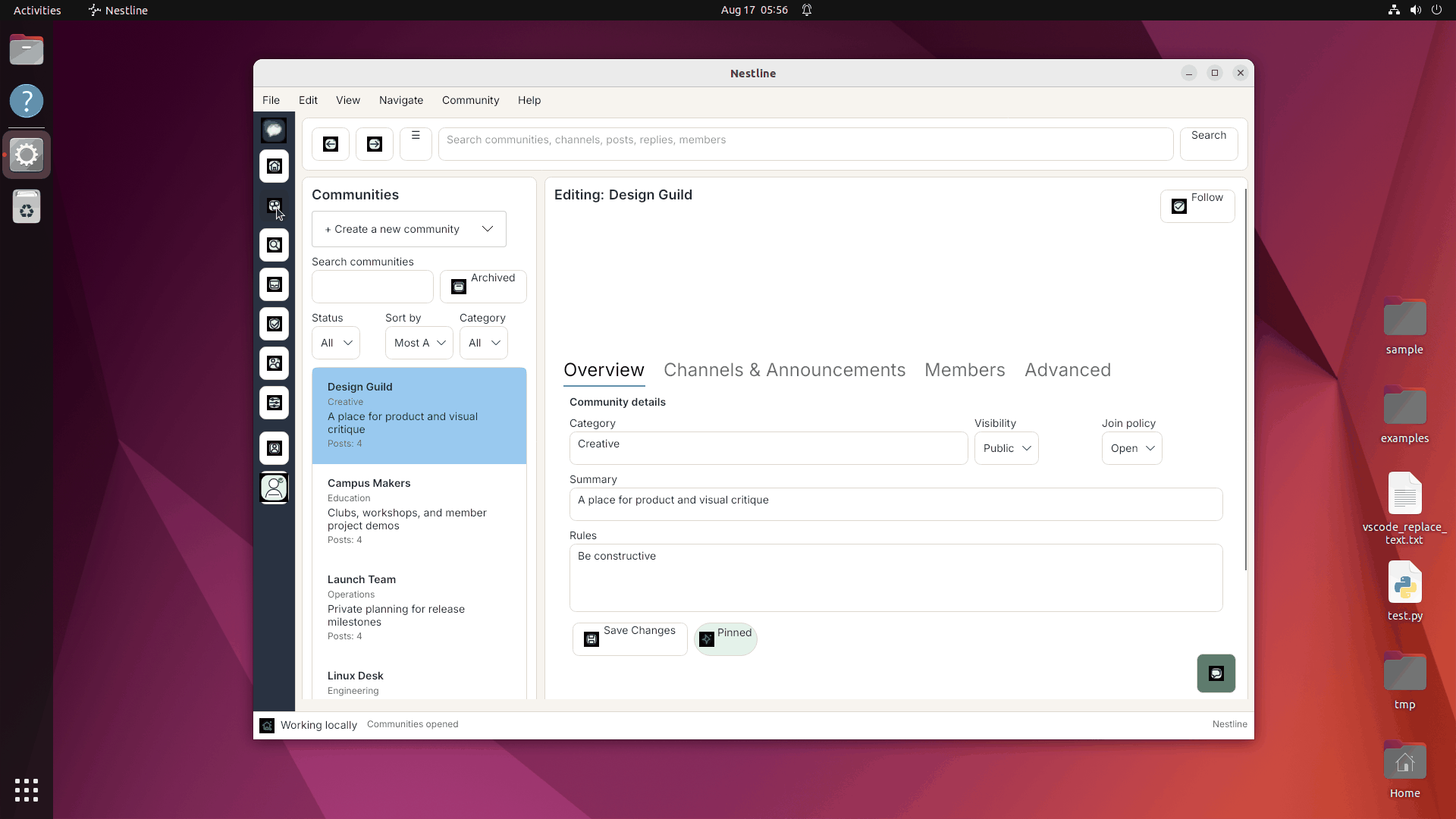}
  \par\vspace{0.6ex}

  {\scriptsize\textbf{Step 2:} Open Communities.\par}
\end{minipage}\hspace{0.012\textwidth}
\begin{minipage}[t]{0.235\textwidth}
  \centering
  \includegraphics[width=\linewidth]{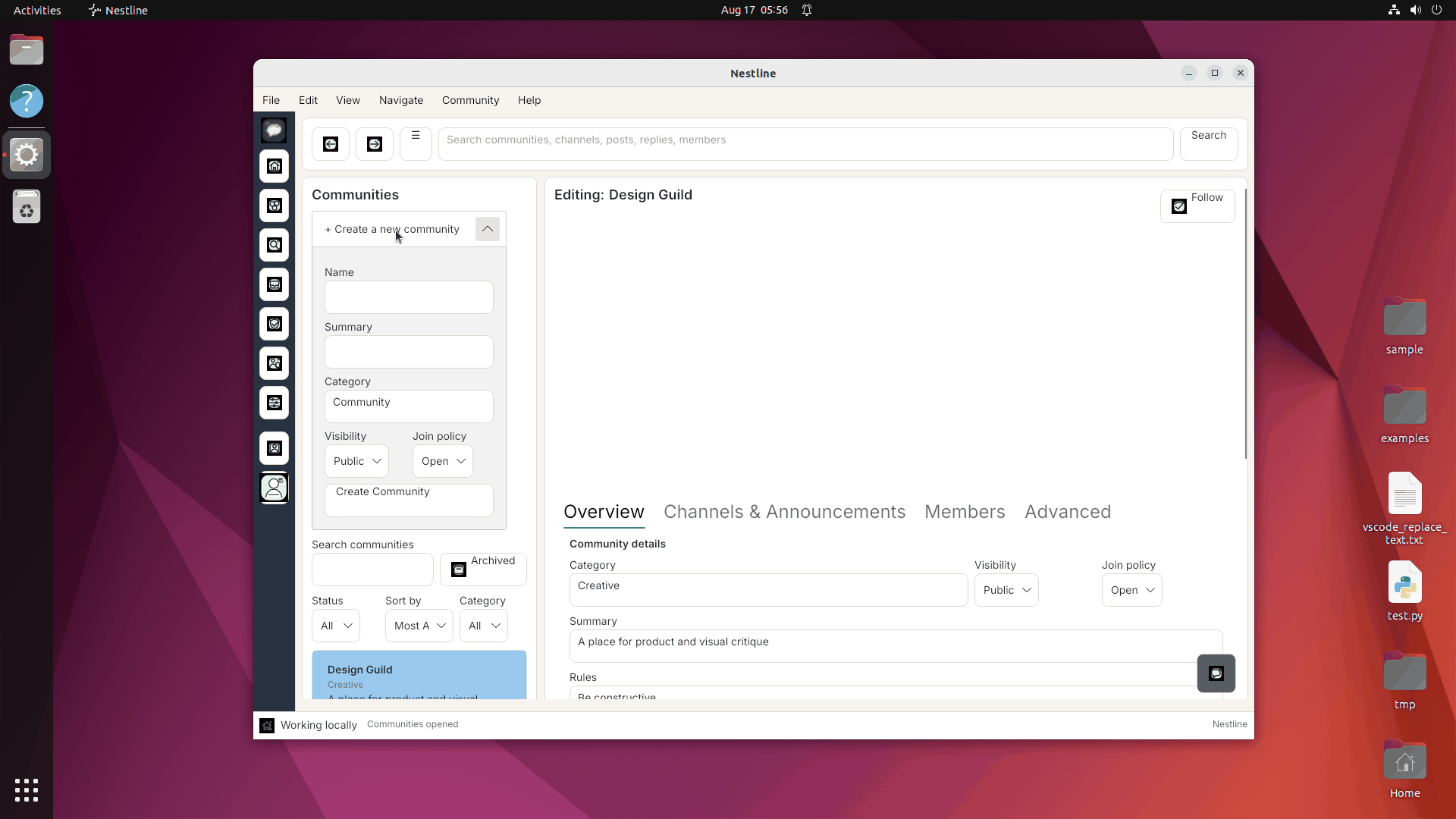}
  \par\vspace{0.6ex}

  {\scriptsize\textbf{Step 3:} Expand creation form.\par}
\end{minipage}\hspace{0.012\textwidth}
\begin{minipage}[t]{0.235\textwidth}
  \centering
  \includegraphics[width=\linewidth]{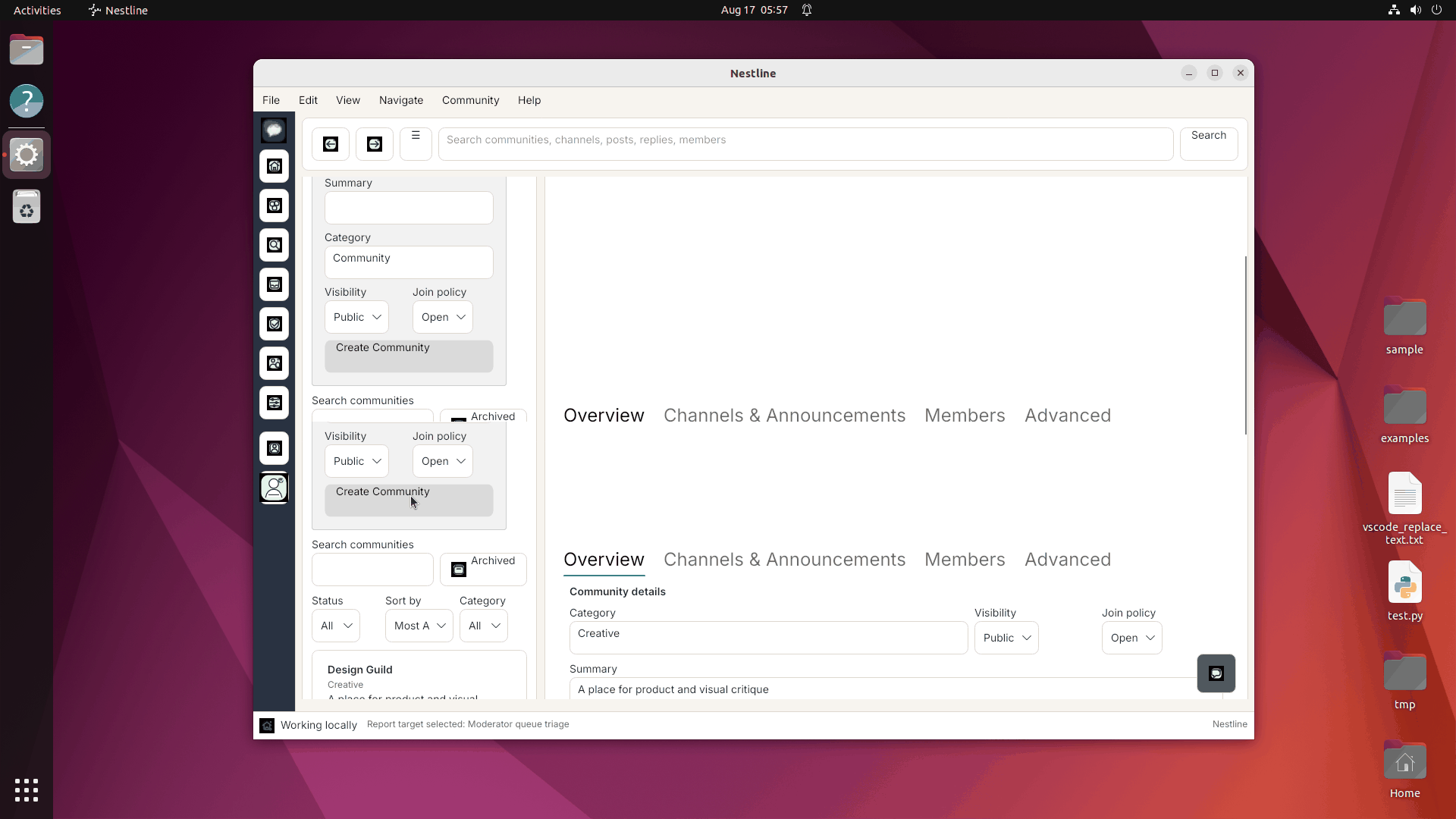}
  \par\vspace{0.6ex}

  {\scriptsize\textbf{Step 4:} Fill fields and submit.\par}
\end{minipage}%
}

\vspace{0.3em}

\makebox[\textwidth][l]{%
\begin{minipage}[t]{0.235\textwidth}
  \centering
  \includegraphics[width=\linewidth]{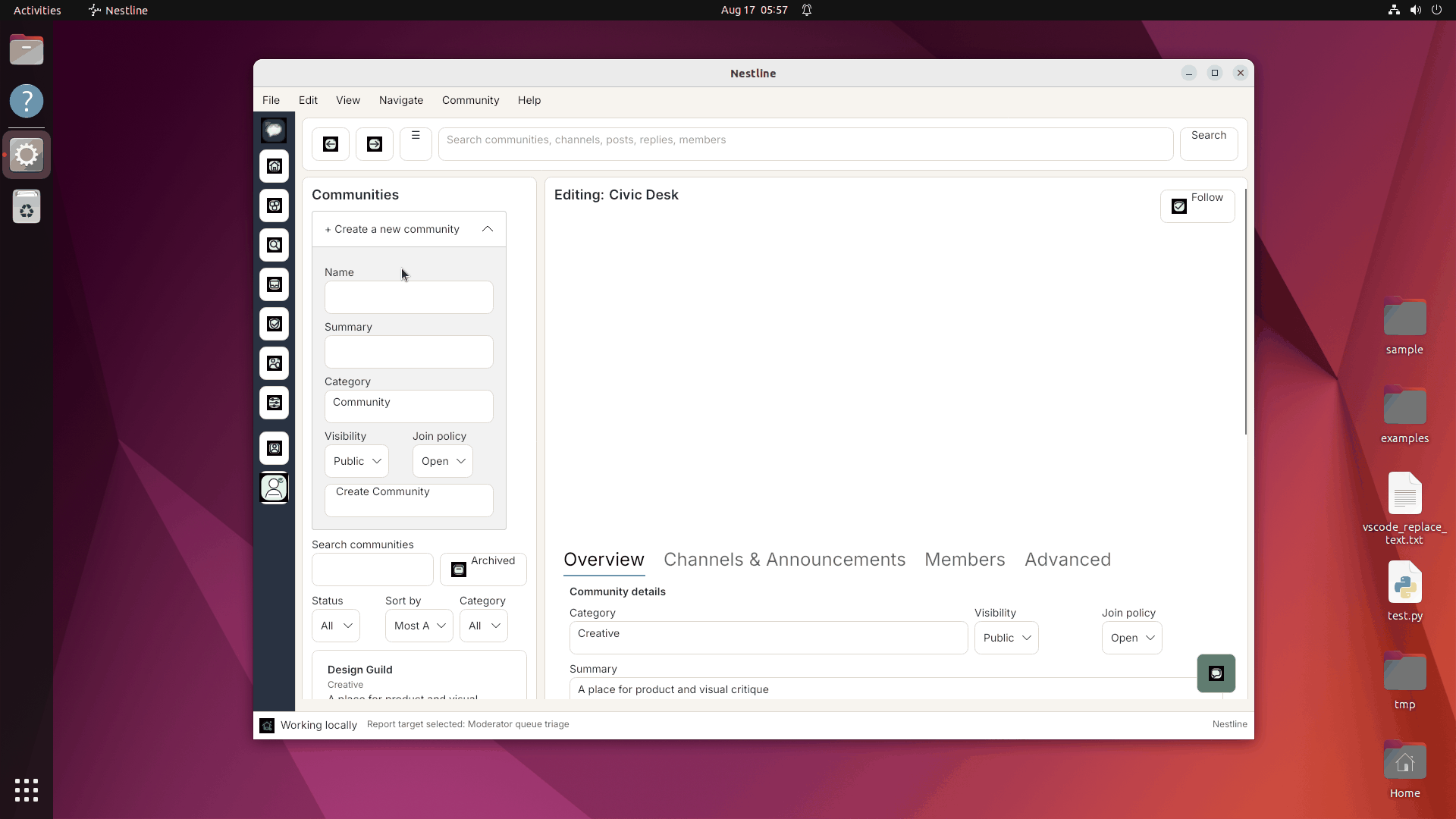}
  \par\vspace{0.6ex}

  {\scriptsize\textbf{Step 5:} Confirm creation.\par}
\end{minipage}\hspace{0.012\textwidth}
\begin{minipage}[t]{0.235\textwidth}
  \centering
  \includegraphics[width=\linewidth]{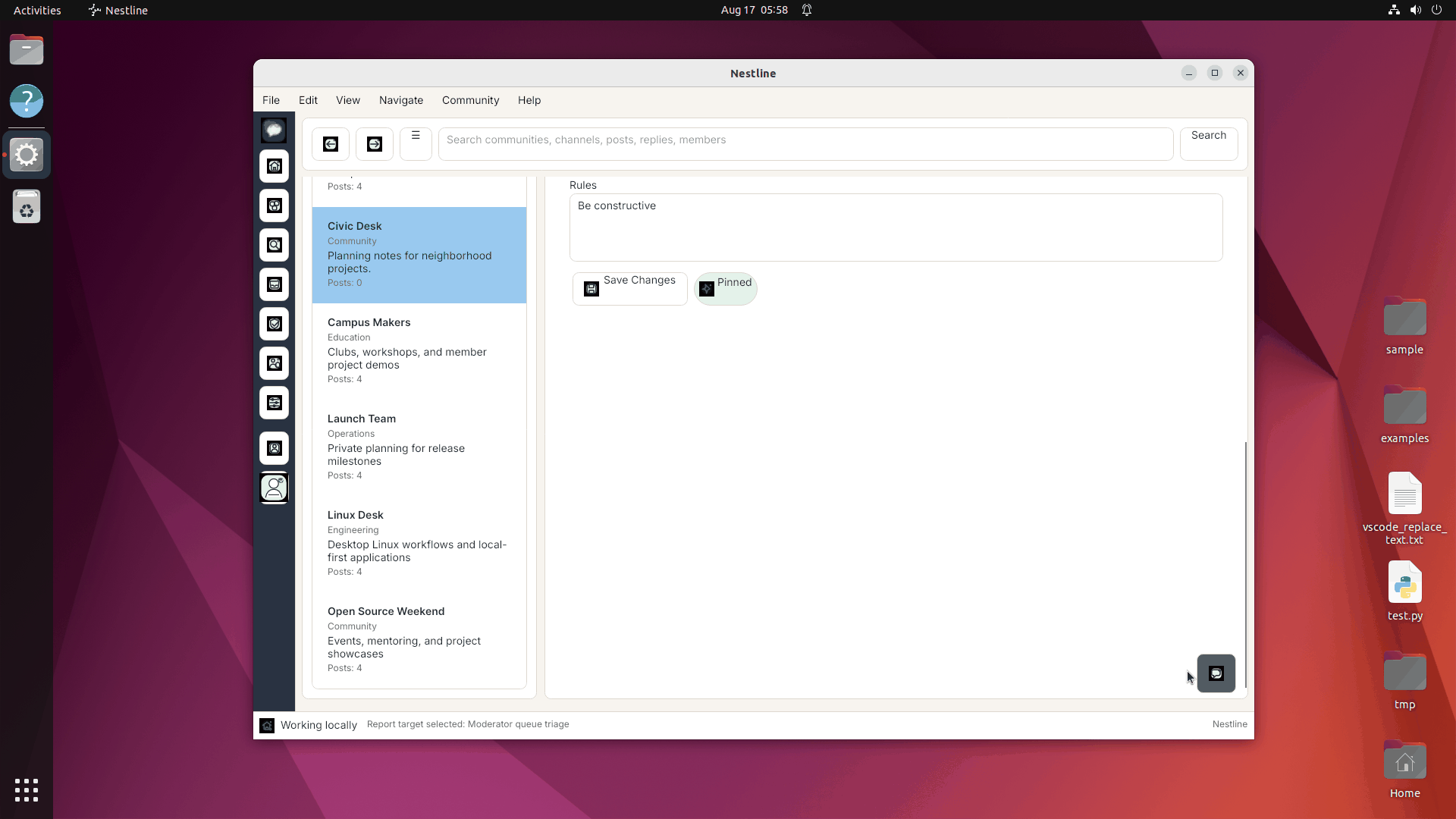}
  \par\vspace{0.6ex}

  {\scriptsize\textbf{Step 6:} Check exact summary.\par}
\end{minipage}\hspace{0.012\textwidth}
\begin{minipage}[t]{0.235\textwidth}
  \centering
  \includegraphics[width=\linewidth]{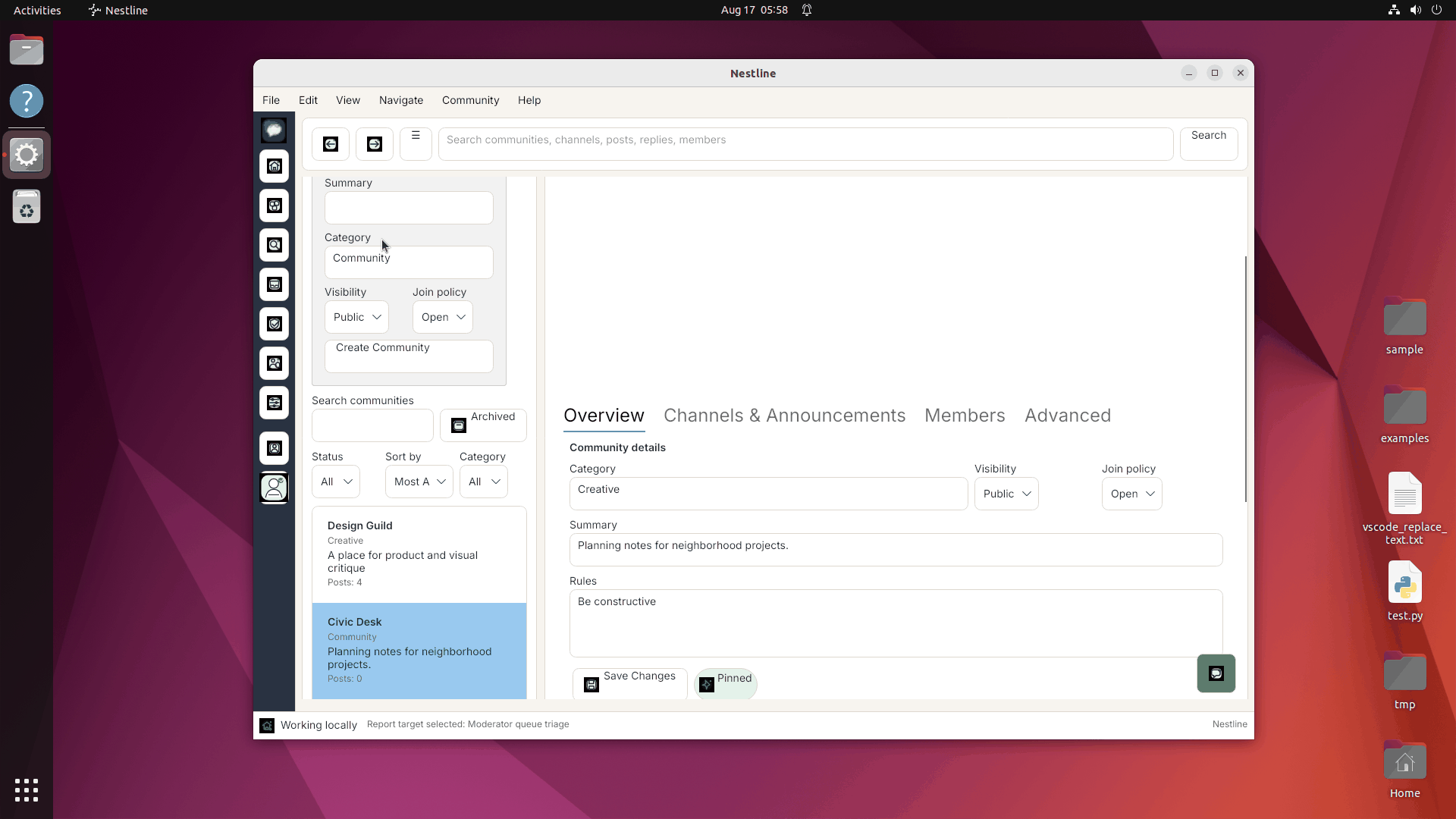}
  \par\vspace{0.6ex}

  {\scriptsize\textbf{Step 7:} Check Public and Open.\par}
\end{minipage}\hspace{0.012\textwidth}
\begin{minipage}[t]{0.235\textwidth}
  \centering
  \includegraphics[width=\linewidth]{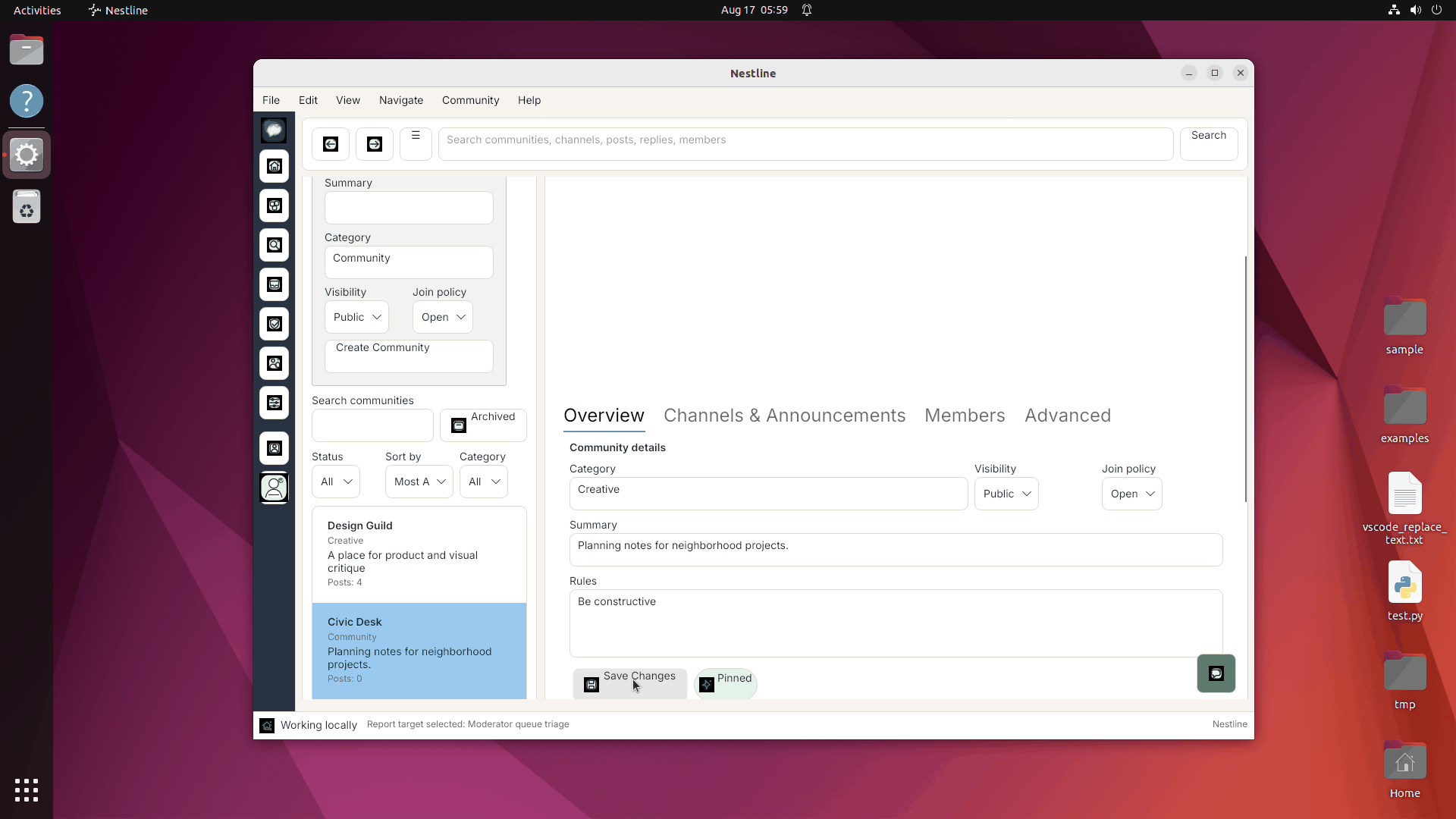}
  \par\vspace{0.6ex}

  {\scriptsize\textbf{Step 8:} Save final settings.\par}
\end{minipage}%
}
\caption{A successful ForumAPP interaction that creates the public, open-join
Civic Desk community with the requested summary.}
\label{fig:forum-app-community-creation-trajectory}
\end{figure}

\noindent\textbf{Instruction:} Please attach project\_brief.txt to the Loom
Council conversation.

\begin{figure}[H]
\centering
\makebox[\textwidth][l]{%
\begin{minipage}[t]{0.188\textwidth}
  \centering
  \includegraphics[width=\linewidth]{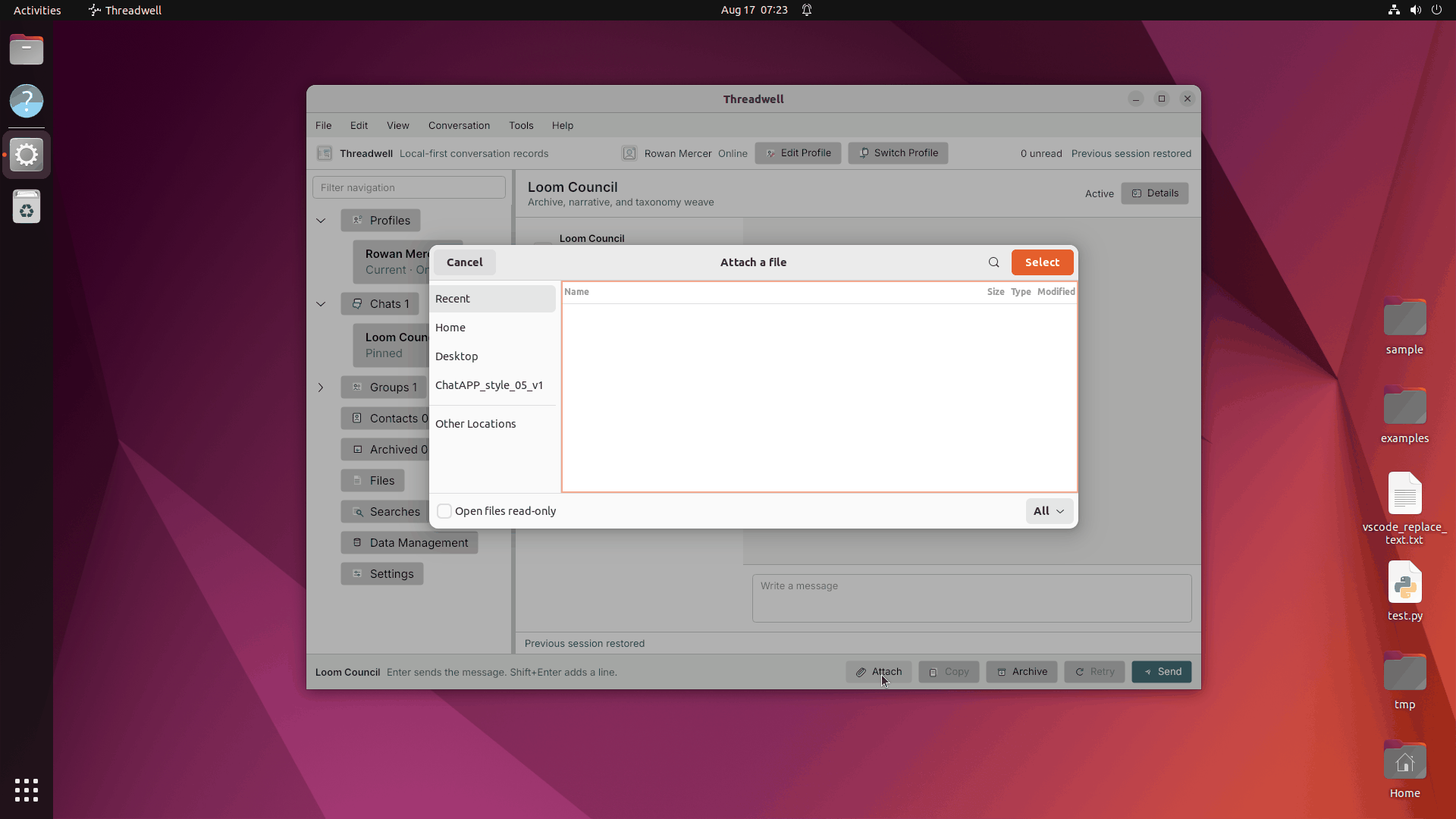}
  \par\vspace{0.6ex}

  {\scriptsize\textbf{Step 1:} Open attachment picker.\par}
\end{minipage}\hspace{0.012\textwidth}
\begin{minipage}[t]{0.188\textwidth}
  \centering
  \includegraphics[width=\linewidth]{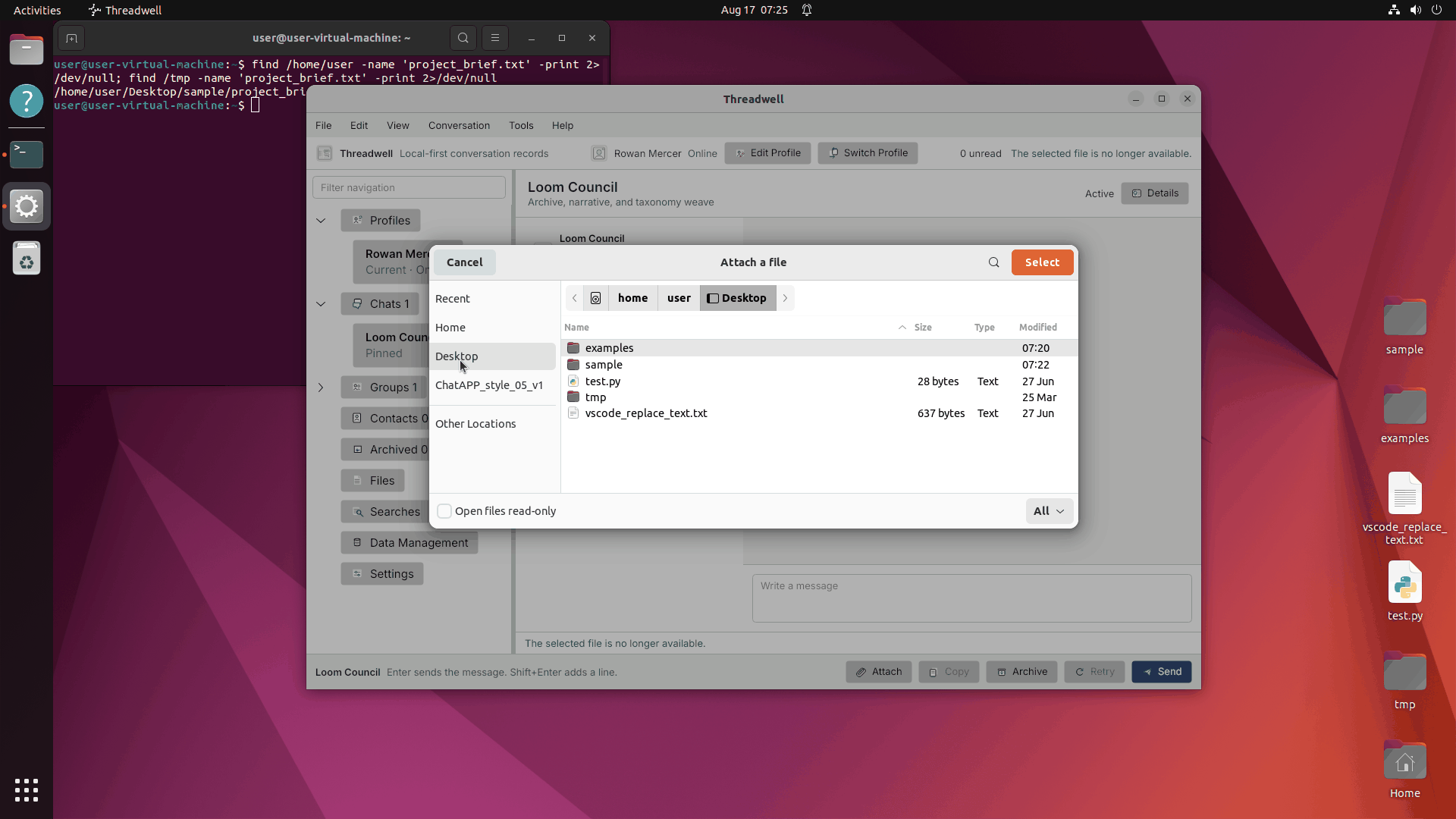}
  \par\vspace{0.6ex}

  {\scriptsize\textbf{Step 2:} Open Desktop.\par}
\end{minipage}\hspace{0.012\textwidth}
\begin{minipage}[t]{0.188\textwidth}
  \centering
  \includegraphics[width=\linewidth]{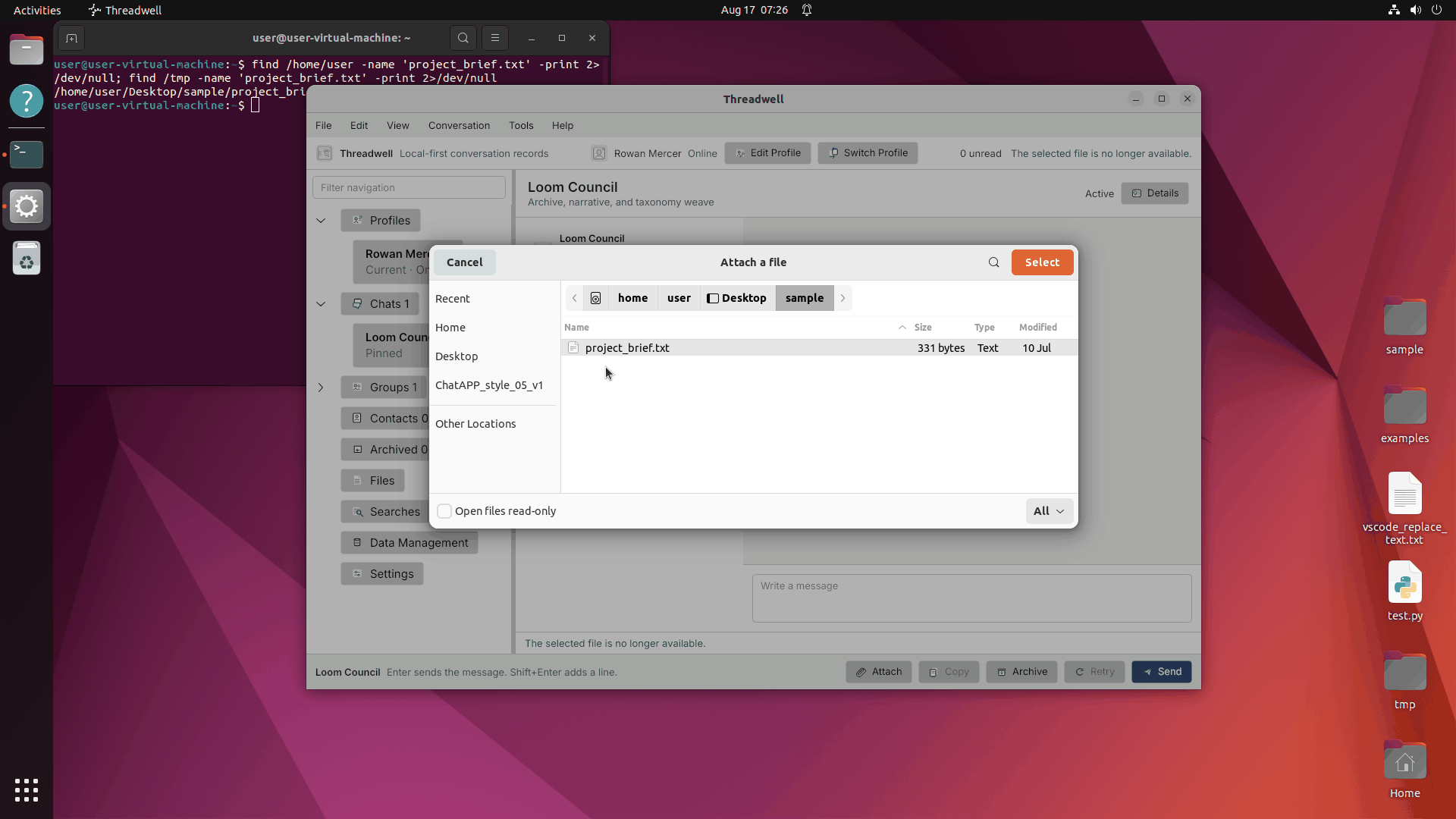}
  \par\vspace{0.6ex}

  {\scriptsize\textbf{Step 3:} Open sample folder.\par}
\end{minipage}\hspace{0.012\textwidth}
\begin{minipage}[t]{0.188\textwidth}
  \centering
  \includegraphics[width=\linewidth]{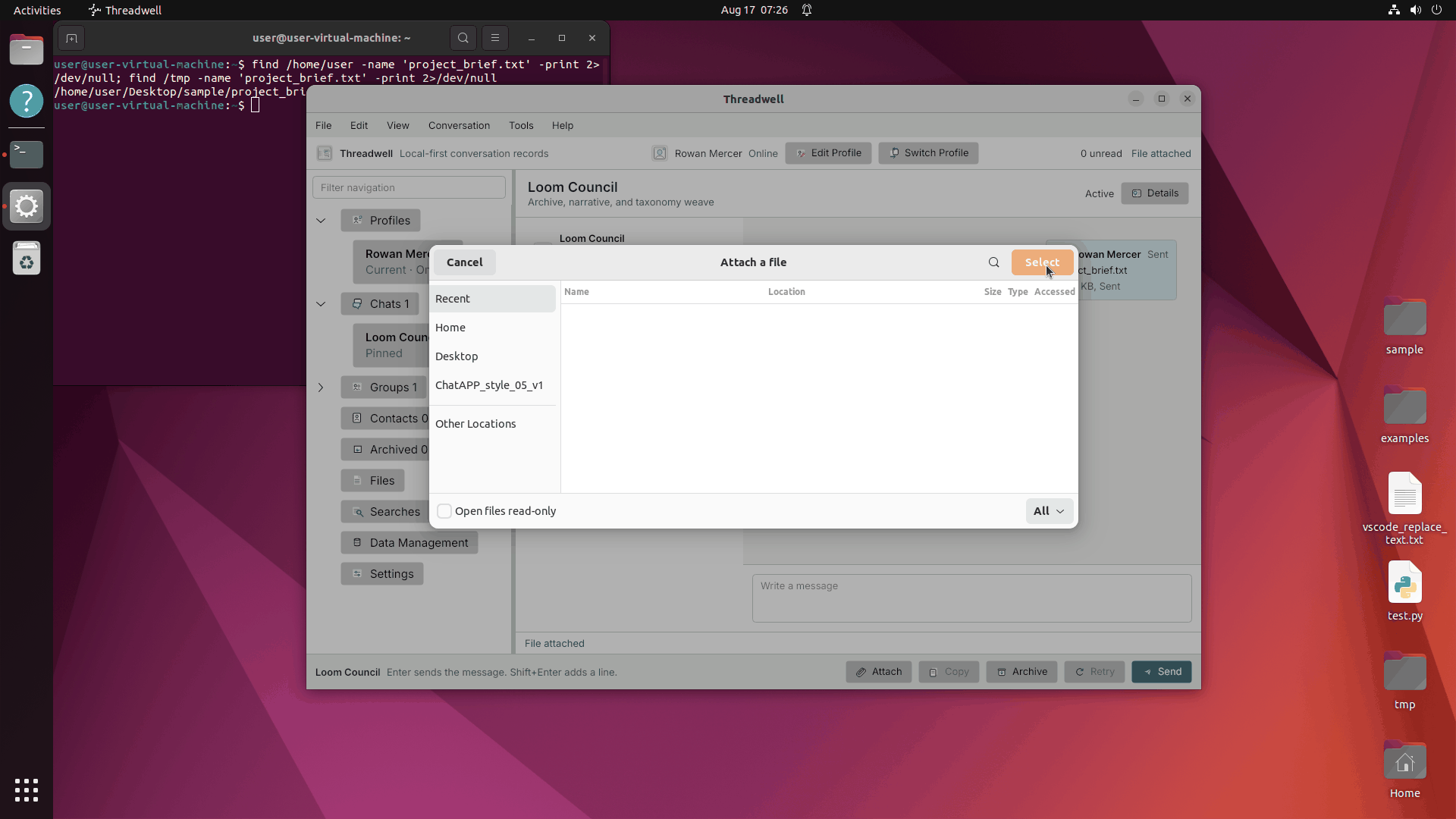}
  \par\vspace{0.6ex}

  {\scriptsize\textbf{Step 4:} Select project\_brief.txt.\par}
\end{minipage}\hspace{0.012\textwidth}
\begin{minipage}[t]{0.188\textwidth}
  \centering
  \includegraphics[width=\linewidth]{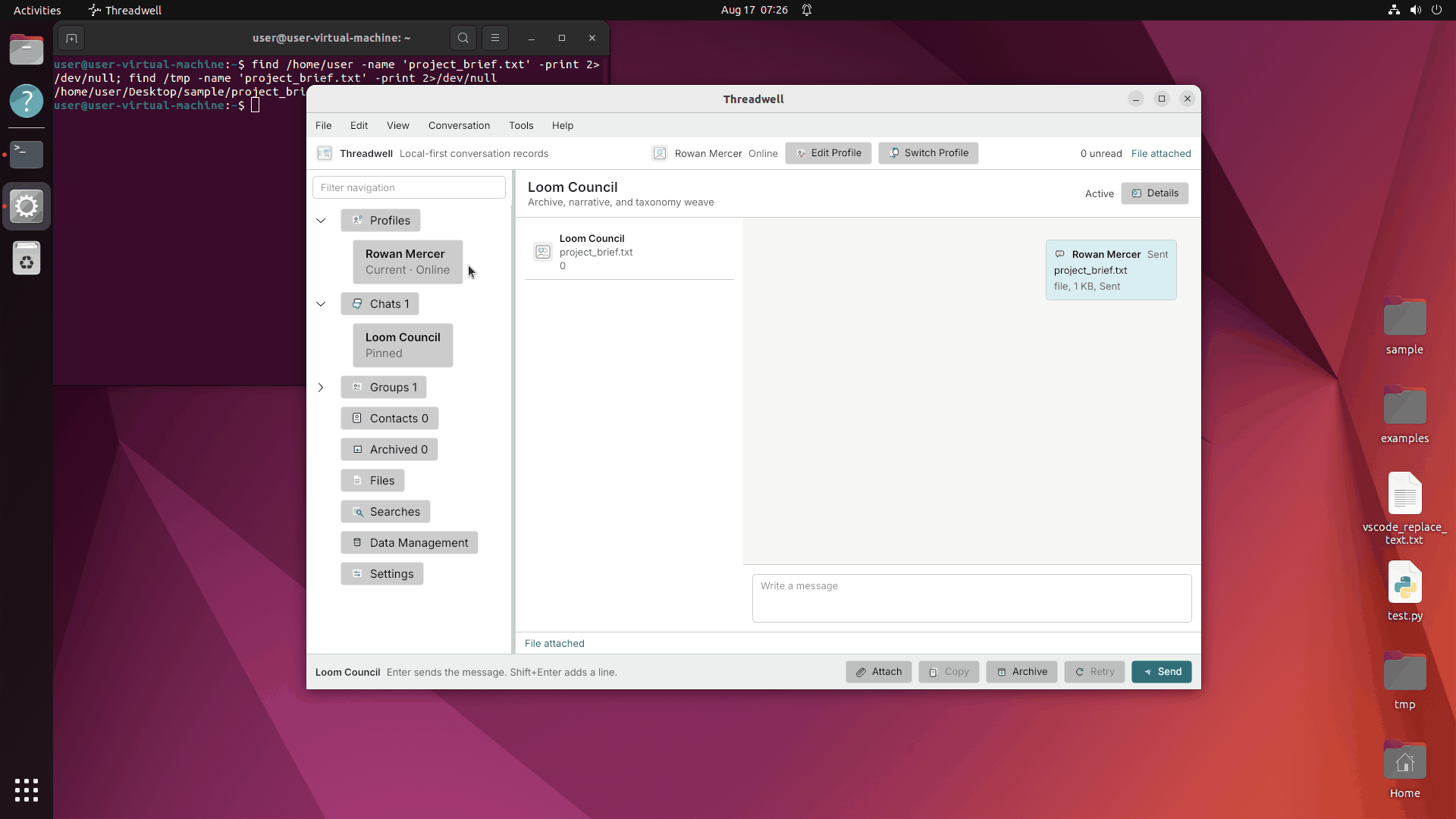}
  \par\vspace{0.6ex}

  {\scriptsize\textbf{Step 5:} Verify Loom Council.\par}
\end{minipage}%
}
\caption{A successful ChatAPP interaction that attaches project\_brief.txt to
the Loom Council conversation.}
\label{fig:chat-app-file-attachment-trajectory}
\end{figure}

\clearpage
\metricdefinitionsappendix

\newcommand{\trajectoryqualityappendix}{%
\onecolumn
\section{Trajectory Scoring and Filtering}
\label{app:trajectory-quality}

Our trajectory-selection procedure asks two complementary questions: whether
the agent reaches the requested final state and whether it follows a reliable
interaction process. After a structural audit, two independent GPT-5.5 judges
score each candidate. Qualified trajectories are converted to Qwen3.5
multimodal format, validated, source-specifically selected, and group-split.

\subsection{Two-stage trajectory scoring}
The outcome and process judges use the following prompts. The user request is
the only task specification;
generated metadata is not ground truth. Outcome scoring uses up to the final
five full assistant messages and their step-aligned post-action screenshots;
process scoring uses every full assistant message and up to ten evenly sampled,
step-aligned screenshots. We retain an attempt only when both scores are at
least 0.8.

\begin{promptbox}{Prompt for Outcome Scoring}
\textbf{System:}\par
You are the OUTCOME judge for a GUI task. Determine whether the agent truly
completed the user's request. The user request is the only task specification;
ignore any generated planning or explanation metadata. You will receive up to
the final five chronological assistant messages. Each message includes the
complete available reasoning and assistant content, followed immediately by
that exact step's paired post-action screenshot when available. Use message
text as context for what the agent attempted, but treat it as an unverified
claim; the paired screenshot is the primary evidence of what actually happened.
Check the final evidence against every material requirement in the user request.
Do not reward an agent merely for declaring success or attempting the right
action. Do not evaluate general path efficiency here except where the final
messages reveal that the requested result was not actually completed. Score
1.0 for clearly complete evidence, 0.7--0.9 for completion with only a minor
deviation, 0.4--0.6 for substantial partial completion, 0.1--0.3 for little
verified progress, and 0.0 for no relevant verified result. Return only JSON:
\texttt{\{"score": <0.0-1.0>, "reason": "<assessment of actual completion
against the user request>"\}}

\medskip
\textbf{User:}\par
\textbf{\# User Request}\par
\texttt{\{user\_request\}}\par
\medskip
\textbf{\# Final Five Assistant Messages (chronological; paired screenshots
follow their exact messages)}\par
\texttt{\{evidence\}}
\end{promptbox}

\begin{promptbox}{Prompt for Process Scoring}
\textbf{System:}\par
You are the PROCESS judge for a GUI task. Judge whether the chronological GUI
trajectory sensibly, safely, and efficiently pursues the user's request. The
user request is the only task specification; ignore generated planning or
explanation metadata. You will receive every chronological assistant message
with its complete available reasoning and assistant content. Up to ten evenly
sampled post-action screenshots are inserted immediately after the exact
messages that produced them. Accept equivalent GUI paths. Consider coherence,
efficiency, error recovery, and unrelated or destructive actions. Explicitly
penalize looping, revisiting the same state without progress, and repeatedly
attempting the same or equivalent action without a meaningful change in
strategy. Do not penalize a limited retry when it is part of sensible error
recovery and the agent adapts based on new evidence. Return only JSON:
\texttt{\{"score": <0.0-1.0>, "reason": "<brief trajectory-quality
assessment against the user request>"\}}

\medskip
\textbf{User:}\par
\textbf{\# User Request}\par
\texttt{\{user\_request\}}\par
\medskip
\textbf{\# Complete Chronological Assistant Messages}\par
\texttt{\{evidence\}}
\end{promptbox}

\subsection{Message conversion, selection, and splitting}
Data from OSWorld and SoftwareGym undergo source-specific selection. For
OSWorld, we retain the highest-scoring qualified attempt from each task
directory; all score-qualified SoftwareGym attempts proceed to conversion.
Selected trajectories from both sources are converted from the teacher schema
to Qwen3.5 multimodal format. Attempts failing strict message-format or
consistency validation are excluded.

After conversion, SoftwareGym caps each frozen semantic group at two trajectories;
OSWorld balances eight domains and difficulty to match corpus and split sizes.
Both use approximately 90/5/5 group-constrained splits without frozen-group
overlap, a guarantee relative to the pipeline's grouping definition.
}

\clearpage
\section{Ablation Protocols and Complete Results}
\label{app:ablation-details}
This section separately examines the style-design procedure, the use of GUI
feedback during implementation repair, and the software source used to collect
training trajectories. Appendix~\ref{app:metric-definitions} defines the
reported metrics, while Appendix~\ref{app:trajectory-quality} details the
trajectory filtering used by the training-source ablation. Selections and
outputs are frozen in versioned manifests.

\subsection{Style-design ablation}
\label{app:style-design-ablation}
The comparison covers Browser, Chat, Code Editor, Forum, Mail, PDF, Rich-text
Editor, Image Editor, Slides, Spreadsheet, Music Player, and Video Player.
Each method generates one set of ten blueprints per category, yielding 120
Full Design blueprints and 120 Simple Design blueprints (240 total). Both
methods receive the same frozen functional requirements and icon inventories.

Full Design generates distinct directions and assigns each to an independent
designer. It enforces variation in information architecture, layout, visual
system, interaction model, and product identity, and rejects structurally
redundant designs. Simple Design receives none of these style-control
instructions; it is told only that the ten final layouts should differ. Both
methods use the same artifact schema and final validators.

For each category and method, we average the normalized tree-edit distance over
the 45 pairs formed by its ten blueprints, using the same corpus-level
structural representation as the production diversity gate. The overall score
is the macro-average of the twelve category scores; Appendix~\ref{app:metric-definitions}
defines the tree representation and distance. Full Design produces more
diverse blueprint structures in every category and raises the macro-average
score from 4.9 to 83.6. Although all Simple blueprints differ at the file
level, each category contains at least one pair with the same normalized
layout tree, and none of the twelve Simple sets passes the production
diversity gate.

\begin{table}[h]
\caption{Full Design consistently produces more diverse blueprint structures
across all twelve categories. Results are on a
0--100 scale; each category contains one set of ten blueprints per method.}
\label{tab:style-ablation-full}
\centering
\small
\setlength{\tabcolsep}{7pt}
\begin{tabular*}{\textwidth}{@{\extracolsep{\fill}}lrrlrr@{}}
\toprule
Category & Full & Simple & Category & Full & Simple \\
\midrule
Browser & 87.5 & 1.8 & Mail & 83.9 & 3.8 \\
Chat & 81.3 & 1.6 & Rich-text Editor & 85.3 & 2.4 \\
Code Editor & 85.8 & 3.6 & Slides & 84.4 & 5.7 \\
Forum & 82.1 & 5.4 & Image Editor & 81.5 & 21.1 \\
Spreadsheet & 78.4 & 2.6 & Video Player & 86.6 & 4.3 \\
Music Player & 81.5 & 3.1 & PDF & 85.5 & 3.4 \\
\midrule
\multicolumn{4}{l}{Macro-average across twelve categories} & 83.6 & 4.9 \\
\bottomrule
\end{tabular*}
\end{table}

\subsection{Evaluation-guided implementation-repair ablation}
\label{app:gui-repair-ablation}
The implementation-repair comparison uses a category-balanced paired set of
three applications from each of the twelve categories ($n=36$), all with
complete artifacts from the shared starting checkpoint through both repair
rounds.

Both branches start from the same $v_1$ source package and use the same two
implementation-repair rounds. We run the same GUI evaluation after every round
for both branches. In the GUI-guided branch, the resulting report is passed to
the Coder for the next repair round; in the source-only branch, the report is
recorded but withheld, so the next repair uses source-code evidence alone.

Scores average GUI task pass rates over the 36 applications while giving each
of the twelve categories equal weight. GUI feedback is decisive:
from the shared 50.4\% starting score, evaluation-guided repair reaches 70.5\%
at $v_2$ and 83.1\% at $v_3$, while source-only repair reaches 50.5\% and then
falls below the shared starting score to 48.8\%. This regression shows that
source-level self-correction alone is insufficient and can even degrade
runtime functionality, making feedback from direct GUI testing essential for
reliable repair. The final 34.3-point advantage is large and consistent (95\%
CI: 28.2--41.0); GUI-guided repair wins on all 36 applications.

\begin{table}[h]
\caption{Evaluation-guided implementation repair substantially improves
functional reliability, whereas source-only repair remains near the shared
starting level. Macro-average GUI task pass rate with 95\% hierarchical
bootstrap confidence intervals.}
\label{tab:gui-ablation-full}
\centering
\small
\begin{tabular*}{\textwidth}{@{\extracolsep{\fill}}lccc@{}}
\toprule
Method & $v_1$ & $v_2$ & $v_3$ \\
\midrule
Evaluation-guided & 50.4 [42.6, 57.9] & 70.5 [64.7, 76.2] & 83.1 [79.2, 86.8] \\
Source-only & 50.4 [42.6, 57.9] & 50.5 [43.1, 57.7] & 48.8 [42.7, 54.7] \\
\bottomrule
\end{tabular*}
\end{table}

\FloatBarrier
\subsection{Training-software-source ablation}
\label{app:training-configuration}
Finally, we train three Qwen3.5-9B models: one on trajectories collected only
from synthesized SoftwareGym applications, one only from native OSWorld
applications, and one on both sources. Before length filtering, the
SoftwareGym-only and OSWorld-only datasets are each matched to 3,715
trajectories and 55,263 serialized examples, with identical
train/validation/test sizes. Both datasets use the same 0.8 outcome and process
thresholds and message-conversion protocol; their post-threshold selection is
source-specific, as specified in Appendix~\ref{app:trajectory-quality}.

For the model trained on both sources, we concatenate the complete frozen
SoftwareGym-only and OSWorld-only datasets. Because the two datasets are
size-matched, each source contributes equally. We preserve their
source-prefixed semantic identifiers and original splits, without additional
attempt sampling or rewriting.

All three models are initialized from Qwen3.5-9B Base and use full-parameter
SFT of the language model for one epoch; the visual encoder and merger are
frozen. The SoftwareGym-only model is trained with 16 A800 GPUs. The
OSWorld-only model and the model trained on both sources are each trained with
8 A800 GPUs. We use bf16, FlashAttention, ZeRO-2, gradient
checkpointing, learning rate
$5\times10^{-6}$, 3\% warmup, cosine decay, sequence length 16,384,
and micro-batch size 1. Gradient accumulation is 2 for the SoftwareGym-only
model and 4 for the OSWorld-only model and the model trained on both sources,
keeping the global batch size at 32. Overlength examples are removed rather
than truncated, leaving 49,445 examples for SoftwareGym-only training, 49,496
for OSWorld-only training, and 98,941 for training on both sources. Evaluation
fixes the context and tool protocol, temperature 1, top-$p$ 0.95, output limit
8,192, 100 interaction steps, and three independent rollouts per task.

Tables~\ref{tab:training-source-osworld-full}
and~\ref{tab:training-source-softwaregym-full} give the complete run-level
results, while Table~\ref{tab:data-composition-ablation} summarizes Pass@3 in
the main text.
Synthesized-software supervision transfers strongly to unseen native
applications and complements native-software data. On OSWorld, the
SoftwareGym-only model obtains 56.97\% Pass@3 without any OSWorld application,
task, or trajectory in training, compared with 61.48\% for the OSWorld-only
model. On SoftwareGym-Bench, the SoftwareGym-only model obtains 40.0\%, whereas
the OSWorld-only model obtains 33.3\%. The model trained on both sources further demonstrates their
complementarity, reaching 47.5\% on SoftwareGym-Bench while retaining 58.61\%
on OSWorld. Matching the base model, pre-tokenization corpus size, effective
batch size, core optimization settings, and evaluation makes the source of the
training trajectories the principal difference between the SoftwareGym-only
and OSWorld-only models.

\begin{center}
\centering
\captionof{table}{Complete OSWorld results for the training-software-source ablation.
Run rows report average evaluator scores; Avg@3 averages the three runs, and
Pass@3 reports task-level success across them.}
\label{tab:training-source-osworld-full}
\scriptsize
\setlength{\tabcolsep}{4.70pt}
\setlength{\aboverulesep}{0pt}
\setlength{\belowrulesep}{0pt}
\renewcommand{\arraystretch}{0.92}
\begin{tabular*}{\textwidth}{@{}p{0.205\textwidth}c*{8}{c}>{\columncolor{overallcolumnbg}[3pt][0pt]}c@{}}
\toprule
\textbf{Training data} & Run & Chrome & GIMP & Calc & Impress & Writer & Thunderbird & VLC & VS Code & \textbf{Overall} \\
\midrule
SoftwareGym & Run 1 & 54.26 & 65.38 & 23.40 & 42.46 & 52.16 & 66.67 & 34.74 & 56.52 & 46.65 \\
 & Run 2 & 45.65 & 73.08 & 25.53 & 42.46 & 52.16 & 53.33 & 50.02 & 60.87 & 46.91 \\
 & Run 3 & 52.09 & 69.23 & 23.40 & 44.59 & 52.16 & 66.67 & 30.39 & 56.52 & 46.75 \\
 & \textbf{Avg@3} & 50.67 & 69.23 & 24.11 & 43.17 & 52.16 & 62.22 & 38.38 & 57.97 & 46.77 \\
 & \textbf{Pass@3} & 58.70 & 84.62 & 31.91 & 53.19 & 65.22 & 73.33 & 58.82 & 60.87 & 56.97 \\
\noalign{\vskip -0.25ex}\hhline{~----------}\noalign{\vskip 0.25ex}
OSWorld & Run 1 & 52.09 & 65.38 & 27.66 & 53.10 & 65.21 & 66.67 & 41.18 & 47.83 & 49.97 \\
 & Run 2 & 49.91 & 73.08 & 34.04 & 46.83 & 52.16 & 60.00 & 47.94 & 39.13 & 48.41 \\
 & Run 3 & 41.30 & 73.08 & 27.66 & 46.80 & 52.16 & 53.33 & 45.81 & 47.83 & 45.81 \\
 & \textbf{Avg@3} & 47.77 & 70.51 & 29.79 & 48.91 & 56.51 & 60.00 & 44.97 & 44.93 & 48.06 \\
 & \textbf{Pass@3} & 54.35 & 84.62 & 44.68 & 68.09 & 69.57 & 73.33 & 64.71 & 52.17 & 61.48 \\
\noalign{\vskip -0.25ex}\hhline{~----------}\noalign{\vskip 0.25ex}
OSWorld + SoftwareGym & Run 1 & 52.09 & 65.38 & 29.79 & 46.80 & 52.16 & 60.00 & 52.94 & 43.48 & 47.93 \\
 & Run 2 & 45.56 & 57.69 & 34.04 & 42.46 & 69.56 & 60.00 & 40.63 & 43.48 & 46.65 \\
 & Run 3 & 47.83 & 57.69 & 27.66 & 42.54 & 56.51 & 66.67 & 52.94 & 52.17 & 46.72 \\
 & \textbf{Avg@3} & 48.49 & 60.26 & 30.50 & 43.93 & 59.41 & 62.22 & 48.84 & 46.38 & 47.10 \\
 & \textbf{Pass@3} & 58.70 & 65.38 & 46.81 & 59.57 & 65.22 & 66.67 & 64.71 & 56.52 & 58.61 \\
\bottomrule
\end{tabular*}
\end{center}

\begin{center}
\centering
\captionof{table}{Complete SoftwareGym-Bench results for the
training-software-source ablation. Run rows report binary success rates;
Avg@3 averages the three runs, and Pass@3 reports task-level success across
them.}
\label{tab:training-source-softwaregym-full}
\scriptsize
\setlength{\tabcolsep}{2.78pt}
\setlength{\aboverulesep}{0pt}
\setlength{\belowrulesep}{0pt}
\renewcommand{\arraystretch}{0.92}
\begin{tabular*}{\textwidth}{@{}p{0.14\textwidth}c*{12}{c}>{\columncolor{overallcolumnbg}[2pt][0pt]}c@{}}
\toprule
\textbf{Training data} & Run & Pic & Mail & PDF & Music & Code & Forum & Writer & Browser & Video & Slides & Table & Chat & \textbf{Overall} \\
\midrule
SoftwareGym & Run 1 & 30.00 & 30.00 & 40.00 & 50.00 & 50.00 & 40.00 & 20.00 & 30.00 & 20.00 & 10.00 & 0.00 & 40.00 & 30.00 \\
 & Run 2 & 30.00 & 10.00 & 50.00 & 40.00 & 20.00 & 20.00 & 20.00 & 40.00 & 20.00 & 20.00 & 10.00 & 30.00 & 25.83 \\
 & Run 3 & 30.00 & 30.00 & 50.00 & 20.00 & 70.00 & 20.00 & 10.00 & 30.00 & 0.00 & 0.00 & 10.00 & 20.00 & 24.17 \\
 & \textbf{Avg@3} & 30.00 & 23.33 & 46.67 & 36.67 & 46.67 & 26.67 & 16.67 & 33.33 & 13.33 & 10.00 & 6.67 & 30.00 & 26.67 \\
 & \textbf{Pass@3} & 40.00 & 40.00 & 60.00 & 50.00 & 70.00 & 40.00 & 30.00 & 60.00 & 20.00 & 20.00 & 10.00 & 40.00 & 40.00 \\
\noalign{\vskip -0.25ex}\hhline{~--------------}\noalign{\vskip 0.25ex}
OSWorld & Run 1 & 20.00 & 20.00 & 50.00 & 30.00 & 30.00 & 20.00 & 0.00 & 10.00 & 20.00 & 10.00 & 0.00 & 20.00 & 19.17 \\
 & Run 2 & 10.00 & 10.00 & 60.00 & 30.00 & 20.00 & 10.00 & 0.00 & 20.00 & 20.00 & 20.00 & 10.00 & 10.00 & 18.33 \\
 & Run 3 & 10.00 & 30.00 & 50.00 & 20.00 & 40.00 & 20.00 & 20.00 & 30.00 & 10.00 & 20.00 & 10.00 & 40.00 & 25.00 \\
 & \textbf{Avg@3} & 13.33 & 20.00 & 53.33 & 26.67 & 30.00 & 16.67 & 6.67 & 20.00 & 16.67 & 16.67 & 6.67 & 23.33 & 20.83 \\
 & \textbf{Pass@3} & 20.00 & 30.00 & 60.00 & 30.00 & 50.00 & 30.00 & 20.00 & 30.00 & 40.00 & 30.00 & 20.00 & 40.00 & 33.33 \\
\noalign{\vskip -0.25ex}\hhline{~--------------}\noalign{\vskip 0.25ex}
OSWorld + SoftwareGym & Run 1 & 50.00 & 40.00 & 50.00 & 40.00 & 40.00 & 30.00 & 10.00 & 50.00 & 10.00 & 20.00 & 50.00 & 30.00 & 35.00 \\
 & Run 2 & 30.00 & 30.00 & 40.00 & 40.00 & 80.00 & 30.00 & 10.00 & 50.00 & 20.00 & 30.00 & 10.00 & 20.00 & 32.50 \\
 & Run 3 & 30.00 & 20.00 & 50.00 & 50.00 & 50.00 & 30.00 & 10.00 & 50.00 & 20.00 & 10.00 & 10.00 & 30.00 & 30.00 \\
 & \textbf{Avg@3} & 36.67 & 30.00 & 46.67 & 43.33 & 56.67 & 30.00 & 10.00 & 50.00 & 16.67 & 20.00 & 23.33 & 26.67 & 32.50 \\
 & \textbf{Pass@3} & 50.00 & 40.00 & 80.00 & 50.00 & 80.00 & 50.00 & 10.00 & 60.00 & 20.00 & 40.00 & 50.00 & 40.00 & 47.50 \\
\bottomrule
\end{tabular*}
\end{center}

\clearpage
\section{Complete Per-run Results for the Main Evaluation}
\label{app:main-per-run-results}
To complement the task-level main-result tables, this appendix reports the
complete average evaluator scores underlying the
main results. For each model, we report all three runs, their arithmetic
mean (Avg@3), and task-level Pass@3 for every application domain. For a domain $d$ with $N_d$ tasks, the
score of run $r$ is $\operatorname{Avg}_{d,r}=100N_d^{-1}
\sum_{i=1}^{N_d}s_{i,r}$, where $s_{i,r}$ is the evaluator reward. These
per-run averages describe individual-run performance and
are distinct from the task-level Pass@3 metric in
Tables~\ref{tab:main-osworld} and~\ref{tab:main-softwaregym}.

\subsection{OSWorld}
Each run covers the same 244 tasks across eight application domains.

\begingroup
\scriptsize
\setlength{\tabcolsep}{4.70pt}
\setlength{\aboverulesep}{0pt}
\setlength{\belowrulesep}{0pt}
\renewcommand{\arraystretch}{0.92}
\begin{longtable}{@{}p{0.205\textwidth}c*{8}{c}>{\columncolor{overallcolumnbg}[3pt][0pt]}c@{}}
\caption{Run 1--3 scores, Avg@3, and Pass@3 (\%) on OSWorld.}
\label{tab:osworld-per-run-full}\\
\toprule
\textbf{Model} & Run & Chrome & GIMP & Calc & Impress & Writer & Thunderbird & VLC & VS Code & \textbf{Overall} \\
\midrule
\endfirsthead
\multicolumn{11}{c}{\tablename~\thetable\ (continued)}\\
\toprule
\textbf{Model} & Run & Chrome & GIMP & Calc & Impress & Writer & Thunderbird & VLC & VS Code & \textbf{Overall} \\
\midrule
\endhead
\bottomrule
\endfoot
Seed 2.1 Pro & Run 1 & 67.26 & 46.15 & 87.23 & 73.41 & 69.56 & 86.67 & 80.49 & 69.57 & 72.59 \\
 & Run 2 & 58.56 & 50.00 & 76.60 & 71.28 & 82.60 & 86.67 & 86.37 & 65.22 & 70.13 \\
 & Run 3 & 65.08 & 61.54 & 70.21 & 64.90 & 73.90 & 86.67 & 74.61 & 65.22 & 68.49 \\
 & \textbf{Avg@3} & 63.63 & 52.56 & 78.01 & 69.86 & 75.35 & 86.67 & 80.49 & 66.67 & 70.41 \\
 & \textbf{Pass@3} & 71.74 & 73.08 & 93.62 & 74.47 & 78.26 & 86.67 & 76.47 & 82.61 & 79.51 \\
\noalign{\vskip -0.25ex}\hhline{~----------}\noalign{\vskip 0.25ex}
Seed 2.0 Pro & Run 1 & 60.73 & 57.69 & 68.09 & 64.90 & 69.16 & 93.33 & 80.49 & 69.57 & 67.64 \\
 & Run 2 & 60.73 & 76.92 & 78.72 & 62.77 & 78.12 & 86.67 & 76.47 & 69.57 & 71.48 \\
 & Run 3 & 65.08 & 84.62 & 76.60 & 60.64 & 60.86 & 86.67 & 80.54 & 47.83 & 68.91 \\
 & \textbf{Avg@3} & 62.18 & 73.08 & 74.47 & 62.77 & 69.38 & 88.89 & 79.17 & 62.32 & 69.34 \\
 & \textbf{Pass@3} & 71.74 & 92.31 & 91.49 & 65.96 & 69.57 & 93.33 & 82.35 & 78.26 & 79.10 \\
\noalign{\vskip -0.25ex}\hhline{~----------}\noalign{\vskip 0.25ex}
Qwen3.8-Flash & Run 1 & 45.52 & 57.69 & 61.70 & 44.79 & 69.56 & 60.00 & 58.74 & 47.83 & 54.09 \\
 & Run 2 & 54.21 & 57.69 & 63.83 & 63.83 & 56.52 & 53.33 & 69.03 & 43.48 & 58.47 \\
 & Run 3 & 34.78 & 42.31 & 57.45 & 63.89 & 60.86 & 40.00 & 75.77 & 52.17 & 52.83 \\
 & \textbf{Avg@3} & 44.84 & 52.56 & 60.99 & 57.50 & 62.31 & 51.11 & 67.85 & 47.83 & 55.13 \\
 & \textbf{Pass@3} & 60.87 & 80.77 & 89.36 & 74.47 & 78.26 & 73.33 & 76.47 & 60.87 & 74.59 \\
\noalign{\vskip -0.25ex}\hhline{~----------}\noalign{\vskip 0.25ex}
GPT-5.4 & Run 1 & 54.21 & 53.85 & 76.60 & 54.29 & 69.43 & 86.67 & 88.16 & 52.17 & 64.10 \\
 & Run 2 & 54.21 & 65.38 & 76.60 & 58.13 & 65.20 & 86.67 & 76.39 & 56.52 & 65.26 \\
 & Run 3 & 45.52 & 69.23 & 72.34 & 68.79 & 73.90 & 80.00 & 84.48 & 60.87 & 66.65 \\
 & \textbf{Avg@3} & 51.31 & 62.82 & 75.18 & 60.40 & 69.51 & 84.44 & 83.01 & 56.52 & 65.34 \\
 & \textbf{Pass@3} & 60.87 & 76.92 & 82.98 & 70.21 & 73.91 & 86.67 & 82.35 & 69.57 & 73.77 \\
\noalign{\vskip -0.25ex}\hhline{~----------}\noalign{\vskip 0.25ex}
GPT-5.5 & Run 1 & 34.78 & 53.85 & 76.60 & 75.22 & 65.21 & 73.33 & 87.28 & 73.91 & 65.24 \\
 & Run 2 & 28.26 & 53.85 & 80.85 & 84.31 & 82.47 & 73.33 & 81.11 & 65.22 & 66.96 \\
 & Run 3 & 28.26 & 61.54 & 76.60 & 75.54 & 73.77 & 66.67 & 93.49 & 65.22 & 64.90 \\
 & \textbf{Avg@3} & 30.43 & 56.41 & 78.01 & 78.36 & 73.82 & 71.11 & 87.29 & 68.12 & 65.70 \\
 & \textbf{Pass@3} & 34.78 & 65.38 & 85.11 & 85.11 & 73.91 & 80.00 & 82.35 & 78.26 & 71.31 \\
\noalign{\vskip -0.25ex}\hhline{~----------}\noalign{\vskip 0.25ex}
Gemini 3.1 Pro & Run 1 & 45.65 & 76.92 & 89.36 & 55.23 & 78.12 & 80.00 & 75.59 & 95.65 & 71.22 \\
 & Run 2 & 43.48 & 84.62 & 91.49 & 64.93 & 82.47 & 80.00 & 81.11 & 82.61 & 73.47 \\
 & Run 3 & 47.83 & 80.77 & 91.49 & 63.80 & 91.29 & 80.00 & 81.11 & 95.65 & 75.73 \\
 & \textbf{Avg@3} & 45.65 & 80.77 & 90.78 & 61.32 & 83.96 & 80.00 & 79.27 & 91.30 & 73.47 \\
 & \textbf{Pass@3} & 52.17 & 84.62 & 97.87 & 70.21 & 86.96 & 80.00 & 82.35 & 95.65 & 79.10 \\
\noalign{\vskip -0.25ex}\hhline{~----------}\noalign{\vskip 0.25ex}
Gemini 3 Flash & Run 1 & 43.48 & 50.00 & 74.47 & 50.98 & 56.51 & 73.33 & 80.94 & 60.87 & 58.90 \\
 & Run 2 & 39.00 & 50.00 & 89.36 & 59.55 & 65.21 & 86.67 & 82.27 & 69.57 & 65.13 \\
 & Run 3 & 45.49 & 53.85 & 87.23 & 55.23 & 65.21 & 86.67 & 69.34 & 60.87 & 63.80 \\
 & \textbf{Avg@3} & 42.66 & 51.28 & 83.69 & 55.25 & 62.31 & 82.22 & 77.52 & 63.77 & 62.61 \\
 & \textbf{Pass@3} & 50.00 & 57.69 & 100.00 & 68.09 & 65.22 & 86.67 & 82.35 & 78.26 & 72.54 \\
\noalign{\vskip -0.25ex}\hhline{~----------}\noalign{\vskip 0.25ex}
Qwen3.7-Plus & Run 1 & 54.35 & 42.31 & 40.43 & 59.49 & 73.90 & 40.00 & 57.87 & 52.17 & 52.37 \\
 & Run 2 & 58.56 & 53.85 & 44.68 & 44.67 & 65.22 & 60.00 & 69.77 & 43.48 & 52.78 \\
 & Run 3 & 58.56 & 53.85 & 46.81 & 51.09 & 65.21 & 60.00 & 64.63 & 43.48 & 54.07 \\
 & \textbf{Avg@3} & 57.16 & 50.00 & 43.97 & 51.75 & 68.11 & 53.33 & 64.09 & 46.38 & 53.08 \\
 & \textbf{Pass@3} & 67.39 & 69.23 & 63.83 & 68.09 & 82.61 & 60.00 & 82.35 & 52.17 & 67.62 \\
\noalign{\vskip -0.25ex}\hhline{~----------}\noalign{\vskip 0.25ex}
EvoCUA-32B & Run 1 & 45.65 & 73.08 & 53.19 & 57.36 & 73.90 & 93.33 & 70.04 & 47.83 & 59.78 \\
 & Run 2 & 45.65 & 76.92 & 55.32 & 59.49 & 69.56 & 73.33 & 45.30 & 52.17 & 58.06 \\
 & Run 3 & 45.65 & 76.92 & 55.32 & 61.61 & 69.56 & 80.00 & 58.75 & 47.83 & 59.40 \\
 & \textbf{Avg@3} & 45.65 & 75.64 & 54.61 & 59.49 & 71.00 & 82.22 & 58.03 & 49.28 & 59.08 \\
 & \textbf{Pass@3} & 50.00 & 84.62 & 61.70 & 70.21 & 73.91 & 93.33 & 76.47 & 60.87 & 67.62 \\
\noalign{\vskip -0.25ex}\hhline{~----------}\noalign{\vskip 0.25ex}
Kimi K2.5 & Run 1 & 32.47 & 53.85 & 17.02 & 29.70 & 30.42 & 40.00 & 33.43 & 52.17 & 33.43 \\
 & Run 2 & 45.65 & 65.38 & 36.17 & 50.98 & 47.82 & 66.67 & 47.06 & 56.52 & 49.57 \\
 & Run 3 & 45.52 & 76.92 & 34.04 & 55.23 & 47.82 & 53.33 & 41.18 & 56.52 & 49.96 \\
 & \textbf{Avg@3} & 41.21 & 65.38 & 29.08 & 45.30 & 42.02 & 53.33 & 40.56 & 55.07 & 44.32 \\
 & \textbf{Pass@3} & 58.70 & 92.31 & 55.32 & 65.96 & 47.83 & 73.33 & 58.82 & 65.22 & 63.52 \\
\noalign{\vskip -0.25ex}\hhline{~----------}\noalign{\vskip 0.25ex}
EvoCUA-8B & Run 1 & 39.13 & 80.77 & 29.79 & 31.83 & 43.47 & 80.00 & 52.86 & 65.22 & 46.70 \\
 & Run 2 & 45.65 & 69.23 & 29.79 & 31.90 & 43.47 & 53.33 & 41.18 & 43.48 & 42.21 \\
 & Run 3 & 47.74 & 69.23 & 25.53 & 33.95 & 26.09 & 60.00 & 17.65 & 47.83 & 39.72 \\
 & \textbf{Avg@3} & 44.17 & 73.08 & 28.37 & 32.56 & 37.67 & 64.44 & 37.23 & 52.17 & 42.88 \\
 & \textbf{Pass@3} & 54.35 & 88.46 & 44.68 & 42.55 & 52.17 & 80.00 & 58.82 & 65.22 & 56.56 \\
\noalign{\vskip -0.25ex}\hhline{~----------}\noalign{\vskip 0.25ex}
Qwen3-VL-235B-A22B-Thinking & Run 1 & 54.21 & 42.31 & 29.79 & 50.98 & 52.16 & 73.33 & 51.77 & 39.13 & 47.01 \\
 & Run 2 & 47.69 & 46.15 & 27.66 & 48.85 & 52.16 & 73.33 & 41.18 & 34.78 & 44.22 \\
 & Run 3 & 47.69 & 42.31 & 29.79 & 48.85 & 47.82 & 80.00 & 52.94 & 34.78 & 45.04 \\
 & \textbf{Avg@3} & 49.87 & 43.59 & 29.08 & 49.56 & 50.71 & 75.56 & 48.63 & 36.23 & 45.42 \\
 & \textbf{Pass@3} & 58.70 & 50.00 & 42.55 & 63.83 & 65.22 & 86.67 & 58.82 & 39.13 & 56.15 \\
\noalign{\vskip -0.25ex}\hhline{~----------}\noalign{\vskip 0.25ex}
DART-GUI-7B & Run 1 & 34.65 & 57.69 & 14.89 & 47.85 & 52.16 & 53.33 & 48.47 & 52.17 & 41.26 \\
 & Run 2 & 32.47 & 65.38 & 19.15 & 45.72 & 43.47 & 66.67 & 49.84 & 56.52 & 42.58 \\
 & Run 3 & 36.82 & 65.38 & 10.64 & 46.72 & 43.47 & 46.67 & 56.16 & 56.52 & 41.16 \\
 & \textbf{Avg@3} & 34.65 & 62.82 & 14.89 & 46.77 & 46.37 & 55.56 & 51.49 & 55.07 & 41.67 \\
 & \textbf{Pass@3} & 43.48 & 76.92 & 23.40 & 55.32 & 56.52 & 73.33 & 47.06 & 60.87 & 50.41 \\
\noalign{\vskip -0.25ex}\hhline{~----------}\noalign{\vskip 0.25ex}
OpenCUA-32B & Run 1 & 41.30 & 57.69 & 21.28 & 40.41 & 39.12 & 46.67 & 35.29 & 47.83 & 39.34 \\
 & Run 2 & 41.30 & 57.69 & 19.15 & 36.16 & 30.42 & 60.00 & 35.29 & 34.78 & 36.88 \\
 & Run 3 & 43.48 & 53.85 & 19.15 & 31.90 & 30.42 & 46.67 & 41.18 & 39.13 & 36.06 \\
 & \textbf{Avg@3} & 42.03 & 56.41 & 19.86 & 36.16 & 33.32 & 51.11 & 37.25 & 40.58 & 37.43 \\
 & \textbf{Pass@3} & 52.17 & 84.62 & 27.66 & 44.68 & 43.48 & 60.00 & 52.94 & 47.83 & 48.77 \\
\noalign{\vskip -0.25ex}\hhline{~----------}\noalign{\vskip 0.25ex}
Qwen3-VL-235B-A22B-Instruct & Run 1 & 45.52 & 19.23 & 12.77 & 36.08 & 56.51 & 73.33 & 40.56 & 34.78 & 35.98 \\
 & Run 2 & 43.34 & 46.15 & 27.66 & 31.83 & 56.51 & 73.33 & 23.53 & 34.78 & 39.30 \\
 & Run 3 & 32.47 & 38.46 & 23.40 & 33.95 & 52.16 & 66.67 & 35.29 & 30.43 & 35.61 \\
 & \textbf{Avg@3} & 40.45 & 34.62 & 21.28 & 33.95 & 55.06 & 71.11 & 33.13 & 33.33 & 36.96 \\
 & \textbf{Pass@3} & 47.83 & 50.00 & 36.17 & 44.68 & 56.52 & 86.67 & 47.06 & 34.78 & 47.13 \\
\noalign{\vskip -0.25ex}\hhline{~----------}\noalign{\vskip 0.25ex}
Qwen3-VL-32B-Instruct & Run 1 & 32.47 & 19.23 & 6.38 & 25.44 & 65.21 & 73.33 & 34.31 & 34.78 & 30.63 \\
 & Run 2 & 39.00 & 50.00 & 21.28 & 42.46 & 69.56 & 73.33 & 46.51 & 39.13 & 42.95 \\
 & Run 3 & 41.17 & 38.46 & 23.40 & 38.21 & 69.56 & 73.33 & 46.51 & 34.78 & 41.31 \\
 & \textbf{Avg@3} & 37.55 & 35.90 & 17.02 & 35.37 & 68.11 & 73.33 & 42.44 & 36.23 & 38.30 \\
 & \textbf{Pass@3} & 41.30 & 53.85 & 23.40 & 40.43 & 69.57 & 86.67 & 41.18 & 39.13 & 44.26 \\
\noalign{\vskip -0.25ex}\hhline{~----------}\noalign{\vskip 0.25ex}
OpenCUA-7B & Run 1 & 32.61 & 42.31 & 2.13 & 23.32 & 17.38 & 33.33 & 29.41 & 30.43 & 24.16 \\
 & Run 2 & 39.13 & 30.77 & 4.26 & 14.97 & 8.70 & 53.33 & 35.29 & 34.78 & 24.19 \\
 & Run 3 & 26.09 & 30.77 & 4.26 & 25.61 & 21.73 & 46.67 & 22.65 & 30.43 & 23.31 \\
 & \textbf{Avg@3} & 32.61 & 34.62 & 3.55 & 21.30 & 15.94 & 44.44 & 29.12 & 31.88 & 23.89 \\
 & \textbf{Pass@3} & 39.13 & 61.54 & 8.51 & 36.17 & 30.43 & 60.00 & 35.29 & 39.13 & 35.25 \\
\noalign{\vskip -0.25ex}\hhline{~----------}\noalign{\vskip 0.25ex}
UI-TARS-1.5-7B & Run 1 & 19.57 & 26.92 & 6.38 & 31.91 & 30.42 & 33.33 & 29.41 & 26.09 & 23.36 \\
 & Run 2 & 15.22 & 30.77 & 12.77 & 34.04 & 47.82 & 33.33 & 29.41 & 34.78 & 27.05 \\
 & Run 3 & 19.57 & 38.46 & 8.51 & 28.15 & 39.12 & 20.00 & 41.18 & 39.13 & 26.32 \\
 & \textbf{Avg@3} & 18.12 & 32.05 & 9.22 & 31.37 & 39.12 & 28.89 & 33.33 & 33.33 & 25.58 \\
 & \textbf{Pass@3} & 26.09 & 42.31 & 12.77 & 36.17 & 43.48 & 40.00 & 41.18 & 52.17 & 33.20 \\
\noalign{\vskip -0.25ex}\hhline{~----------}\noalign{\vskip 0.25ex}
Qwen3.5-9B Base & Run 1 & 23.91 & 38.46 & 12.77 & 27.66 & 43.47 & 46.67 & 35.29 & 26.09 & 28.28 \\
 & Run 2 & 26.09 & 34.62 & 21.28 & 31.83 & 26.08 & 46.67 & 17.65 & 30.43 & 28.26 \\
 & Run 3 & 15.22 & 42.31 & 8.51 & 36.08 & 52.16 & 46.67 & 29.41 & 30.43 & 28.67 \\
 & \textbf{Avg@3} & 21.74 & 38.46 & 14.18 & 31.86 & 40.57 & 46.67 & 27.45 & 28.99 & 28.40 \\
 & \textbf{Pass@3} & 28.26 & 53.85 & 27.66 & 48.94 & 56.52 & 66.67 & 47.06 & 30.43 & 41.39 \\
\noalign{\vskip -0.25ex}\hhline{~----------}\noalign{\vskip 0.25ex}
Qwen3.5-9B SFT (Ours) & Run 1 & 54.26 & 65.38 & 23.40 & 42.46 & 52.16 & 66.67 & 34.74 & 56.52 & 46.65 \\
 & Run 2 & 45.65 & 73.08 & 25.53 & 42.46 & 52.16 & 53.33 & 50.02 & 60.87 & 46.91 \\
 & Run 3 & 52.09 & 69.23 & 23.40 & 44.59 & 52.16 & 66.67 & 30.39 & 56.52 & 46.75 \\
 & \textbf{Avg@3} & 50.67 & 69.23 & 24.11 & 43.17 & 52.16 & 62.22 & 38.38 & 57.97 & 46.77 \\
 & \textbf{Pass@3} & 58.70 & 84.62 & 31.91 & 53.19 & 65.22 & 73.33 & 58.82 & 60.87 & 56.97 \\
\end{longtable}
\endgroup

\subsection{SoftwareGym-Bench}
We next report SoftwareGym-Bench. Each run covers 120 tasks, with ten tasks
from each of the twelve software categories. All models have complete
three-run records.

\begingroup
\scriptsize
\setlength{\tabcolsep}{2.78pt}
\setlength{\aboverulesep}{0pt}
\setlength{\belowrulesep}{0pt}
\renewcommand{\arraystretch}{0.92}
\begin{longtable}{@{}p{0.14\textwidth}c*{12}{c}>{\columncolor{overallcolumnbg}[2pt][0pt]}c@{}}
\caption{Run 1--3 scores, Avg@3, and Pass@3 (\%) on SoftwareGym-Bench.}
\label{tab:softwaregym-per-run-full}\\
\toprule
\textbf{Model} & Run & Pic & Mail & PDF & Music & Code & Forum & Writer & Browser & Video & Slides & Table & Chat & \textbf{Overall} \\
\midrule
\endfirsthead
\multicolumn{15}{c}{\tablename~\thetable\ (continued)}\\
\toprule
\textbf{Model} & Run & Pic & Mail & PDF & Music & Code & Forum & Writer & Browser & Video & Slides & Table & Chat & \textbf{Overall} \\
\midrule
\endhead
\bottomrule
\endfoot
GPT-5.4 & Run 1 & 60.00 & 80.00 & 80.00 & 40.00 & 80.00 & 60.00 & 30.00 & 70.00 & 70.00 & 50.00 & 60.00 & 70.00 & 62.50 \\
 & Run 2 & 80.00 & 80.00 & 70.00 & 40.00 & 70.00 & 60.00 & 50.00 & 90.00 & 60.00 & 50.00 & 40.00 & 80.00 & 64.17 \\
 & Run 3 & 70.00 & 90.00 & 90.00 & 50.00 & 70.00 & 60.00 & 30.00 & 80.00 & 50.00 & 50.00 & 60.00 & 70.00 & 64.17 \\
 & \textbf{Avg@3} & 70.00 & 83.33 & 80.00 & 43.33 & 73.33 & 60.00 & 36.67 & 80.00 & 60.00 & 50.00 & 53.33 & 73.33 & 63.61 \\
 & \textbf{Pass@3} & 80.00 & 90.00 & 90.00 & 50.00 & 80.00 & 80.00 & 50.00 & 90.00 & 70.00 & 60.00 & 70.00 & 80.00 & 74.17 \\
\noalign{\vskip -0.25ex}\hhline{~--------------}\noalign{\vskip 0.25ex}
GPT-5.5 & Run 1 & 70.00 & 90.00 & 90.00 & 50.00 & 60.00 & 70.00 & 40.00 & 80.00 & 50.00 & 40.00 & 70.00 & 70.00 & 65.00 \\
 & Run 2 & 70.00 & 60.00 & 90.00 & 40.00 & 70.00 & 60.00 & 40.00 & 90.00 & 50.00 & 40.00 & 60.00 & 70.00 & 61.67 \\
 & Run 3 & 60.00 & 90.00 & 80.00 & 40.00 & 50.00 & 60.00 & 20.00 & 80.00 & 80.00 & 20.00 & 70.00 & 70.00 & 60.00 \\
 & \textbf{Avg@3} & 66.67 & 80.00 & 86.67 & 43.33 & 60.00 & 63.33 & 33.33 & 83.33 & 60.00 & 33.33 & 66.67 & 70.00 & 62.22 \\
 & \textbf{Pass@3} & 70.00 & 90.00 & 100.00 & 50.00 & 90.00 & 70.00 & 60.00 & 90.00 & 80.00 & 40.00 & 70.00 & 70.00 & 73.33 \\
\noalign{\vskip -0.25ex}\hhline{~--------------}\noalign{\vskip 0.25ex}
Gemini 3.1 Pro & Run 1 & 60.00 & 100.00 & 80.00 & 40.00 & 70.00 & 90.00 & 50.00 & 40.00 & 70.00 & 40.00 & 30.00 & 70.00 & 61.67 \\
 & Run 2 & 80.00 & 80.00 & 90.00 & 30.00 & 80.00 & 70.00 & 30.00 & 50.00 & 60.00 & 20.00 & 50.00 & 70.00 & 59.17 \\
 & Run 3 & 70.00 & 90.00 & 80.00 & 50.00 & 70.00 & 60.00 & 10.00 & 40.00 & 60.00 & 20.00 & 40.00 & 70.00 & 55.00 \\
 & \textbf{Avg@3} & 70.00 & 90.00 & 83.33 & 40.00 & 73.33 & 73.33 & 30.00 & 43.33 & 63.33 & 26.67 & 40.00 & 70.00 & 58.61 \\
 & \textbf{Pass@3} & 80.00 & 100.00 & 90.00 & 50.00 & 80.00 & 90.00 & 60.00 & 70.00 & 70.00 & 40.00 & 50.00 & 70.00 & 70.83 \\
\noalign{\vskip -0.25ex}\hhline{~--------------}\noalign{\vskip 0.25ex}
Gemini 3 Flash & Run 1 & 40.00 & 80.00 & 50.00 & 40.00 & 80.00 & 60.00 & 30.00 & 80.00 & 40.00 & 30.00 & 40.00 & 50.00 & 51.67 \\
 & Run 2 & 60.00 & 90.00 & 50.00 & 40.00 & 60.00 & 60.00 & 10.00 & 100.00 & 60.00 & 30.00 & 40.00 & 70.00 & 55.83 \\
 & Run 3 & 80.00 & 70.00 & 60.00 & 40.00 & 70.00 & 50.00 & 40.00 & 80.00 & 50.00 & 30.00 & 50.00 & 70.00 & 57.50 \\
 & \textbf{Avg@3} & 60.00 & 80.00 & 53.33 & 40.00 & 70.00 & 56.67 & 26.67 & 86.67 & 50.00 & 30.00 & 43.33 & 63.33 & 55.00 \\
 & \textbf{Pass@3} & 90.00 & 90.00 & 70.00 & 40.00 & 80.00 & 80.00 & 60.00 & 100.00 & 60.00 & 40.00 & 50.00 & 70.00 & 69.17 \\
\noalign{\vskip -0.25ex}\hhline{~--------------}\noalign{\vskip 0.25ex}
Qwen3.8-Flash & Run 1 & 60.00 & 50.00 & 30.00 & 40.00 & 50.00 & 30.00 & 40.00 & 50.00 & 50.00 & 10.00 & 30.00 & 50.00 & 40.83 \\
 & Run 2 & 60.00 & 50.00 & 40.00 & 40.00 & 80.00 & 40.00 & 40.00 & 70.00 & 40.00 & 20.00 & 20.00 & 50.00 & 45.83 \\
 & Run 3 & 30.00 & 50.00 & 50.00 & 60.00 & 70.00 & 30.00 & 30.00 & 70.00 & 40.00 & 10.00 & 40.00 & 50.00 & 44.17 \\
 & \textbf{Avg@3} & 50.00 & 50.00 & 40.00 & 46.67 & 66.67 & 33.33 & 36.67 & 63.33 & 43.33 & 13.33 & 30.00 & 50.00 & 43.61 \\
 & \textbf{Pass@3} & 60.00 & 80.00 & 70.00 & 60.00 & 80.00 & 50.00 & 60.00 & 90.00 & 70.00 & 30.00 & 60.00 & 70.00 & 65.00 \\
\noalign{\vskip -0.25ex}\hhline{~--------------}\noalign{\vskip 0.25ex}
Kimi K2.5 & Run 1 & 50.00 & 50.00 & 50.00 & 40.00 & 50.00 & 30.00 & 30.00 & 50.00 & 60.00 & 10.00 & 10.00 & 70.00 & 41.67 \\
 & Run 2 & 40.00 & 70.00 & 80.00 & 30.00 & 70.00 & 40.00 & 30.00 & 60.00 & 40.00 & 30.00 & 20.00 & 40.00 & 45.83 \\
 & Run 3 & 50.00 & 50.00 & 80.00 & 40.00 & 70.00 & 40.00 & 0.00 & 70.00 & 50.00 & 20.00 & 20.00 & 50.00 & 45.00 \\
 & \textbf{Avg@3} & 46.67 & 56.67 & 70.00 & 36.67 & 63.33 & 36.67 & 20.00 & 60.00 & 50.00 & 20.00 & 16.67 & 53.33 & 44.17 \\
 & \textbf{Pass@3} & 70.00 & 90.00 & 90.00 & 40.00 & 80.00 & 60.00 & 50.00 & 70.00 & 60.00 & 30.00 & 40.00 & 70.00 & 62.50 \\
\noalign{\vskip -0.25ex}\hhline{~--------------}\noalign{\vskip 0.25ex}
Seed 2.0 Pro & Run 1 & 60.00 & 60.00 & 60.00 & 20.00 & 60.00 & 40.00 & 30.00 & 70.00 & 50.00 & 20.00 & 10.00 & 80.00 & 46.67 \\
 & Run 2 & 30.00 & 80.00 & 70.00 & 40.00 & 60.00 & 30.00 & 50.00 & 50.00 & 30.00 & 30.00 & 30.00 & 40.00 & 45.00 \\
 & Run 3 & 50.00 & 70.00 & 70.00 & 30.00 & 60.00 & 30.00 & 50.00 & 60.00 & 30.00 & 20.00 & 10.00 & 60.00 & 45.00 \\
 & \textbf{Avg@3} & 46.67 & 70.00 & 66.67 & 30.00 & 60.00 & 33.33 & 43.33 & 60.00 & 36.67 & 23.33 & 16.67 & 60.00 & 45.56 \\
 & \textbf{Pass@3} & 60.00 & 90.00 & 80.00 & 50.00 & 80.00 & 50.00 & 60.00 & 70.00 & 50.00 & 30.00 & 30.00 & 80.00 & 60.83 \\
\noalign{\vskip -0.25ex}\hhline{~--------------}\noalign{\vskip 0.25ex}
Seed 2.1 Pro & Run 1 & 40.00 & 40.00 & 50.00 & 50.00 & 60.00 & 40.00 & 30.00 & 50.00 & 30.00 & 30.00 & 30.00 & 40.00 & 40.83 \\
 & Run 2 & 50.00 & 60.00 & 60.00 & 30.00 & 50.00 & 50.00 & 30.00 & 60.00 & 20.00 & 20.00 & 10.00 & 40.00 & 40.00 \\
 & Run 3 & 40.00 & 60.00 & 60.00 & 40.00 & 40.00 & 60.00 & 20.00 & 50.00 & 50.00 & 30.00 & 20.00 & 60.00 & 44.17 \\
 & \textbf{Avg@3} & 43.33 & 53.33 & 56.67 & 40.00 & 50.00 & 50.00 & 26.67 & 53.33 & 33.33 & 26.67 & 20.00 & 46.67 & 41.67 \\
 & \textbf{Pass@3} & 50.00 & 80.00 & 80.00 & 50.00 & 70.00 & 70.00 & 50.00 & 70.00 & 50.00 & 50.00 & 30.00 & 70.00 & 60.00 \\
\noalign{\vskip -0.25ex}\hhline{~--------------}\noalign{\vskip 0.25ex}
Qwen3.5-9B (OSWorld + SoftwareGym) & Run 1 & 50.00 & 40.00 & 50.00 & 40.00 & 40.00 & 30.00 & 10.00 & 50.00 & 10.00 & 20.00 & 50.00 & 30.00 & 35.00 \\
 & Run 2 & 30.00 & 30.00 & 40.00 & 40.00 & 80.00 & 30.00 & 10.00 & 50.00 & 20.00 & 30.00 & 10.00 & 20.00 & 32.50 \\
 & Run 3 & 30.00 & 20.00 & 50.00 & 50.00 & 50.00 & 30.00 & 10.00 & 50.00 & 20.00 & 10.00 & 10.00 & 30.00 & 30.00 \\
 & \textbf{Avg@3} & 36.67 & 30.00 & 46.67 & 43.33 & 56.67 & 30.00 & 10.00 & 50.00 & 16.67 & 20.00 & 23.33 & 26.67 & 32.50 \\
 & \textbf{Pass@3} & 50.00 & 40.00 & 80.00 & 50.00 & 80.00 & 50.00 & 10.00 & 60.00 & 20.00 & 40.00 & 50.00 & 40.00 & 47.50 \\
\noalign{\vskip -0.25ex}\hhline{~--------------}\noalign{\vskip 0.25ex}
Qwen3.5-9B SFT (Ours) & Run 1 & 30.00 & 30.00 & 40.00 & 50.00 & 50.00 & 40.00 & 20.00 & 30.00 & 20.00 & 10.00 & 0.00 & 40.00 & 30.00 \\
 & Run 2 & 30.00 & 10.00 & 50.00 & 40.00 & 20.00 & 20.00 & 20.00 & 40.00 & 20.00 & 20.00 & 10.00 & 30.00 & 25.83 \\
 & Run 3 & 30.00 & 30.00 & 50.00 & 20.00 & 70.00 & 20.00 & 10.00 & 30.00 & 0.00 & 0.00 & 10.00 & 20.00 & 24.17 \\
 & \textbf{Avg@3} & 30.00 & 23.33 & 46.67 & 36.67 & 46.67 & 26.67 & 16.67 & 33.33 & 13.33 & 10.00 & 6.67 & 30.00 & 26.67 \\
 & \textbf{Pass@3} & 40.00 & 40.00 & 60.00 & 50.00 & 70.00 & 40.00 & 30.00 & 60.00 & 20.00 & 20.00 & 10.00 & 40.00 & 40.00 \\
\noalign{\vskip -0.25ex}\hhline{~--------------}\noalign{\vskip 0.25ex}
Qwen3.7-Plus & Run 1 & 30.00 & 10.00 & 30.00 & 30.00 & 40.00 & 20.00 & 0.00 & 20.00 & 30.00 & 20.00 & 10.00 & 30.00 & 22.50 \\
 & Run 2 & 20.00 & 20.00 & 30.00 & 30.00 & 50.00 & 20.00 & 20.00 & 30.00 & 30.00 & 10.00 & 20.00 & 30.00 & 25.83 \\
 & Run 3 & 30.00 & 20.00 & 40.00 & 20.00 & 30.00 & 40.00 & 30.00 & 10.00 & 30.00 & 30.00 & 10.00 & 20.00 & 25.83 \\
 & \textbf{Avg@3} & 26.67 & 16.67 & 33.33 & 26.67 & 40.00 & 26.67 & 16.67 & 20.00 & 30.00 & 20.00 & 13.33 & 26.67 & 24.72 \\
 & \textbf{Pass@3} & 30.00 & 40.00 & 50.00 & 40.00 & 60.00 & 40.00 & 40.00 & 30.00 & 30.00 & 30.00 & 20.00 & 50.00 & 38.33 \\
\noalign{\vskip -0.25ex}\hhline{~--------------}\noalign{\vskip 0.25ex}
EvoCUA-32B & Run 1 & 20.00 & 20.00 & 40.00 & 30.00 & 30.00 & 20.00 & 0.00 & 40.00 & 30.00 & 20.00 & 10.00 & 30.00 & 24.17 \\
 & Run 2 & 20.00 & 20.00 & 30.00 & 30.00 & 50.00 & 30.00 & 10.00 & 10.00 & 10.00 & 30.00 & 10.00 & 30.00 & 23.33 \\
 & Run 3 & 40.00 & 10.00 & 40.00 & 40.00 & 40.00 & 20.00 & 10.00 & 10.00 & 30.00 & 20.00 & 0.00 & 40.00 & 25.00 \\
 & \textbf{Avg@3} & 26.67 & 16.67 & 36.67 & 33.33 & 40.00 & 23.33 & 6.67 & 20.00 & 23.33 & 23.33 & 6.67 & 33.33 & 24.17 \\
 & \textbf{Pass@3} & 40.00 & 20.00 & 60.00 & 40.00 & 60.00 & 30.00 & 10.00 & 40.00 & 40.00 & 30.00 & 10.00 & 40.00 & 35.00 \\
\noalign{\vskip -0.25ex}\hhline{~--------------}\noalign{\vskip 0.25ex}
Qwen3.5-9B (OSWorld) & Run 1 & 20.00 & 20.00 & 50.00 & 30.00 & 30.00 & 20.00 & 0.00 & 10.00 & 20.00 & 10.00 & 0.00 & 20.00 & 19.17 \\
 & Run 2 & 10.00 & 10.00 & 60.00 & 30.00 & 20.00 & 10.00 & 0.00 & 20.00 & 20.00 & 20.00 & 10.00 & 10.00 & 18.33 \\
 & Run 3 & 10.00 & 30.00 & 50.00 & 20.00 & 40.00 & 20.00 & 20.00 & 30.00 & 10.00 & 20.00 & 10.00 & 40.00 & 25.00 \\
 & \textbf{Avg@3} & 13.33 & 20.00 & 53.33 & 26.67 & 30.00 & 16.67 & 6.67 & 20.00 & 16.67 & 16.67 & 6.67 & 23.33 & 20.83 \\
 & \textbf{Pass@3} & 20.00 & 30.00 & 60.00 & 30.00 & 50.00 & 30.00 & 20.00 & 30.00 & 40.00 & 30.00 & 20.00 & 40.00 & 33.33 \\
\noalign{\vskip -0.25ex}\hhline{~--------------}\noalign{\vskip 0.25ex}
Qwen3-VL-235B-A22B-Thinking & Run 1 & 20.00 & 40.00 & 20.00 & 30.00 & 30.00 & 30.00 & 0.00 & 10.00 & 30.00 & 0.00 & 20.00 & 20.00 & 20.83 \\
 & Run 2 & 10.00 & 10.00 & 40.00 & 30.00 & 0.00 & 30.00 & 10.00 & 20.00 & 20.00 & 0.00 & 10.00 & 20.00 & 16.67 \\
 & Run 3 & 30.00 & 30.00 & 20.00 & 20.00 & 10.00 & 30.00 & 10.00 & 20.00 & 20.00 & 10.00 & 10.00 & 30.00 & 20.00 \\
 & \textbf{Avg@3} & 20.00 & 26.67 & 26.67 & 26.67 & 13.33 & 30.00 & 6.67 & 16.67 & 23.33 & 3.33 & 13.33 & 23.33 & 19.17 \\
 & \textbf{Pass@3} & 30.00 & 40.00 & 40.00 & 30.00 & 40.00 & 50.00 & 20.00 & 20.00 & 30.00 & 10.00 & 30.00 & 40.00 & 31.67 \\
\noalign{\vskip -0.25ex}\hhline{~--------------}\noalign{\vskip 0.25ex}
Qwen3.5-9B Base & Run 1 & 20.00 & 10.00 & 20.00 & 20.00 & 10.00 & 20.00 & 20.00 & 10.00 & 10.00 & 20.00 & 20.00 & 10.00 & 15.83 \\
 & Run 2 & 10.00 & 10.00 & 30.00 & 10.00 & 10.00 & 30.00 & 0.00 & 10.00 & 0.00 & 0.00 & 10.00 & 30.00 & 12.50 \\
 & Run 3 & 20.00 & 30.00 & 20.00 & 30.00 & 10.00 & 10.00 & 0.00 & 10.00 & 10.00 & 10.00 & 10.00 & 20.00 & 15.00 \\
 & \textbf{Avg@3} & 16.67 & 16.67 & 23.33 & 20.00 & 10.00 & 20.00 & 6.67 & 10.00 & 6.67 & 10.00 & 13.33 & 20.00 & 14.44 \\
 & \textbf{Pass@3} & 20.00 & 30.00 & 40.00 & 30.00 & 10.00 & 40.00 & 20.00 & 10.00 & 20.00 & 30.00 & 20.00 & 40.00 & 25.83 \\
\noalign{\vskip -0.25ex}\hhline{~--------------}\noalign{\vskip 0.25ex}
EvoCUA-8B & Run 1 & 10.00 & 10.00 & 10.00 & 20.00 & 10.00 & 20.00 & 0.00 & 10.00 & 10.00 & 0.00 & 0.00 & 20.00 & 10.00 \\
 & Run 2 & 20.00 & 20.00 & 10.00 & 30.00 & 10.00 & 10.00 & 0.00 & 0.00 & 10.00 & 10.00 & 10.00 & 40.00 & 14.17 \\
 & Run 3 & 10.00 & 30.00 & 20.00 & 20.00 & 10.00 & 10.00 & 0.00 & 0.00 & 20.00 & 10.00 & 0.00 & 30.00 & 13.33 \\
 & \textbf{Avg@3} & 13.33 & 20.00 & 13.33 & 23.33 & 10.00 & 13.33 & 0.00 & 3.33 & 13.33 & 6.67 & 3.33 & 30.00 & 12.50 \\
 & \textbf{Pass@3} & 20.00 & 30.00 & 20.00 & 30.00 & 20.00 & 20.00 & 0.00 & 10.00 & 20.00 & 10.00 & 10.00 & 40.00 & 19.17 \\
\noalign{\vskip -0.25ex}\hhline{~--------------}\noalign{\vskip 0.25ex}
OpenCUA-32B & Run 1 & 10.00 & 20.00 & 30.00 & 10.00 & 0.00 & 10.00 & 0.00 & 0.00 & 10.00 & 0.00 & 10.00 & 10.00 & 9.17 \\
 & Run 2 & 10.00 & 20.00 & 30.00 & 0.00 & 20.00 & 0.00 & 0.00 & 10.00 & 10.00 & 0.00 & 0.00 & 30.00 & 10.83 \\
 & Run 3 & 20.00 & 0.00 & 30.00 & 10.00 & 0.00 & 0.00 & 0.00 & 0.00 & 10.00 & 10.00 & 0.00 & 30.00 & 9.17 \\
 & \textbf{Avg@3} & 13.33 & 13.33 & 30.00 & 6.67 & 6.67 & 3.33 & 0.00 & 3.33 & 10.00 & 3.33 & 3.33 & 23.33 & 9.72 \\
 & \textbf{Pass@3} & 20.00 & 20.00 & 40.00 & 10.00 & 20.00 & 10.00 & 0.00 & 10.00 & 20.00 & 10.00 & 10.00 & 40.00 & 17.50 \\
\noalign{\vskip -0.25ex}\hhline{~--------------}\noalign{\vskip 0.25ex}
OpenCUA-7B & Run 1 & 10.00 & 10.00 & 10.00 & 30.00 & 10.00 & 20.00 & 0.00 & 0.00 & 0.00 & 0.00 & 0.00 & 20.00 & 9.17 \\
 & Run 2 & 10.00 & 10.00 & 20.00 & 20.00 & 0.00 & 20.00 & 0.00 & 0.00 & 0.00 & 10.00 & 0.00 & 20.00 & 9.17 \\
 & Run 3 & 10.00 & 10.00 & 10.00 & 20.00 & 0.00 & 0.00 & 0.00 & 0.00 & 0.00 & 10.00 & 0.00 & 30.00 & 7.50 \\
 & \textbf{Avg@3} & 10.00 & 10.00 & 13.33 & 23.33 & 3.33 & 13.33 & 0.00 & 0.00 & 0.00 & 6.67 & 0.00 & 23.33 & 8.61 \\
 & \textbf{Pass@3} & 10.00 & 20.00 & 20.00 & 30.00 & 10.00 & 30.00 & 0.00 & 0.00 & 0.00 & 20.00 & 0.00 & 30.00 & 14.17 \\
\noalign{\vskip -0.25ex}\hhline{~--------------}\noalign{\vskip 0.25ex}
DART-GUI-7B & Run 1 & 10.00 & 0.00 & 30.00 & 10.00 & 10.00 & 10.00 & 0.00 & 0.00 & 10.00 & 0.00 & 0.00 & 10.00 & 7.50 \\
 & Run 2 & 20.00 & 10.00 & 20.00 & 20.00 & 10.00 & 10.00 & 0.00 & 0.00 & 10.00 & 0.00 & 0.00 & 10.00 & 9.17 \\
 & Run 3 & 10.00 & 0.00 & 10.00 & 20.00 & 20.00 & 0.00 & 0.00 & 0.00 & 10.00 & 10.00 & 0.00 & 10.00 & 7.50 \\
 & \textbf{Avg@3} & 13.33 & 3.33 & 20.00 & 16.67 & 13.33 & 6.67 & 0.00 & 0.00 & 10.00 & 3.33 & 0.00 & 10.00 & 8.06 \\
 & \textbf{Pass@3} & 20.00 & 10.00 & 30.00 & 30.00 & 20.00 & 10.00 & 0.00 & 0.00 & 10.00 & 10.00 & 0.00 & 10.00 & 12.50 \\
\noalign{\vskip -0.25ex}\hhline{~--------------}\noalign{\vskip 0.25ex}
Qwen3-VL-235B-A22B-Instruct & Run 1 & 10.00 & 10.00 & 30.00 & 0.00 & 0.00 & 20.00 & 0.00 & 10.00 & 0.00 & 0.00 & 0.00 & 0.00 & 6.67 \\
 & Run 2 & 10.00 & 10.00 & 20.00 & 20.00 & 0.00 & 20.00 & 0.00 & 0.00 & 0.00 & 0.00 & 0.00 & 0.00 & 6.67 \\
 & Run 3 & 20.00 & 10.00 & 10.00 & 10.00 & 10.00 & 10.00 & 0.00 & 0.00 & 0.00 & 10.00 & 0.00 & 0.00 & 6.67 \\
 & \textbf{Avg@3} & 13.33 & 10.00 & 20.00 & 10.00 & 3.33 & 16.67 & 0.00 & 3.33 & 0.00 & 3.33 & 0.00 & 0.00 & 6.67 \\
 & \textbf{Pass@3} & 20.00 & 10.00 & 30.00 & 20.00 & 10.00 & 30.00 & 0.00 & 10.00 & 0.00 & 10.00 & 0.00 & 0.00 & 11.67 \\
\noalign{\vskip -0.25ex}\hhline{~--------------}\noalign{\vskip 0.25ex}
Qwen3-VL-32B-Instruct & Run 1 & 10.00 & 10.00 & 10.00 & 0.00 & 20.00 & 0.00 & 0.00 & 0.00 & 20.00 & 0.00 & 0.00 & 40.00 & 9.17 \\
 & Run 2 & 10.00 & 10.00 & 10.00 & 0.00 & 20.00 & 0.00 & 0.00 & 10.00 & 20.00 & 10.00 & 0.00 & 30.00 & 10.00 \\
 & Run 3 & 10.00 & 10.00 & 10.00 & 0.00 & 20.00 & 0.00 & 0.00 & 0.00 & 20.00 & 0.00 & 0.00 & 30.00 & 8.33 \\
 & \textbf{Avg@3} & 10.00 & 10.00 & 10.00 & 0.00 & 20.00 & 0.00 & 0.00 & 3.33 & 20.00 & 3.33 & 0.00 & 33.33 & 9.17 \\
 & \textbf{Pass@3} & 10.00 & 10.00 & 10.00 & 0.00 & 20.00 & 0.00 & 0.00 & 10.00 & 20.00 & 10.00 & 0.00 & 40.00 & 10.83 \\
\noalign{\vskip -0.25ex}\hhline{~--------------}\noalign{\vskip 0.25ex}
UI-TARS-1.5-7B & Run 1 & 20.00 & 10.00 & 20.00 & 20.00 & 0.00 & 0.00 & 0.00 & 0.00 & 20.00 & 0.00 & 0.00 & 10.00 & 8.33 \\
 & Run 2 & 10.00 & 0.00 & 20.00 & 10.00 & 10.00 & 0.00 & 0.00 & 0.00 & 20.00 & 0.00 & 0.00 & 10.00 & 6.67 \\
 & Run 3 & 10.00 & 0.00 & 20.00 & 10.00 & 10.00 & 0.00 & 0.00 & 0.00 & 20.00 & 0.00 & 0.00 & 10.00 & 6.67 \\
 & \textbf{Avg@3} & 13.33 & 3.33 & 20.00 & 13.33 & 6.67 & 0.00 & 0.00 & 0.00 & 20.00 & 0.00 & 0.00 & 10.00 & 7.22 \\
 & \textbf{Pass@3} & 20.00 & 10.00 & 20.00 & 20.00 & 10.00 & 0.00 & 0.00 & 0.00 & 20.00 & 0.00 & 0.00 & 10.00 & 9.17 \\
\end{longtable}
\endgroup
\clearpage

\clearpage
\trajectoryqualityappendix

\end{document}